\documentclass{article}
\usepackage[centertags]{amsmath}
\usepackage{hyperref}
\usepackage{color}
\usepackage{amsfonts}
\usepackage{amssymb}
\usepackage{amsthm}
\usepackage{cite}
\usepackage{fullpage}
\usepackage{graphicx}
\usepackage{pgf}
\usepackage{graphics}
\usepackage{tikz}
\usetikzlibrary{angles, calc, quotes}
\usetikzlibrary{shapes.geometric}
\usepackage{rotating}
\usepackage{asymptote}
\usetikzlibrary{math} 
\usepackage{bm}\usepackage{authblk}
\usetikzlibrary{decorations.pathreplacing,calligraphy,backgrounds}
\usetikzlibrary{decorations.shapes}
\usetikzlibrary{patterns}
\usepackage{mathtools}
\usetikzlibrary{intersections,decorations.markings}

\numberwithin{equation}{section}

\newcommand{\ii}{{\rm i}}
\newcommand{\dd}{{\rm d}}

\newcommand\tikzhyperbola[6][thick]{
    \draw [#1, rotate around={#2: (0, 0)}, shift=#3]
        plot [variable = \t, samples=1000, domain=-#6:#6] ({#4 / cos( \t )}, {#5 * tan( \t )});
}

\newcommand\tikzhyperbolaPART[7][thick]{
    \draw [#1, rotate around={#2: (0, 0)}, shift=#3]
        plot [variable = \t, samples=1000, domain=-#6:#7] ({#4 / cos( \t )}, {#5 * tan( \t )});
}

\usetikzlibrary{shapes.misc}

\tikzset{cross/.style={cross out, draw=black, minimum size=2*(#1-\pgflinewidth), inner sep=0pt, outer sep=0pt},
cross/.default={2.2pt}}

\title{Zamolodchikov recurrence relation and modular properties of effective coupling in \texorpdfstring{$\mathcal{N}=2$}{N=2} SQCD}

\date{\vspace{-5ex}}
\author{}

\begin{document}

\maketitle


\vspace{6pt}
\begin{center}	
	{\textsl
	Aleksei Bykov$^{\dagger}$\footnote{\scriptsize \tt aleksei.bykov@alumni.uniroma2.eu}},Ekaterina Sysoeva$^{\ddagger}$\footnote{\scriptsize \tt sysoeva.caterina@gmail.com} \\
\vspace{1cm}
$\dagger$\textit{\small  Universit\`a degli Studi di Firenze, Dipartimento di Matematica e Informatica ``U. Dini'' \\  Via
S. Marta 3,  I-50139 Firenze, Italy. } \\ $\ddagger$\textit{\small Universit\`a di Torino, Dipartimento di Fisica, Via P. Giuria 1, I-10125 Torino, Italy \\ I.N.F.N.- sezione di Torino, Via P. Giuria 1, I-10125 Torino, Italy }
\vspace{6pt}
\end{center}

\begin{center}
\textbf{Abstract}
\end{center}

\vspace{4pt} {\small

 \noindent In this work, we present a recurrence relation for the instanton partition function of the $\mathcal{N}=2$ SYM $SU(N)$ gauge theory with $2N$ fundamental multiplets. The main  difficulty lies in determining the asymptotic behaviour of the partition function in the regime of large vacuum expectation values of the Higgs field. Using the saddle point method and the $qq$-characters technique, we demonstrate that, in this limit, the partition function is governed by the Quantum Seiberg-Witten curves, as in the Nekrasov-Shatashvili limit, up to a normalisation constant. With the asymptotic behaviour found, we are able to write the recurrence relation for the partition function and to find the effective infrared coupling constant. The resulting effective constant is an inverse of a modular function with respect to a certain triangle group, and the asymptotic itself is a product of modular functions and forms with respect to triangle groups.

}

\newpage

\tableofcontents

\section{Introduction}

Fifteen years ago Poghossian \cite{Poghossian2}, based on the Alday-Gaotto-Tachikawa (AGT) duality \cite{AGT}, translated the Zamolodchikov recurrence relation for the 4-point conformal blocks in 2d CFT \cite{Zamolodchikov} into the terms of the equivariant instanton partition function in $SU(2)$ gauge theory with 4 fundamental matter multiplets introduced by Nekrasov \cite{Nekrasov}. Since then, the recurrence relation for the instanton partition function was generalised for the gauge theories with other types of multiplets and higher rank gauge theories \cite{Poghossian2,Poghossian3,SysoevaBykov} and has proven to be a powerful tool for addressing computational difficulties in $\mathcal{N}=2$ SYM $SU(2)$ gauge theory \cite{Bershtein, Bonelli}.
\par However, to date, the named recurrence relation has not been written for the case of the $SU(N)$ gauge theory in the presence of $2N$ fundamental multiplets (with the exception of $N=2$ and $N=3$ cases \cite{Poghossian2,Poghossian3}) due to the difficulties arising in defining the asymptotic behaviour of the partition function in the limit of large vacuum expectation values of the scalar field $a_u$, $u=1,\ldots,N$. This choice of matter multiplets is interesting to study, since it realises one of the possible conformal $N=2$ supersymmetric four-dimensional gauge theories \cite{KohSuperconformalZoo}. Besides, the recurrence relation for the partition function in the theory with $2N$ fundamental multiplets can potentially be translated back to the language of the conformal blocks in the 2d CFT by the AGT duality. In the present work, we address the problem of determining the asymptotic behaviour of the instanton partition function and, consequently, deriving a recurrence relation for it.
\par The analysis of the asymptotic behaviour of the partition function bears certain similarities to the study of the non-equivariant limit in \cite{NekrasovOkunkov} and of the Nekrasov-Shatashvili limit in \cite{PoghossianQSW, Fucitoetal}, but also exhibits substantial differences.
\par As in those papers, the main difficulty is choosing an effective way to sum the contributions of infinitely many Young diagrams. Following the ideas used in \cite{NekrasovOkunkov, PoghossianQSW, Fucitoetal}, we encode the information about the Young diagrams into an auxiliary function $\omega(x)$, and then show that, in the saddle point approximation, it satisfies a certain functional equation, its periods are related to the vacuum parameters, and its asymptotic behaviour is determined by the effective number of instantons. The latter allows one to find the instanton partition function in the limit of large vacuum expectation values of the Higgs field. Surprisingly enough, this regime is described by the Quantum (or deformed) Seiberg-Witten curves, just like the Nekrasov-Shatashvili limit, up to a normalisation constant\footnote{Usually, in the literature, the term Quantum  Seiberg-Witten equation refers to a certain type of Baxter equation (see, e.g.,  \cite{Poploitov,Zenkevich,Ito}). We, however, use this term for a non-linear functional equation derived in \cite{PoghossianQSW, Fucitoetal}, which was shown to be equivalent, in a certain sense, to the mentioned Baxter equation.}. 
\par The main difference is that, contrary to the qualitative expectations, for $N>2$, the partition function is dominated by a contribution corresponding to a finite, rather than an asymptotically large, number of instantons. Consequently, for finite values of $\epsilon_{1,2}$, one cannot simply replace the Young diagrams with continuous shapes as in \cite{NekrasovOkunkov} or by tuples of real-length strips as in \cite{PoghossianQSW, Fucitoetal}. Instead, the saddle point method must be modified to account for the non-negligible discrete structure of the diagrams. Moreover, even the adjusted saddle point method does not recover the asymptotic behaviour completely for $N>3$. Nevertheless, the discrepancy between the asymptotic behaviour predicted by the saddle point method and that obtained from direct computations of the partition function is small. It reduces to a minor correction that is independent of the Higgs field vevs, the multiplet masses, and the equivariant parameters, and depends only on the bare coupling constant and the rank of the theory. The saddle point method remains relatively accurate, despite the fact that the dominant instanton number is not large, for two reasons.  Firstly, the saddle point equation coincides with the exact first $qq$-character equation \cite{qqchar} up to the correlators of certain observables. Secondly, with the higher $qq$-characters, one can show that those correlators are small in the limit we are considering. In particular, the second $qq$-character equation yields a relevant correction to the saddle point approximation.
\par 
The second important difference in our study comes from its practical orientation. To write the recurrence relation, we need an explicit expression for the asymptotic behaviour of the partition function and hence we actually have to find an approximate solution of the Quantum Seiberg-Witten equations. In order to do this, we had to analyse the convergence of its solutions to the classical Seiberg-Witten curve in the non-equivariant limit. The question turned out to be non-trivial and led to the appearance of an appendix where we study convergence of generalised continued fractions perturbed by a noise. It was written with the sole purpose to justify some technical claims we make in the main part of the paper, but we believe that this appendix can find broader applications in future. For example, we explain there the reason why one of the two ways to write a solution of the Quantum Seiberg-Witten equation as a continued fraction gives the desired solution, while the other does not, continuing the discussion of this issue started in \cite{PoghossianH}. 
\par
In our case the deformation parameter of the Quantum Seiberg-Witten equation is small, and can be treated perturbatively. The leading order is described by the usual algebraic SW curve, and only first two orders of the perturbation theory turn out to be relevant.  We show that in the region of interest the deformed Seiberg-Witten equation in each order of the perturbation theory reduces to a differential equation, which then can be further reduced to an algebraic one.
\par
Using these techniques we arrive to the central result of this paper
 \begin{eqnarray}  \label{eqn:asympt-corr-intro}
        Z\sim \left(\frac{q}{D q_{\mathrm{IR}}
        } \right)^{-\frac{w_1}{\epsilon_1\epsilon_2}}  f_2^{\frac{N}{4 \epsilon_1 \epsilon_2}\left(\tilde{T}_0^2-2\tilde{T}_1-\tilde{T}_0 \epsilon+\frac{\epsilon^2(N+1)}{6} \right)} (1- q)^{-\frac{1}{\epsilon_1 \epsilon_2}\left( \frac{N-1}{2N} \tilde{T}_0^2+\tilde{T}_0\epsilon-\tilde{T}_1-N \epsilon^2\right)} Z_c ,
    \end{eqnarray} \label{eqn:correction-intro}
    \begin{equation}
        Z_c =\frac{f_2^{\frac{N(N+1)}{24}}}{\prod_{l=1}^{N-1}(f_2^{(l)})^{\frac{1}{4}}}.
    \end{equation}
Here $w_1=\sum_{u<v}a_u a_v$, $    \tilde{T}_0= \sum_{i} m_i$, $\tilde{T}_1= \sum_{i<j} m_{i} m_{j}$, $f_2$ is a weight two modular form of a certain triangle group, $q_{\mathrm{IR}}$ is a function of the ultraviolet (or bare) coupling constant $q$ such that inverse to $\tau_{\rm IR}=\ln( q_{\rm IR})/2 \pi \ii$ is a modular function with respect to the same triangle group, and $D$ is a constant depending on the number of colours $N$ only, whose role is to cancel the perturbative one-loop contributions (see \ref{eqn:logD}). Finally, $Z_c$ is the mentioned above correction to the saddle point approximation and $f_2^{(l)}$ are weight two modular forms of triangle groups satisfying $f_2^{(1)}=f_2$, $f_2^{(l)}=f_2^{(N-l)}$. 
\par The striking feature of (\ref{eqn:asympt-corr-intro}) is that the $a_u$-dependent factor looks just like the classical contribution. Thus, it is natural to interpret $q_{\mathrm{IR}}$ as the effective infrared coupling constant, as it was done in the $N=2,3$ cases \cite{Poghossian2,Poghossian3}. This result seems to contradict the known fact that for $N\geq 4$ the effective coupling matrix is never proportional to the classical one \cite{MinahanNemeschansky}. In \cite{Lerda} the structure of the coupling matrix was studied in details for the so-called special vacuum, a regime of the theory closely related to the asymptotic we consider. Based on the analysis of the perturbative contribution and the action of the $S$-duality group, it was predicted that the coupling matrix can be parametrised by $\lfloor\frac{N}{2}\rfloor$ coupling constants. The constants were chosen in such a way that they transform independently under the action of the $S$-duality group. The analysis of the monodromy group of the coupling constant appearing in (\ref{eqn:asympt-corr-intro}) hints that it is one of the constants of \cite{Lerda}. 
 \par
\par To resolve this apparent contradiction, we have studied the relation between our asymptotic expansion and the coupling matrix. We show that the latter cannot be recovered from the former, since taking the derivatives with respect to vevs mixes the orders of the asymptotic expansion, and information of only the leading order of the asymptotic, in general, is not enough to reproduce the whole coupling matrix. On the other hand, we find that 
the effective infrared coupling constant $\tau_{\mathrm{IR}}$ is a first (rather than a second) derivative of the prepotential. In other words, it is more natural to interpret it as a combination of the dual variables $a_u^D$ than as a coupling constant. We show, however, that in the asymptotic regime it can be recovered from the usual coupling matrix via a contraction procedure. 
\par
With the asymptotic expansion (\ref{eqn:asympt-corr-intro}) and our previous results \cite{SysoevaBykov} at hand, we write down a Zamolodchikov-like recurrence relation. It presents the instanton partition function as a power series with respect to the effective coupling constant $q_{\mathrm{IR}}$. This indicates that this constant may have some meaning even beyond the special vacuum.

\vspace{10pt}

\par The paper is organised as follows:
\begin{itemize}

    \item In Section \ref{sec:instZ}, we define the theory we are working with and clarify our notation for the Nekrasov instanton partition function.

    \item In Section \ref{sec:asympt}, we describe our way to approach the limit of large vevs of the Higgs field, find the asymptotic behaviour of the instanton partition function in this limit and introduce the infrared effective coupling constant.

    \item In Section \ref{sec:comparison}, we compare the results found with saddle point approach with already known in literature and with direct computations for higher rank theory partition functions and find the discrepancy for $N \geq 4$.

    \item In Section \ref{sec:q-character-corr} we recognise the first $qq$-character equation in the QSW curve and find the missing factor in the asymptotics.

    \item In Section \ref{sec:recrel}, we present the recurrence relation and see that the instanton partition function can be resummed and written as a series in terms of the effective coupling constant.

    \item Section \ref{sec:concl} is a summary with a short discussion of the result.

    \item Appendix \ref{app-rec} is where we study the generalised continued fractions perturbed by noise and their convergence.

    \item In Appendix \ref{app-pertalg} we show that the quantum Seiberg-Witten equation can be approximated by an algebraic one in any order of the perturbation theory.
    
    \item Appendix \ref{app-comp} is a technical appendix with computations of appearing integrals.

\end{itemize}

\section{Instanton partition function} \label{sec:instZ}

We consider the $\mathcal{N}=2$ topologically twisted gauge theory with gauge group $SU(N)$ on $\mathbb{R}^4$ in the presence of $2N$ fundamental multiplets.
\par  The object of our interest is the instanton partition function of this theory derived in \cite{Nekrasov}. 
\par The partition function is presented as a sum over the instanton sectors
\begin{equation}  \label{zinstsum}
    Z=\sum_{k=0}^\infty q_0^k Z_k({\bf a}) ,
\end{equation}
where the components of the vector ${\bf a}=(a_1,\ldots,a_N)\in\mathbb{C}^N$ with $\sum_{u=1}^{N}a_u=0$ are the vacuum expectation values of the scalar field of the vector multiplet $q_0=e^{2\pi\ii \tau_0}$, and we refer both to $q_0$ and $\tau_0$ as the coupling constant.
\par In the presence of $2N$ fundamental hypermultiplets, the contribution of the $k$-sector can be written as an integral
\begin{eqnarray} \label{Zk}
        Z_k^{(2N,0)}({\bf a})&=&\frac{\epsilon^k}{(2 \pi {\rm i} \epsilon_1 \epsilon_2 )^k} \oint \prod_{i=1}^k {\rm d} \phi_i\frac{ \prod_{f=1}^{2N} (\phi_i-{m}_{f}) }{\prod_{u=1}^N \left[ (\phi_i-a_u)(a_u-\phi_i+\epsilon)\right]} \prod_{ j <i}\frac{\phi_{ij}^2(\phi_{ij}^2-\epsilon^2)}{(\phi_{ij}^2-\epsilon_1^2)(\phi_{ij}^2-\epsilon_2^2)} = \nonumber \\
        &=& \frac{1}{(2 \pi {\rm i})^k} \oint \prod_{i=1}^k {\rm d} \phi_i\frac{ \prod_{f=1}^{2N} (\phi_i-{m}_{f}) }{\prod_{u=1}^N \left[ (\phi_i-a_u)(a_u-\phi_i+\epsilon)\right]} {\prod_{ j, \, i}}'\frac{\phi_{ij}(\phi_{ij}+\epsilon)}{(\phi_{ij}+\epsilon_1)(\phi_{ij}+\epsilon_2)},
\end{eqnarray}
where $\epsilon_1$, $\epsilon_2$ characterise the non-trivial geometry of $\Omega$-background, $\epsilon=\epsilon_1+\epsilon_2$, $\phi_{ij}=\phi_{i}-\phi_{j}$ and the prime means that the zeroes in the product are omitted.
\par The poles of the integrand in (\ref{Zk}) located inside the integration contour are parametrized by $N$ Young diagrams $\vec{Y}=(Y_1, \, \ldots, \, Y_N)$ with the total number of boxes equal to the number of instantons $|\vec{Y}|= k$. The poles of the integral corresponding to $\vec{Y}$ are located at the points
\begin{equation} \label{statpoints}
    \phi^{\vec{Y}}_I=a_I-\epsilon_1(\alpha_I-1)-\epsilon_2(\beta_I-1),
\end{equation}
where $I$ labels a box belonging to one of the Young diagrams $Y_u \in \vec{Y}$, $(\alpha_I, \, \beta_I)$ are coordinates of the box $I$ in $Y_u$ and $a_I=a_u$.
\par Formally, we can turn a fundamental multiplet into an antifundamental one by redefining the mass and the sign of the parameter $q$.
\begin{eqnarray} \label{Zkfaf}
        Z_k^{(N,N)}({\bf a})&=&(-1)^{kN} Z_k^{(2N,0)}({\bf a})= \nonumber \\ &=&\frac{\epsilon^k}{(2 \pi {\rm i} \epsilon_1 \epsilon_2 )^k} \oint \prod_{i=1}^k {\rm d} \phi_i\frac{ \prod_{f=1}^{N} (-\phi_i+\epsilon+\overline{m}_{f})  \prod_{f=1}^{N} (\phi_i-{m}_{f}) }{\prod_{u=1}^N \left[ (\phi_i-a_u)(a_u-\phi_i+\epsilon)\right]} \prod_{ j <i}\frac{\phi_{ij}^2(\phi_{ij}^2-\epsilon^2)}{(\phi_{ij}^2-\epsilon_1^2)(\phi_{ij}^2-\epsilon_2^2)}
\end{eqnarray}
\begin{equation} \label{eqn:massredef}
    \overline{m}_f={m}_{f+N}-\epsilon, \, f=1,\ldots, N.
\end{equation}
\par The analysis of the asymptotic behaviour in the large $a$ limit simplifies for a specific choice of masses. Namely, if we assume
\begin{equation}\label{eqn:goodMass}
    \overline{m}_f=m_f, \, f=1,\ldots,N.
\end{equation}

\par With this assumption, we rewrite (\ref{Zkfaf}) as
\begin{equation} \label{ZkfafPQ}
        Z_k^{(2N,0)}({\bf a})= (-1)^{kN}\frac{\epsilon^k}{(2 \pi {\rm i} \epsilon_1 \epsilon_2 )^k} \oint \prod_{i=1}^k {\rm d} \phi_i \frac{ Q(\phi_i )Q(\phi_i -\epsilon)}{P_0(\phi_i )P_0(\phi_i -\epsilon)} \prod_{ j <i}\frac{\phi_{ij}^2(\phi_{ij}^2-\epsilon^2)}{(\phi_{ij}^2-\epsilon_1^2)(\phi_{ij}^2-\epsilon_2^2)},
\end{equation}
where
\begin{equation}
    P_0(x)=\prod_{u=1}^N(x-a_u),\ Q(x)=\prod_{i=1}^N(x-m_i).
\end{equation}
It is convenient to introduce the following notation for the coefficients of $P_0$ and $Q$:
\begin{equation} \label{symPoly}
    P_0(x)=x^N+x^{N-2}w_1-x^{N-3} w_2 +\ldots + (-1)^N w_{N-1}. 
\end{equation}
\begin{equation} \label{symQoly}
   Q(x)=x^N-T_0 x^{N-1}+T_1 x^{N-2}+\ldots+(-1)^N T_{N-1}.
\end{equation}
Here we take into account that $a_1+\ldots+a_N=0$, so there is no  term linear in $x$ in $P_0$. 
We recall that the coefficients $w_i$ and $T_i$ are the elementary symmetric polynomials of the variables $a_i$ and $m_i$ respectively:
\begin{align}
    w_0&= 0  &   T_0&= \sum_{i=1}^{N} m_i  \nonumber \\
     w_1&= \sum_{u_1<u_2} a_{u_1} a_{u_2}   &    T_1&= \sum_{i_1<i_2} m_{i_1} m_{i_2} \label{eqn:wTdefinition} \\
    w_2&= \sum_{u_1<u_2<u_3} a_{u_1} a_{u_2}a_{u_3}  &   T_2&= \sum_{i_1<i_2<i_3} m_{i_1} m_{i_2}m_{i_3}  \nonumber \\
    &\ldots   &  &\ldots \nonumber \\
     w_{N-1}&=a_1 a_2 \ldots a_N  &  T_{N-1}&=m_1 m_2 \ldots m_N \nonumber
\end{align}
\par We note that the variables $a_i$ (respectively, $m_i$) can be reconstructed from the coefficients $w_i$ (respectively, $T_i$) as roots of the polynomial $P_0$ (respectively, $Q$) up to a permutation. In other words, $w_i$ and $T_i$ can be understood as more natural arguments of the partition function, which automatically take into account the Weyl and flavour permutation symmetries. Such variables have also appeared in the context of AGT duality in \cite{Poghossian3}. 
\par 
The polynomial $Q$ we use differs from the one usually used in the literature. The latter, which we denote by $\tilde{Q}$, is related to ours as 
\begin{equation}\label{eqn:Qfactor}
    \tilde{Q}(x)=Q(x)Q(x-\epsilon).
\end{equation}
The assumption on the mass spectrum (\ref{eqn:goodMass}) was introduced in order to allow this factorisation. However, as we will see in Subsection \ref{subsec:arbmass}, this assumption is not necessary for the further derivation, and we start with it just to simplify computations.
\par For the sake of brevity, we denote $Z_k\triangleq Z_k^{(N,N)}=(-1)^{kN} Z_k^{(2N,0)}$ and we introduce a parameter $q$
\begin{equation} \label{eqn:newq}
    q \triangleq (-1)^N q_0 .
\end{equation}
We will compute
\begin{equation}
    Z\triangleq Z^{(2N,0)}=\sum_k q^k Z_k = \sum_k q_0^k Z_k^{(2N,0)}.
\end{equation}
To compare our results with the direct computations we will use the already integrated form of the partition function\footnote{In \cite{SysoevaBykov} we wrote a different sign for the masses of antifundamental multiplets in (\ref{eqn:ZkNak}). The current choice is consistent with (\ref{Zkfaf}).}
\begin{equation} \label{eqn:ZkNak}
          Z_k^{(2N-l,l)}=\sum_{\underset{|\vec{Y}|=k}{\vec{Y}}}\frac{\prod_{u,v=1}^{2N-l}Z_{\varnothing,Y_v}(m_u,a_v) \prod_{u,v=1}^{l} Z_{Y_u,\varnothing}(a_u,-m_v)}{\prod_{u,v=1}^{N}Z_{Y_u,Y_v}(a_u,a_v)},
\end{equation}
where
\begin{eqnarray}
    Z_{Y_u,Y_v}(a_u,a_v)= \prod_{(i,j)\in Y_u}(a_{vu}+\epsilon_1(i-\tilde{l}_{Y_v,j})-\epsilon_2(j-1-l_{Y_u,i})) \nonumber \\
     \prod_{(i,j)\in Y_v}(a_{vu}-\epsilon_1(i-1-\tilde{l}_{Y_u,j})+\epsilon_2(j-l_{Y_v,i})),
\end{eqnarray}
$a_{vu}=a_v-a_u$ and $l_{Y,i}$ is the length of the $i$-th row of diagram $Y$, $\tilde{l}_{Y,i}$ is the length of the $i$-th column of diagram $Y$.

\section{Asymptotic behaviour of instanton partition function at large vevs} \label{sec:asympt}

\subsection{How we approach infinity}
First of all, we define how exactly we approach infinity. As explained in \cite{SysoevaBykov}, writing the recurrence relation requires the knowledge of the behaviour of the partition function in the limit $w_{N-1}\to \infty$ with all the rest of $w_i$ kept fixed. 
\par It is convenient to introduce a large parameter $A$, so that
\begin{equation}    \label{limit}
    w_{N-1}=-(-A)^N, \quad A \rightarrow \infty.
\end{equation}
\par
Let us understand what this means in terms of the original variables $a_u$ by finding approximate solutions of the equation $P_0(x)=0$. The relevant in the limit $A\to \infty$ part of the corresponding equation is\footnote{This can be derived by the standard Newton asymptotic analysis or just by guessing the asymptotic behaviour of the roots $x\sim A$ and verifying that all $N$ roots can be found in this way.}
\begin{equation}
    x^N-A^N\approx 0.
\end{equation}
We see that the limit (\ref{limit}) corresponds to placing all $a_u$ at the vertices of a regular $N$-sided polygon and sending its diameter to
infinity,
\begin{equation}
    a_u \approx A \, e^{\frac{2 \pi \ii u}{N}} +\mathcal{O}\left( A^{-1}\right) .
\end{equation}
More precisely, using the perturbation theory up to the second order, we get
\begin{equation} \label{au}
     a_u = A \, e^{\frac{2 \pi \ii u}{N}}  -\frac{w_1}{N A}  e^{-\frac{2 \pi \ii u}{N}} +\mathcal{O}\left( A^{-2}\right).
\end{equation}
\par Note that (\ref{limit}) does not define $A$ uniquely. Taking another value of $A$ clearly corresponds to a shift of the enumeration of $a_u$, 
\begin{equation}\label{eqn:Atrans}
    A\to A \, e^{\frac{2\pi \ii d}{N}},\ a_u\to a_{u+d},
\end{equation} 
where $d=1,\ldots,N-1$ and we identify $a_u$ with $a_{u+N}$. This is nothing but the residual Weyl symmetry $\mathbb{Z}_n$, preserving the preferred enumeration of $a_u$ in the special vacuum.
From this, for example, one can immediately see that the next orders of (\ref{au}) are proportional to $A^{-k}e^{-\frac{2\pi \ii u k}{N}}$. In other words, the order in the expansion with respect to $A$ determines which Fourier harmonic with respect to $u$ can appear.
\par The purpose of this section is to find the asymptotic behaviour $Z^{(\infty)}$ defined as
\begin{equation*}
    Z=Z^{(\infty)}(1+\mathcal{O}(A^{-1})).
\end{equation*}
\par We can immediately note that since $Z^{(\infty)}$ cannot change due to the shift of enumeration, a stronger claim 
\begin{equation*}
    Z=Z^{(\infty)}(1+\mathcal{O}(w_{N-1}^{-1}))
\end{equation*}
holds automatically.


\subsection{Saddle point approximation}\label{subsec:sadptapp}

\subsubsection{Preliminaries}
In the limit $A\to \infty$, the contribution to the partition function of the $k$-instanton sector behaves as
\begin{equation} \label{Zksim}
    Z_k \sim A^{2k}.
\end{equation}
Indeed, the poles of the integrand (\ref{statpoints}) are located at $\overline{\phi}_I=a_u(1+\mathcal{O}(A^{-1}))\sim A$ for $I\in Y_u$. From this follows that, in the leading order, almost all brackets in the numerator of (\ref{Zkfaf}) are cancelled by the brackets in the denominator written below them. The only exception appears due to the fact that for $I\in Y_u$ the difference $\overline{\phi}_I-a_u$ does not contain the large parameter $A$. As there are precisely two such factors for each instanton, there is an uncompensated factor of $A^{2k}$ in the numerator.
\par 
As a consequence of (\ref{Zksim}) one may expect that the partition function is dominated by the contributions with a large number of instantons $k$. This would make our problem similar to the non-equivariant limit \cite{NekrasovOkunkov} and the Nekrasov-Shatashvili limit \cite{PoghossianQSW,Fucitoetal}, for which the asymptotic behaviour was found using the saddle point approximation. In this subsection we present an attempt to apply the same method to our problem. As we shall see in the end, 
the number of instantons of the dominant contributions is, contrary to the na\"ive argument above, \emph{not large}, which means that the saddle point approximation \emph{is not justified}. 
\par 
\par Nevertheless, the outcome of the saddle point analysis, taking the finiteness (and thus discreteness) of the Young diagrams into account, is relatively accurate. Let us discuss how this can be done.
\par
Firstly, we cannot replace the summation
over the Young diagrams by a functional integral over the profile functions as in \cite{NekrasovOkunkov}, or by an integral over the renormalised lengths of the  rows of the diagrams as in \cite{PoghossianQSW,Fucitoetal}. Instead, we work directly with the integral (\ref{ZkfafPQ}) over $\phi_i$. This is somewhat close to the method mentioned in Subsubsection 4.5.1 of \cite{NekrasovOkunkov}, where it was suggested to use the density of $\phi_i$ as a variable of integration, similarly to the theory of large random matrices.  However, in our case we work with finitely many variables $\phi_i$, so we cannot introduce a continuous density function. Moreover, we have to manually include the condition that the number of instantons $k$ is also, in a sense, stable. Due to the discrete nature of our problem, we cannot compute the derivatives with respect to $k$, and instead, we perform the `discrete variation', requiring $Z_k$ not to change when we add or remove one instanton. 
\par 
We emphasize that this `saddle point summation' is the trickiest step, which clearly does not work in many cases\footnote{It is enough to consider a summand which have a very sharp maximum that stays between two integer values of the summation index.}. We will justify the application of this approach for our setting in Section \ref{sec:q-character-corr}. 
\par  The straightforward application of the saddle point method to the computation of the Nekrasov partition function is always  complicated by finding the correct normalisation factor. To avoid the problem, one usually looks for logarithmic derivatives of the partition function instead, which can be physically interpreted as expectation values of certain `observables' and computed directly from the fixed point data. In the non-equivariant limit, the standard approach is to look for 
\begin{equation*}
    a_i^D=\epsilon_1\epsilon_2\frac{\partial \ln Z}{\partial a_i} ,
\end{equation*}
because it gives precisely the same description as in the Seiberg-Witten formalism \cite{SeibergWitten}. We prefer 
  to follow the approach of \cite{PoghossianQSW} and look for
\begin{equation} \label{Zviakeff}
    \frac{\partial \ln Z}{\partial \ln q}=\frac{\sum_{k=0}^\infty k \, q^k Z_k({\bf a})}{\sum_{k=0}^\infty q^k Z_k({\bf a})} = k_{\rm eff}(q) ,
\end{equation}
which we refer to as the effective number of instantons, since the expression in the middle can be naturally interpreted as the average number of instantons. As long as the saddle point approximation applies, the number of instantons at the saddle point coincides with the effective one in the leading order,
\begin{equation}
    k_0 = k_{\rm eff}+\mathcal{O}\left(A^{-1}\right).
\end{equation}
Together with the obvious condition $Z|_{q=0}=1$, the function $k_{\rm eff}(q)$ is enough to reconstruct the partition function. 

\par Therefore, our goal is to find the number of instantons in the saddle point configuration. Note that although originally the number of instantons $k$ is an integer, the average number of instantons, in general, is not. It should not come as a surprise and just means that the saddle point is somewhere between two instanton configurations. 
\paragraph{Remark} Formally, the limit $A\to \infty$ can be traded for the non-equivariant limit by a rescaling. Namely, for the obvious dimensional reasons, the partition function does not change if we substitute
\begin{eqnarray}\label{eqn:rescale}
    a_u\to a_u/A,\ m_f\to m_f/A,\ \epsilon_{1,2}\to \epsilon_{1,2}/A.
\end{eqnarray}
However, one has to be careful with this kind of argument, because, as it always happens with the limit of a dimensional parameter being small, we should always ask ourselves with respect to which fixed parameters they are small. For the analysis of Nekrasov and Okounkov to be applicable, $\epsilon_1\epsilon_2$ must be small in comparison with certain combinations of $a_u$, so that the effective instanton number would be large. But, as we shall see, in our case, the combination we should compare with is not $w_{N-1} \sim A^N$, but $w_1$, which under the rescaling (\ref{eqn:rescale}) becomes small,
\begin{equation}
    w_1\to w_1/A^2.
\end{equation}
So, although the rescaling by $A$ is indeed useful in what comes next, it does not simply reduce the $A\to \infty$ asymptotic to the non-equivariant limit

  \subsubsection{The energy function}
  
  \par We start by rewriting (\ref{ZkfafPQ}) as
  \begin{eqnarray} \label{Zkexp}
      q^k Z_k({\bf a})&=&  \oint  \,  {\rm exp} \left[- \mathcal{H}_k(\bm{\phi})\right]
      \prod_{i=1}^k {\rm d} \phi_i , \nonumber
  \end{eqnarray}
where
\begin{eqnarray} \label{action}
     &&-\mathcal{H}_k(\bm{\phi}) = k \ln\left(\frac{q\epsilon}{2\pi\ii \epsilon_1\epsilon_2}\right)+\sum_{i}\ln(Q(\phi_i)Q(\phi_i-\epsilon)) \\ &&-\ln(P_0(\phi_i)P_0(\phi_i-\epsilon))+{\sum_{i\neq j}} \ln \left(\frac{\phi_{ij}(\phi_{ij}+\epsilon)}{(\phi_{ij}+\epsilon_1)(\phi_{ij}+\epsilon_2)}\right) . \nonumber
\end{eqnarray}
This makes the problem formally equivalent to finding a partition function of a thermodynamic system with coordinates $\phi_i$ and an energy function $\mathcal{H}_k$. Since the number of coordinates is precisely the number of instantons, it is convenient to refer to $\phi_i$ as the positions of the instantons. For example, when we say that we add or remove an instanton at $z\in\mathbb{C}$ we mean that we add or remove the instanton $i$ with $\phi_i=z$. These coordinates are not related in any sense to the physical location of the instantons on the space $\mathbb{C}^2$ where the gauge theory lives, which is always the origin of $\mathbb{C}^2$.
\par
For future use, we introduce the function
\begin{equation}
    \mu(y)=\frac{(y+\epsilon_1/A)(y+\epsilon_2/A)}{y(y+\epsilon/A)}=\mu\left(-\frac{\epsilon}{A}-y\right).
\end{equation}
In the thermodynamic interpretation, it can be used to write the exponential of the energy of interaction of two instantons, one at $\phi_i$ and another at $\phi_j$, as
\begin{equation}
   \mu\left(\frac{\phi_{ij}}{A}\right)\mu\left(\frac{\phi_{ji}}{A}\right)= \mu\left(\frac{\phi_{ij}}{A}\right)\mu\left(\frac{\phi_{ij}}{A}-\frac{\epsilon}{A}\right).
\end{equation}
This allows us to write an effective energy of one instanton, say, the last one, in the field of all the others as
\begin{equation}\label{eqn:eff-energ}
    \exp\left(-\mathcal{E}_{k}(\phi_k,{\bm\phi'})\right)=\frac{ \exp\left(-\mathcal{H}_{k}({\bm \phi})\right)}{ \exp\left(-\mathcal{H}_{k-1}({\bm\phi'})\right)}=\frac{1}{2\pi\ii}\omega\left(\frac{\phi_k}{A}, \, {\bm \phi'}\right)\omega\left(\frac{\phi_k}{A}-\frac{\epsilon}{A}, \, {\bm \phi'}\right) \frac{\epsilon}{\epsilon_1\epsilon_2},
\end{equation}
where ${\bm \phi'}=(\phi_1,\ldots,\phi_{k-1})$ and
\begin{equation}\label{eqn:omegadef}
    \omega(y, \, {\bm \phi})= \sqrt{q}\frac{Q(Ay)}{P_0(Ay)}\prod_j \mu_j\left(y-\frac{\phi_j}{A}\right)^{-1}=\sqrt{q}\frac{Q(Ay)}{P_0(Ay)}\prod_j\frac{\left(y-\frac{\phi_j}{A}\right)\left(y-\frac{\phi_j-\epsilon}{A}\right)}{\left(y-\frac{\phi_j-\epsilon_1}{A}\right)\left(y-\frac{\phi_j-\epsilon_2}{A}\right)}.
\end{equation}
The function $\omega(\cdot,{\bm \phi})$  has a lot of zeroes and poles. One may worry if the effective energy (\ref{eqn:eff-energ}) is well-defined. It turns out that in the configurations of interest it is not. Indeed, we already know that (\ref{ZkfafPQ}) can be computed as a sum of residues at the poles ${\bm \phi}^{\vec{Y}}$ defined in (\ref{statpoints}). Then the saddle point should also be one of these poles, where the energy is minimal (and in fact equal to minus infinity), and the saddle point equation should determine which of the poles has the largest residue. So, the function of interest is the renormalised effective energy, defined as
\begin{eqnarray}\label{eqn:eff-enery-r-def}
    \exp\left(-\tilde{\mathcal{E}}_{k}(\phi_k,{\bm\phi'})\right)=2\pi\ii \frac{{\rm Res}_{{\bm \phi}={\bm \phi}^{\vec{Y}}} \exp\left(-\mathcal{H}_{k}({\bm \phi})\right)}{{\rm Res}_{{\bm \phi}'=
    {\bm \phi}^{\vec{Y}'}} \exp\left(-\mathcal{H}_{k-1}({\bm\phi'})\right)}=\frac{q^{|\vec{Y}|}Z_{\vec{Y}}}{q^{|\vec{Y}'|}Z_{\vec{Y}'}}\\
    \nonumber
    \left(\omega_{\vec{Y}'}(y)\omega_{\vec{Y}'}\left(y-\frac{\epsilon}{A}\right) \frac{\epsilon\left(A y-{\phi_k}\right)}{\epsilon_1\epsilon_2}\right)_{y\to\frac{\phi_k}{A}},
\end{eqnarray}
where 
\begin{equation}
    \omega_{\vec{Y}}(y)=\omega(y,{\bm \phi}^{\vec{Y}}),
\end{equation}
and $\vec{Y}'$ is the tuple $\vec{Y}$ with the $k$-th cell removed. We note that the latter operation makes sense only if $k$ is a convex corner of some of the diagrams in $\vec{Y}$.  It is convenient to recognize in the last factor the function $\mu$, and to rewrite the renormalised effective energy as
\begin{align}\label{eqn:eff-energ-r}
    \exp\left(-\tilde{\mathcal{E}}_{k}(\phi_k,{\bm\phi'})\right)=\left(\omega_{\vec{Y}'}(y)\omega_{\vec{Y}'}\left(y-\frac{\epsilon}{A}\right)\mu\left(y-\frac{\phi_k}{A}\right)^{-1}\right)_{y\to\frac{\phi_k}{A}}=\\
    \left(\omega_{\vec{Y}}(y)\omega_{\vec{Y}'}\left(y-\frac{\epsilon}{A}\right)\right)_{y\to\frac{\phi_k}{A}}=\left(\omega_{\vec{Y}}(y)\omega_{\vec{Y}}\left(y-\frac{\epsilon}{A}\right)
    \mu\left(y-\frac{\phi_k+\epsilon}{A}\right)
    \right)_{y\to\frac{\phi_k}{A}}. \nonumber
\end{align}
These three forms will be of crucial value at different points of the argument.
\par 
Before stating the saddle point equation, let us list some properties of the function $\omega_{\vec{Y}}$.
\par 
We begin by analysing its analytical structure and behaviour in the limit $A\to \infty$. In this limit, the prefactor behaves as
\begin{equation*}
    \frac{Q(Ay)}{P_0(Ay)}\sim \left (1-y^{-N}\right)^{-1} +\mathcal{O}\left(A^{-1}\right).
\end{equation*}
 and $|\phi^{\vec{Y}}_i/A| \to 1$, so 
 \begin{equation}\label{eqn:mu-limit}
     \mu\left(y-\frac{\phi^{\vec{Y}}_i}{A}\right)\to 1,
 \end{equation}
 unless $y$ is close to
 \begin{equation}\label{eqn:specpnts}
     \frac{\phi_i}{A},\frac{\phi_i-\epsilon_1}{A},\frac{\phi_i-\epsilon_2}{A},\frac{\phi_i-\epsilon}{A}\approx \frac{a_u}{A}\approx e^{\frac{2\pi\ii u}{N}}.
 \end{equation}
 \par 
 We conclude that $\omega_{\vec{Y}}$ is well behaved in the $A\to \infty$ limit, holomorphic except in a neighbourhood of $y=0$, and special points (\ref{eqn:specpnts}), where it can have a pole or a zero. The singular behaviour near $y=0$ comes from $Q(A y)$ in the denominator.
 \par
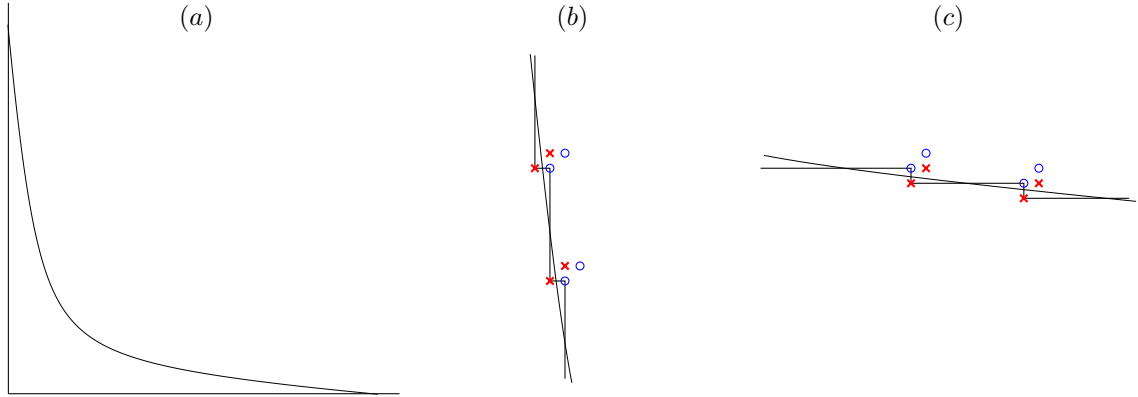
\begin{figure}
    \centering
    \begin{tikzpicture}

    \node[] at (2.5,5) {{$(a)$}};

        \draw (0.0,0.0) -- (5.2,0.0);
        \draw (0,0) -- (0,5.2);

        \coordinate (center) at (0.2,0.2);
        \def\angle{45}
        \def\bigaxis{1}
        \def\smallaxis{1.15}
        \tikzhyperbola[line width = 0.2pt, color=black]{\angle}{(center)}{\bigaxis}{\smallaxis}{71.7}

    \node[] at (7.5,5) {{$(b)$}};

        \draw[black] (7,4.5)--(7,3)--(7.2,3)--(7.2,1.5)--(7.4,1.5)--(7.4,0.2);

        \draw[blue] (7.2,3) circle (1.5pt); 
        \draw[blue] (7.4,3.2) circle (1.5pt);
        \draw[blue] (7.4,1.5) circle (1.5pt); 
        \draw[blue] (7.6,1.7) circle (1.5pt);

        \draw (7,3) node[cross,red,thick] {};
        \draw (7.2,3.2) node[cross,red,thick] {};
        \draw (7.2,1.5) node[cross,red,thick] {};
        \draw (7.4,1.7) node[cross,red,thick] {};

        \coordinate (centerb) at (7.47,-2.3);
        \def\angleb{45}
        \def\bigaxisb{1}
        \def\smallaxisb{1.2}
        \tikzhyperbolaPART[line width = 0.2pt, color=black]{\angleb}{(centerb)}{\bigaxisb}{\smallaxisb}{-55}{77}

    \node[] at (12.5,5) {{$(c)$}};

        \draw[black] (10,3)--(12,3)--(12,2.8)--(13.5,2.8)--(13.5,2.6)--(14.9,2.6);

        \draw[blue] (12.2,3.2) circle (1.5pt); 
        \draw[blue] (12,3) circle (1.5pt);
        \draw[blue] (13.5,2.8) circle (1.5pt); 
        \draw[blue] (13.7,3) circle (1.5pt);

        \draw (12,2.8) node[cross,red,thick] {};
        \draw (12.2,3) node[cross,red,thick] {};
        \draw (13.5,2.6) node[cross,red,thick] {};
        \draw (13.7,2.8) node[cross,red,thick] {};

        \coordinate (centerc) at (7.6,3.15);
        \def\anglec{45}
        \def\bigaxisc{1}
        \def\smallaxisc{1.2}
        \tikzhyperbolaPART[line width = 0.2pt, color=black]{\angleb}{(centerc)}{\bigaxisc}{\smallaxisc}{78}{-55}

    \end{tikzpicture}
	
     \caption{Boundary of a large Young diagram (a), and its zoomed pieces, with slope more (b) and less (c) than one. On the latter two pictures are shown the poles (red crosses) and zeroes (blue circles) of the function in the left-hand sides of (\ref{eqn:omegaomega}). } 
    
     \label{fig:zeroespoles}
 \end{figure}

 As for the poles and zeroes around $e^{\frac{2\pi \ii u}{N}}$, a more careful analysis, very similar to what was done in \cite{SysoevaBykov}\footnote{See also \cite{ABCD}}, shows that the poles of $\omega$ have the form
 \begin{equation} \label{eqn:poles}
     A^{-1}(a_u-\epsilon_1 (\alpha_I-1)-\epsilon_2(\beta_I-1)),
 \end{equation}
 where $I$ is a convex corner of $Y_u$, and the zeroes are
 \begin{equation} \label{eqn:zeroes}
     A^{-1}(a_u-\epsilon_1 (\alpha_I)-\epsilon_2(\beta_I)),
 \end{equation}
  where $I$ is a concave corner of $Y_u$. We note that this implies that the renormalised energy (\ref{eqn:eff-energ-r}), as we expect, has regular behaviour near $y=\phi_k/A$. Indeed, one can verify that 
\begin{equation}
    \omega_{\vec{Y}}(y)\omega_{\vec{Y}}\left(y-\frac{\epsilon}{A}\right)
\end{equation}
  has zeroes at $\overline{\phi}_k$, $\overline{\phi}_k-\epsilon/A$ and poles at $\overline{\phi}_k-\epsilon_1/A$, $\overline{\phi}_k+\epsilon_2/A$ (respectively,
 $\overline{\phi}_k+\epsilon_1/A$, $\overline{\phi}_k-\epsilon_2/A$), when the slope is smaller (respectively, larger) than one. This is schematically shown on Fig. \ref{fig:zeroespoles} (b, c).

  For the sake of definiteness, let us assume that the corner corresponding to $k$ lies on the part of curve with slope less than one. Then, near $y=\overline{\phi}_k$, 
\begin{equation}\label{eqn:not-so-const}
      \omega_{\vec{Y}}(y)\omega_{\vec{Y}}\left(y-\frac{\epsilon}{A}\right)\approx \Omega_k \frac{\left(y-\frac{\phi_k}{A}\right)\left(y-\frac{\phi_k-\epsilon}{A}\right) }{\left(y-\frac{\phi_k-\epsilon_1}{A}\right)\left(y-\frac{\phi_k+\epsilon_2}{A}\right)},
\end{equation}
where $\ln(\Omega_k)$ has regular beahvior near $y=\phi_k/A$.  
We note that the zeroes and poles of(\ref{eqn:not-so-const}) do not coincide exactly with the ones of the function $\mu\left(y-\frac{\phi_k}{A}\right)$, but when $y\to\phi_k/A$ we have
\begin{equation}
     \frac{\left(y-\frac{\phi_k}{A}\right)\left(y-\frac{\phi_k-\epsilon}{A}\right) }{\left(y-\frac{\phi_k-\epsilon_1}{A}\right)\left(y-\frac{\phi_k+\epsilon_2}{A}\right)}\approx -\left(y-\frac{\phi_k}{A}\right)\frac{\epsilon}{\epsilon_1\epsilon_2}\approx -\mu\left(y-\frac{\phi_k}{A}\right).
\end{equation}
We conclude that $-\tilde{E}_{k}(\phi_k,\phi_0)\approx \ln (\Omega_k)$ can be understood as a regularisation of (\ref{eqn:not-so-const}). We note that, just as $\mu$, the singular factor in (\ref{eqn:not-so-const}) tends to identity in the $A\to\infty$ limit.
\par
We are also interested in the asymptotic behaviour as $y\to \infty$. 
\begin{eqnarray} \label{omegaexp-phi}
    \omega(y, \, {\bm \phi})=\sqrt{q}\frac{1-\frac{T_0}{Ay}+\frac{T_1}{A^2y^2}}{1+\frac{w_1}{A^2y^2}}\left( 1-\frac{\epsilon_1\epsilon_1}{A^2y^2} k\right)+\mathcal{O}\left(\frac{1}{y^3}\right)= \nonumber \\ 
    \sqrt{q}\left(1-\frac{T_0}{A y}+\frac{T_1-w_1-k \, \epsilon_1 \epsilon_2}{A^2y^2} \right)+\mathcal{O}\left(\frac{1}{y^3}\right). 
\end{eqnarray}
Note that the number of instantons $k$ appears in the subsubleading term, so it can in principle be recovered from $\omega$.
\par 
Finally, we introduce the average over $\bm{\phi}$ (or, equivalently, over $\vec{Y}$), defined as
\begin{equation}\label{eqn:omega-avg}
    \omega(y)=\frac{1}{Z}\sum_k \oint  \,  {\rm exp} \left[ \mathcal{H}_k(\bm{\phi})\right]\omega(y,\bm{\phi})
      \prod_{i=1}^k {\rm d} \phi_i. 
\end{equation}
 \par Note that $\omega$ depends on the choice of $A$, so it is not invariant under the transformation (\ref{eqn:Atrans}). Instead, $\omega(y,\bm{\phi})$, $\omega(y)$ are invariant under simultaneous transformations
\begin{equation} \label{eqn:Atrans-omega}
    \begin{cases}
    A\to A e^{\frac{2\pi \ii d}{N}} ,    \\
    y \to y e^{-\frac{2\pi\ii d}{N}} .
    \end{cases}
\end{equation}
As usual, in the saddle point approximation, we can think of $\omega$ as $\omega(y,\overline{{\bm \phi}})$ with $\overline{\bm \phi}={\bm \phi}^{\overline{\vec{Y}}}$ being the saddle point value of ${\bm \phi}$, corresponding to the saddle point tuple of Young diagrams $\overline{\vec{Y}}$.
\par 
From the analysis above, we see that in the large $A$ limit we expect $\omega$ to be a holomorphic function with a singularity at $y\to 0$ due to the polynomial $Q(Ay)$ and some poles located near the points $e^{\frac{2\pi \ii  u}{N}}$. If, as our na\"ive estimate showed, the typical diagrams contributing to the average were large, the poles would not be located precisely at $e^{\frac{2\pi\ii u}{N}}$, as predicted by (\ref{eqn:specpnts}), because in general for some $I\in Y_{u}$, $\phi^{\vec{Y}}_I-a_u$ could be of order $A$. Together with the distance between the poles going to zero as $A^{-1}$, and existence of an $A$-independent asymptotic at a generic point, this would lead to the poles gluing into continuous cuts of finite length, as happens in the non-equivariant and Nekrasov-Shatashvili limits. As we shall see, in our regime this happens only for $N=2$. For larger $N$, the poles still stick together into a single singularity, but its size is of order of $A^{-1}$. For simplicity, we refer to this singularity as a cut anyway, even if it has zero length. We will denote the cut, located near (or at) $e^{\frac{2\pi\ii u}{N}}$ by $\mathcal{C}_u$.
\par
From (\ref{omegaexp-phi}) we see that in the $y\to \infty$ limit,
\begin{eqnarray} \label{omegaexp}
    \omega(y)=
    \sqrt{q}\left(1-\frac{T_0}{A y}+\frac{T_1-w_1-k_{\rm eff} \, \epsilon_1 \epsilon_2}{A^2y^2} \right)+\mathcal{O}\left(\frac{1}{y^3}\right). 
\end{eqnarray}
\paragraph{Remark.}
The function $\omega$ is a direct analogue of the function $w$ in \cite{PoghossianQSW, Fucitoetal}, and homonymous function in \cite{NekrasovOkunkov}. Moreover, even more similar function was introduced in \cite{qqchar}. We discuss the relation between the discrete saddle point method and the $qq$-characters used of \cite{qqchar} in Section \ref{sec:q-character-corr}.  
The notable difference in our definition is the rescaling $y=x/A$, natural for a study of the limit $A\to \infty$. 
\subsubsection{The saddle point equation}
To begin with, let us write the saddle point equation with respect to variations of $\phi_i$. It is given by
\begin{eqnarray} \label{saddlepoint}
    0&=&-\partial_{\phi_i}\mathcal{H}_k(\bm{\phi})=\frac{Q'(\phi_i)}{Q(\phi_i)}+\frac{Q'(\phi_i-\epsilon)}{Q(\phi_i-\epsilon)}-\frac{P_0'(\phi_i)}{P_0(\phi_i)}-\frac{P_0'(\phi_i-\epsilon)}{P_0(\phi_i-\epsilon)} \\
    &+&{\sum_j}'\frac{2}{\phi_{ij}}+{\sum_j}\left(\frac{1}{\phi_{ij}+\epsilon}+\frac{1}{\phi_{ij}-\epsilon}-\frac{1}{\phi_{ij}+\epsilon_1}-\frac{1}{\phi_{ij}-\epsilon_1}-\frac{1}{\phi_{ij}+\epsilon_2}-\frac{1}{\phi_{ij}-\epsilon_2}\right) . \nonumber
\end{eqnarray}
Here, the prime, analogously to (\ref{Zk}), means that the singular term is omitted. It is natural to expect that the saddle point equation can be rewritten via the effective energy of one instanton that we just have determined, and thus via the function $\omega$. Indeed, assuming that the average $\omega(y)$ is dominated by the saddle point, we have 
\begin{equation} \label{eqn:pre-omegavp}
    A \frac{\dd}{\dd y} \ln \left(\omega(y+\ii  \delta, \, {\bm \phi})\omega\left(y-\frac{\epsilon}{A}-\ii \delta, \, {\bm \phi}\right) \right) \bigg|_{\begin{smallmatrix} 
    \delta \to 0 \end{smallmatrix}}  =
    \mathcal{H}'(Ay)+{\rm v.p.} \, \frac{2}{Ay-\phi_i} 
\end{equation}
and the saddle point equation
(\ref{saddlepoint}) can be informally written as
\begin{equation} \label{omegavp}
    A \frac{\dd}{\dd y} \ln \left({\omega}(y+\ii  \delta){\omega}\left(y-\frac{\epsilon}{A}-\ii \delta\right) \right) \bigg|_{\begin{smallmatrix} Ay \to \phi_i & \\ \delta \to 0 \end{smallmatrix}} = {\rm v.p.} \, \frac{2}{Ay-\phi_i} \bigg|_{A y \to \phi_i} .
\end{equation}
This equation is somewhat confusing due to the singular expression in the right-hand side. To deal with it, let us pretend for now that the cuts $\mathcal{C}_u$ do not shrink to a point. We then claim that
\begin{equation}
     A \frac{\dd}{\dd y} \ln \left({\omega}(y+\ii  \delta){\omega}\left(y-\frac{\epsilon}{A}-\ii \delta\right) \right)=0
\end{equation}
on the cut $\mathcal{C}_u$. To motivate this, consider the integrals of the form
\begin{equation}\label{eqn:testfint}
     \int_{\mathcal{C}_u} f(y) A \frac{\dd}{\dd y} \ln \left({\omega}(y+\ii  \delta){\omega}\left(y-\frac{\epsilon}{A}-\ii \delta\right) \right) \dd y,
\end{equation}
where $f$ is a test function. In the limit $A\to \infty$, the cut $\mathcal{C}_u$ is densely filled by the points of the form $\phi_i/A$. Thus, for any given $\Delta>0$, by taking $A$ large enough, $f(y)$ can be approximated with any desired precision by functions of the form
\begin{eqnarray*}
    e^{-\frac{\left|y-\frac{\phi_i}{A}\right|^2}{\Delta^2}}.
\end{eqnarray*}
Then (\ref{eqn:pre-omegavp}) and (\ref{saddlepoint}) imply that (\ref{eqn:testfint}) can be made as small as we want, so it vanishes in the limit. 
\par We conclude that
 \begin{equation}\label{eqn:omegaomega}
    {\omega}(y+\ii  \delta){\omega}\left(y-\frac{\epsilon}{A}-\ii \delta\right)=c_u
\end{equation}
on each cut $\mathcal{C}_u$ for some, in general, $u$-dependent constant $c_u\in\mathbb{C}$.
 \par We now intend to show that $c_u=1$ for all values of $u$. Indeed, from (\ref{eqn:eff-energ-r}) we see that
 \begin{equation}\label{eqn:cuemu}
     c_u\approx  \exp\left(-
     \tilde{\mathcal{E}}_{k}(\phi_k,{\bm\phi'})\right) \mu\left(y-\frac{\phi_k+\epsilon}{A}\right)^{-1},
 \end{equation}
 where we assume that we have re-enumerated the instantons in such a way that $y$ is close to $\phi_k/A$. We recall that $\tilde{\mathcal{E}}_{k}$ is the effective energy of the $k$-th instanton. Again, we have a serious problem here, because the renormalised effective energy is expected to be regular, while $\mu$ has zeroes and poles. 
 By (\ref{eqn:mu-limit}), it is tempting to just remove the function $\mu$ in the vicinity of $\phi_k/A$. The only way out of this is to recall that (\ref{eqn:omegaomega}) was initially derived via the smearing procedure (\ref{eqn:testfint}), and thus, for finite values of $A$, it should again be smeared out by a smooth function to make sense. Then we can neglect the contribution of the poles and zeroes themselves and use 
 (\ref{eqn:cuemu}), concluding that the saddle point equation (\ref{eqn:omegaomega}) just means that the renormalised effective energy of the instanton is constant along the cut. We note that this is the natural condition for the `mechanical' stability of the instantons treated as a thermodynamic system. Moreover, in the true saddle point we should have
 \begin{equation}
      \tilde{\mathcal{E}}_{k}(\phi_k,{\bm\phi'})=0.
 \end{equation}
 Indeed, by definition, $ \tilde{\mathcal{E}}_{k}$ measures the energy that is added to the system when we add one instanton (or that is taken from the system, when we remove an instanton). In a saddle point configuration, this should be zero. We note that this argument is valid even if the cuts shrink to a point. Therefore, the saddle point equation in terms of the function $\omega(y)$ can be written as
\begin{equation} \label{onacut}
     {\omega}(y \pm \ii 0){\omega}\left(y-\frac{\epsilon}{A}\mp \ii 0\right)  = 1
\end{equation}

\vspace{1 ex}

\vspace{1 ex}

\noindent \textbf{Remark.} The largeness of $A$ was used only to show that $\mu(y)\to 1$. We note that $\mu(y)\to 1$ also holds in the non-equivariant and Nekrasov-Shatashvili limits. Moreover, $\mu(y)\to 1$ is necessary for  (\ref{eqn:cuemu}) to be compatible with (\ref{eqn:omegaomega}), so it seems that if any other regime treatable with the saddle point approximation is found in future, $\mu(y)\to 1$ will still hold in that regime and the saddle point equation will be (\ref{onacut}). In Section \ref{sec:q-character-corr}, we will see this equation once again. 

\subsection{Quantum Seiberg-Witten curve}

Consider a function
\begin{equation} \label{function}
    \omega\left(y-\frac{\epsilon}{A}\right) +\frac{1}{\omega(y)} .
\end{equation}
It is analytic on all the cuts $\mathcal{C}_u$ due to (\ref{onacut}), hence, all its possible singularities come from $Q(A y)$ appearing in $\omega(y)$. Therefore, we can write it as
\begin{equation}
    \omega\left(y-\frac{\epsilon}{A}\right) +\frac{1}{\omega(y)} = A^N\frac{g(y)}{Q(A y)}
\end{equation}
with $g(y)$ being a holomorphic function on $\mathbb{C}$. 
\par 
We observe that $g(y)$ is holomorphic everywhere and grows no faster than $y^N$ at $y \to \infty$, so it is a polynomial of degree $N$. Taking into account the asymptotic behaviour (\ref{omegaexp}), we write it as
\begin{equation} \label{structure}
    \ \omega\left(y-\frac{\epsilon}{A}\right) +\frac{1}{\omega(y)}  =  A^N\left(\sqrt{q}+\frac{1}{\sqrt{q}} \right)\frac{P_N(y) }{Q(A y)} .
\end{equation}
Here
\begin{equation}
    P_N(y)= y^N-p_0 y^{N-1}+p_1 y^{N-2}+\ldots+p_{N-1},
\end{equation}
and the first two coefficients are given by (\ref{omegaexp})
\begin{equation}\label{eqn:p0}
    p_0=\frac{2q T_0}{A(1+q)},
\end{equation}
\begin{equation}\label{eqn:p1-keff}
    p_1=\frac{q(T_0^2+2T_1-\epsilon T_0)+(w_1+k_{0} \epsilon_1 \epsilon_2)(1-q)}{A^2(1+q)}.
\end{equation}


\par In light of the remark in the previous subsection it is not surprising that (\ref{structure}), up to a rescaling, coincides with the deformation of Seiberg-Witten curves found in \cite{PoghossianQSW} in the Nekrasov-Shatashvili limit.
\par
Although we are interested in the $A\to \infty$ limit, we are not omitting the small $\epsilon/A$ term in (\ref{structure}). 
We shall see that, to get the leading order of the partition function, it is necessary  to solve (\ref{structure}) up to the subsubleading order.

\subsection{Periods of \texorpdfstring{$\omega(y)$}{w(y)}}\label{ssec:omegapers}

\par Taking into account that we are interested in the case of $a_u\gg m_f$ for each $u$ and $f$, we can choose contours $\mathcal{A}_u$  for $u=1,\ldots,N$ encircling the poles and zeroes of $\omega(y, {\bm \phi})$ around $a_u$ and leaving $m_f$ outside of it. Then
\begin{equation} \label{eqn:periods}
    \oint_{\mathcal{A}_u} y \frac{\dd \omega(y, {\bm \phi})}{\omega(y, {\bm \phi})}=   \frac{2 \pi \ii}{A} \left(\sum_j(\phi_j+\phi_j+\epsilon-\phi_j-\epsilon_1-\phi_j-\epsilon_2)-a_u\right) =-2\pi \ii \frac{a_u}{A}.
\end{equation}
The integrals in the left-hand side of (\ref{eqn:periods}) are called the periods of $\omega(y, {\bm \phi})$.
\par Since (\ref{eqn:periods}) is true for any value of $\bm{\phi}$, it also holds at the saddle point (or, in other words, it holds for the average $\omega(y)$ at the leading order of the fixed point approximation).
\begin{equation} \label{eqn:Au-and-au}
    \oint_{\mathcal{A}_u} y \frac{\dd \omega(y)}{\omega(y)}= -2\pi \ii \frac{a_u}{A}.
\end{equation}
\par With the deformed SW curve (\ref{structure}) and the constraints (\ref{eqn:Au-and-au}), we have everything we need to extract $k_0$, however, technically it is not easy. There are two tricks that will simplify the task for us.
\par Firstly, we can redefine
\begin{equation}\label{eqn:new-omega}
    \tilde{\omega}(y)=\omega(y)e^{\frac{\dd}{\dd y}\alpha(y)},
\end{equation}
where $\alpha(y)$ is holomorphic in some open region containing the contours $\mathcal{A}_u$, and $\alpha=\mathcal{O}(A^{-1})$. 
Then we have
\begin{eqnarray} \label{eqn:omegaredef}
\oint_{\mathcal{A}_u}y \frac{\dd \tilde{\omega}(y)}{\tilde{\omega}(y)}&=& \oint_{\mathcal{A}_u}y \, \dd \ln(\tilde{\omega}(y))=\oint_{\mathcal{A}_u}y \, \dd \ln(\omega(y))+\oint_{\mathcal{A}_u}y \alpha''(y) dy \\
   &=& \oint_{\mathcal{A}_u}y \frac{\dd {\omega}(y)}{{\omega}(y)}-\oint_{\mathcal{A}_u} \alpha'(y) dy=\oint_{\mathcal{A}_u}y \frac{\dd {\omega}(y)}{{\omega}(y)}. \nonumber
\end{eqnarray}
Here we use the fact that both $\alpha(y)$ and its derivative have no cuts intersecting the contour $\mathcal{A}_u$, so all boundary terms due to integration by parts disappear. We conclude that the change of variables (\ref{eqn:new-omega}) does not affect the periods (\ref{eqn:Au-and-au}) and we can use it to simplify (\ref{structure}). 
\par Secondly, we change variables in (\ref{eqn:omegaredef}). Namely, we take $z=\tilde{\omega}(y)$ as a new variable. Then $y$ becomes a function of $z$ defined by the equation
\begin{equation}\label{eqn:magic-period}
    \tilde{\omega}(y(z))=z.
\end{equation}
As we shall see, the multivalued function $y(z)$ in the region of interest has $N$ branches, which we denote by $y_u$ ($u=1,\ldots,N$), and all the contours $\mathcal{A}_u$ can be defined as images of the same contour $\mathcal{Z}$ under the functions $y_u$. We conclude that (\ref{eqn:Au-and-au}) is equivalent to
\begin{equation}
    \oint_{\mathcal{Z}}y_u(z) \frac{\dd z}{z}=-2\pi \ii \frac{a_u}{A}.
\end{equation}

\subsection{Applicability of the perturbation theory}
Let us introduce a function
\begin{equation}
    R(y)=A^N\left(\sqrt{q}+\frac{1}{\sqrt{q}} \right)\frac{P_N(y) }{Q(A y)}.
\end{equation}
Equation (\ref{structure}) now has the form
\begin{equation} \label{eqn:Rstructure}
    \omega\left(y-\frac{\epsilon}{A}\right) +\frac{1}{\omega(y)}  =  R(y) .
\end{equation}
To choose one of the many possible solutions of (\ref{eqn:Rstructure}) we need to add a boundary condition. This role is played by the asymptotic expansion (\ref{omegaexp}) from which we retain only the leading order
\begin{equation}\label{eqn:omegaexp-lead}
    \omega(y)=\sqrt{q}+\mathcal{O}(y^{-1}).
\end{equation}
\par 
At the leading order of the $A\to \infty$ limit, (\ref{eqn:Rstructure}) becomes just an algebraic equation, and it seems natural to treat this equation with the perturbation theory methods. For that, keeping in mind that $\omega$ is an analytic function, we write the Taylor series
\begin{equation}
    \omega\left(y-\frac{\epsilon}{A}\right)=\sum_{n=0}^{\infty}\frac{\omega^{(n)}(y)}{n!} \left(-\frac{\epsilon}{A}\right)^{n} ,
\end{equation}
turn (\ref{eqn:Rstructure}) into
\begin{equation} \label{eqn:structureTaylor}
    \omega(y)+\frac{1}{\omega(y)}+\sum_{n=1}^{\infty}\frac{\omega^{(n)}(y)}{n!} \left(-\frac{\epsilon}{A}\right)^{n}=R(y)
\end{equation}
and look for $\omega(y)$ in the form
\begin{equation} \label{eqn:omegapertseries}
    \omega(y)= \sum_{k=0}^\infty \omega_k (y) \left( \frac{\epsilon}{A} \right)^k.
\end{equation}
Then in the leading order, as expected, we get an algebraic equation\footnote{In this subsection we ignore the fact that $R$ itself depends on $A$ for simplicity. It does not affect the outcome.}
\begin{equation}\label{eqn:Rstructure-lead}
    \omega_0(y)+\frac{1}{\omega_0(y)}=R(y),
\end{equation}
and the corrections $\omega_k$ can be found order by order in the usual way.
\par
Although this approach looks natural, we should be careful not to lose solutions of (\ref{eqn:Rstructure})  when we omit the higher derivatives in (\ref{eqn:structureTaylor}). By doing this we have changed drastically the type of the equation: from a non-linear difference equation, rewritten as a differential equation of infinite order, we got a simple algebraic equation. To proceed in this manner we have to show that the solution $\omega(y)$ of (\ref{eqn:Rstructure}) selected by the asymptotic behaviour (\ref{eqn:omegaexp-lead}) can be approximated by some solution $\omega_0(y)$ of (\ref{eqn:Rstructure-lead}). Note that it requires to identify the appropriate branch of the multivalued function $\omega_0$, which includes choosing the cuts connecting the branching points among infinitely many options.
\par 

\par In Appendix \ref{app-rec} we study equations of the type (\ref{eqn:Rstructure}) 
with asymptotic condition (\ref{eqn:omegaexp-lead})
for slow varying functions $R(y)$. This applies to the case on hands, because the singularities of $R(y)$ are concentrated around the zeroes of $Q(yA)$, \textit{i.e.} near the origin and far away from the region of interest where the contours $\mathcal{A}_u$ are.
\par From Appendix \ref{app-rec} we know that the qualitative behaviour of the solution of (\ref{eqn:Rstructure}-\ref{eqn:omegaexp-lead}) depends on whether $|q|$ is bigger, smaller or equal to one. Since we start from the series in terms of positive powers of $q$ (\ref{zinstsum}), it is natural to limit ourselves to the case $|q|<1$. Then, by the results of the appendix, (\ref{eqn:Rstructure}-\ref{eqn:omegaexp-lead}) has a unique solution. Moreover,  there exist precisely one branch of $\omega_0(y)$ that can be chosen as the initial approximation of the perturbation theory for most of the values of the argument. Namely, the cuts of $\omega_0$ are defined by the equation $|\omega_0(y)|=1$, and, aside from the cuts, $\omega_0$ is uniquely defined by the requirement $|\omega_0(y)|<1$.
\par  
More precisely, the approximation
\begin{equation} \label{approximation}
        \omega(y)=\omega_0(y)+\mathcal{O}(A^{-1})
\end{equation}
may fail only in small neighbourhoods of singularities of $\omega_0$, which includes the already mentioned cuts, as well as the zeroes of $Q(Ay)$. Hypothetically, for some special choice of the coefficients $p_i$, there can exist other pathological points where (\ref{approximation}) is not valid, but the contours $\mathcal{A}_u$ can be chosen so that the contribution of such points to the periods is negligible. We refer to the end of Appendix \ref{app-rec}, and especially to discussion of Fig. \ref{fig:shadows} for further details.      

\subsection{Leading order and the contours}
We start with the leading order of (\ref{eqn:structureTaylor}). 
\begin{equation}\label{eqn:stucture-lead}
    \omega(y)+\frac{1}{\omega(y)}= \omega_0(y)+\frac{1}{\omega_0(y)} + \mathcal{O}(A^{-1})=\frac{P_N(y)}{y^N}\kappa+\mathcal{O}(A^{-1}),
\end{equation}
 where we expanded $Q(Ay)$ in the right-hand side and introduced $\kappa=\sqrt{q}+\frac{1}{\sqrt{q}}$.
 \par 
 By the symmetry property (\ref{eqn:Atrans-omega}), the $A$-independent leading order of $\omega$ should be invariant under the transformation of the argument $y\to y e^{\frac{2\pi\ii d}{N}}$. This implies that $P_N$ shares the same property,
 \par
  \begin{equation}
  P_N(y   e^{\frac{2\pi \ii}{N}})=P_N(y)+\mathcal{O}(A^{-1})    
  \end{equation}
  and hence
 \begin{equation} \label{eqn:PN}
     P_N(y)=y^N-U^N+\mathcal{O}(A^{-1})
 \end{equation}
for some $U\in \mathbb{C}$.

 \par Let us identify the cuts of $\omega_0(y)$. Since they are defined by the equation $|\omega_0(y)|=1$, we can parametrize them as curves $[0,2\pi]\to\mathbb{C}$, $\theta\mapsto y^{(0)}(e^{\ii \theta})$, where $y^{(0)}(z)$ is a solution of the equation
 \begin{equation}
     1-\frac{U^N}{y^{(0)}(z)^N}=\frac{\left(z+\frac{1}{z}\right)}{\kappa}.
 \end{equation}
The solution is an $N$-valued function. Its branches are
\begin{equation}\label{eqn:y-cut}
    y_u^{(0)}(z)=U \rho(z)^{-1/N}e^{\frac{2 \pi u \ii}{ N}},\quad \rho(z)=\left(1-\frac{\left(z+\frac{1}{z}\right)}{\kappa}\right),
\end{equation}
and $u=1,\ldots,N$. 
Here we assume that the branch of the $N$-th root is fixed (we take the standard branch for definiteness). 
\par
\begin{figure}
    \centering
   
    \begin{tikzpicture}
        \tikzmath{\n = 9; }

        \node[draw=none,minimum size=4cm,regular polygon,regular polygon sides={\n}] (b) {};
     
        \node[rotate=-90] at (0,0) {
             \begin{tikzpicture} 
                \foreach \x in {1,2,...,{\n}}
                    \draw[rotate=90+360*(\x-1)/\n,black] (b.corner \x) arc (0:360:0.5cm and 0.03cm) ;  
             \end{tikzpicture}
        };

        \coordinate[label=center: $\mathcal{C}_N$] (o) at (b.corner 8);

         \coordinate[label=center: $\mathcal{C}_1$] (o) at (b.corner 9);

        \coordinate[label=center: $\mathcal{C}_2$] (o) at (b.corner 1);

        \coordinate[label=center: $\mathcal{C}_3$] (o) at (b.corner 2);

        \coordinate[label=below: $\ldots$] (o) at (b.corner 3);

        \coordinate[label=center: $\mathcal{C}_{N-1}$] (o) at (b.corner 7);

        \coordinate[label=above: $\ldots$] (o) at (b.corner 6);
    \end{tikzpicture}
    
     \caption{Cuts of $\omega_0(y)$}     
    
     \label{fig:cuts}
\end{figure}
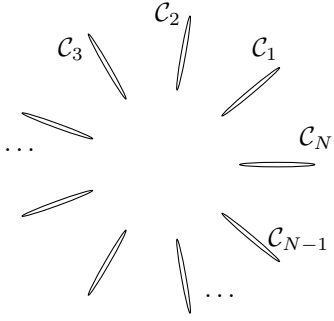

 We recall that we expect $\omega_0$ to have $N$ cuts $\mathcal{C}_u$, located near the edges of the $N$-polygon $a_u/A=e^{u\frac{2\pi\ii}{N}}$. Since the images of $y_u^{(0)}$ differ exactly by a rotation by $\frac{2\pi}{N}$, up to a choice of phase of $U$, we can identify the image of $y_u^{(0)}$ with $\mathcal{C}_u$ for each $u=1,\ldots,N$ (see Fig. \ref{fig:cuts}).

\par This identification is particularly simple in the limit $q\to 0$ (or equivalently $\kappa\to \infty$). In this case (\ref{eqn:y-cut}) becomes
 \begin{equation}
     y^{(0)}_u=U e^{\frac{2 \pi \ii u}{ N}}+\mathcal{O}(q).
 \end{equation}
So the $N$ cuts shrink to $N$ points on a $U$-circle. 
  \par On the other hand, with $q \to 0$ only the contribution with the instanton number equal to zero survives. From (\ref{eqn:omegadef}) we see that in this case $\omega_0$ has $N$ single poles at $y=a_u/A$, $u=1,\ldots,N$. We conclude that 
 \begin{equation}
    U=-1+\mathcal{O}(q),\ q\to 0.
 \end{equation}
 
 \par For finite values of $q$ we have to use (\ref{eqn:Au-and-au}) to find $U$, but the identification of the cuts $\mathcal{C}_u$ with the images of the unit circle under the branches $y_u$ persists. 
 \par The function $y_u(z)$ defined by (\ref{eqn:y-cut}) admits an analytical continuation to a neighbourhood of a unit circle, say for $|z|\in (r_0,r_1)$, where $0<r_0<1<r_1$. Indeed, it is enough to require $r_0,r_1$ be such that
 \begin{equation}\label{eqn:r0r1}
     \left|z+\frac{1}{z}\right|<\kappa,\, \forall z: r_0<|z|<r_1,
 \end{equation}
 which is possible for $|\kappa|>1$, i.e. for not too large $|q|$.
 Then, for $|z|\in (r_0,r_1)$ we have
 \begin{equation}
     \omega_0(y_u^{(0)}(z))+\frac{1}{\omega_0(y_u^{(0)}(z))}=R(y_0^{(0)}(z))=z+\frac{1}{z},
 \end{equation}
 since the function $y\to \omega_0(y)+\frac{1}{\omega_0(y)}$ is analytical in the neighbourhood of the cuts $\mathcal{C}_u$ of $y_u$. Then, since the branch of $\omega_0$ is selected by $|\omega_0|<1$, for $r_0<|z|\leq 1$ we get 
    \begin{equation}\label{eqn:omega-of-y}
         \omega_0(y_u^{(0)}(z))=z.
    \end{equation}
 Let us for each $r\in (r_0,1)$ consider the curves $\mathcal{A}_u^{r}$ defined by the parametrisation $\theta\mapsto y_u(r e^{\ii \theta})$. We claim that any of these curves can be taken as the   contour $\mathcal{A}_u$. Indeed, $\mathcal{A}_u^{r}$ is a family of closed curves without self-intersections, continuously depending on the parameter $r$. As long as $r<1$, $\mathcal{A}_u$ does not intersect $\mathcal{C}_u$, but in the limit $r\to 1$ we get a curve that goes twice (back and forth) along the cut $\mathcal{C}_u$. Finally, from (\ref{eqn:omega-of-y}) 
 \begin{equation}
     \lim_{r\to 1} \omega_0(y_u^{(0)}(r e^{\pm \ii \theta}))=e^{\pm \ii \theta},
 \end{equation}
 which means that the curves $\mathcal{A}_u^r$ with $r$ substantially close to one pass along both borders of the cut. Then by elementary geometric topology, for any $r\in (r_0,1)$,  $\mathcal{A}_u^r$ encircle the contour.  
 \par 
Putting $\mathcal{A}_u=\mathcal{A}_u^r$ into (\ref{eqn:Au-and-au}), we get
\begin{equation}\label{eqn:Au-lead}
    \oint_{\mathcal{Z}}y_u^{(0)}(z)\frac{dz}{z}=-2\pi \ii e^{\frac{2\pi \ii u}{N}},
\end{equation}
where 
\begin{equation}
    \mathcal{Z}=\big\{z\in\mathbb{C}\big||z|=r\big\},
\end{equation}
and we are free to choose any $r\in (r_0,1)$. In other words, at least at the leading order, (\ref{eqn:Au-and-au}) can indeed be presented in the desired form (\ref{eqn:magic-period}). Let us now explain why it holds beyond the leading order. 
Note that the condition (\ref{eqn:r0r1}) defining $r_0$ is independent of $A$. So, assuming $A$ is large enough, we can always choose $r$ so that the contour $\mathcal{A}_u$ stays within the region where the perturbation theory for $\omega(y)$ works (see Appendix \ref{app-rec} for details). In particular, we can chose $r_0'\in (r_0,1)$ so that $\mathcal{A}_u^r$ stays away from the singularities of $\omega(y)$ whenever $r_0<r<r'_0$. It follows that the contours $\mathcal{A}_u^r$ with $r\in (r_0,r'_0)$ can be used for the next orders of perturbation theory, because they still encircle the singularities of $\omega$. 
\par 
Define a function 
 $y_u$ satisfying 
\begin{equation}
    \tilde{\omega}(y_u(z))=z
\end{equation}
for $|z|\in (r_0',r_0)$. Since in this region $\omega_0$ approximates $\omega$ for large $A$,
\begin{equation}\label{eqn:y0y}
    y_u(z)=y_u^{(0)}(z)+\mathcal{O}(A^{-1}).
\end{equation}
Thus, the contours $\tilde{\mathcal{A}}^r_u$  defined by parametrisation $\theta\mapsto y_u(r e^{\ii \theta})$ are just small deformations of the contours ${\mathcal{A}}^r_u$, and hence enjoy the same topological properties. In particular, they can be used as the contours $\mathcal{A}_u$. This means that (\ref{eqn:magic-period}) holds at all orders of the perturbation theory and all we have to do is find perturbative corrections to (\ref{eqn:y0y}).

\subsection{Next orders of the saddle point equation} \label{subsec:next}
We will use the redefinition of $\omega(y)$ (\ref{eqn:omegaredef}) to simplify the expanded saddle point equation (\ref{eqn:structureTaylor}) order by order. Namely, we will choose $\alpha(y)$ in such a way that at each order of perturbation theory in terms of $1/A$ the highest derivative cancels.
\par At first order it can be done by setting
\begin{equation}\label{eqn:alpha-lead}
    \alpha(y)=\frac{\epsilon}{2A} \ln{(1-\tilde{\omega}(y)^2)}+\mathcal{O}(A^{-2}).
\end{equation}
Dependence of $\alpha$ on a yet unknown function $\tilde{\omega}$ is not important because the system of equations composed of (\ref{eqn:new-omega}) and (\ref{eqn:alpha-lead}) can be solved perturbatively order by order. The branch of the logarithm can be chosen to be the standard one, having a cut at the negative real semi-axis. Then $\alpha$ is holomorphic in the region where $\omega$ is. With (\ref{eqn:alpha-lead}), (\ref{eqn:structureTaylor}) becomes
\begin{equation}
    \tilde{\omega}(y)+\frac{1}{\tilde{\omega}(y)}=R(y)+\mathcal{O}(A^{-2}).
\end{equation}
In other words, up to a replacement of the function $\omega$ by $\tilde{\omega}$ according to (\ref{eqn:new-omega}) and (\ref{eqn:alpha-lead}), not affecting the periods,  the shift of the argument by $\epsilon/A$ in the left-hand sides of (\ref{eqn:Rstructure}) can be ignored not only at the leading, but also at the subleading order with respect to $\epsilon/A$.
\par At second order we set
\begin{equation}\label{eqn:alpha-sublead}
    \alpha(y)=\frac{\epsilon}{2A} \ln{(1-\tilde{\omega}(y)^2)}-\left(\frac{\epsilon}{A}\right)^2 \frac{\tilde{\omega}(y)(1+\tilde{\omega}(y)^2)\tilde{\omega}'(y)}{2(1-\tilde{\omega}_0(y)^2)^2} +\mathcal{O}(A^{-3}).
\end{equation}
This converts (\ref{eqn:structureTaylor}) into
\begin{equation}\label{eqn:Rstructure-mod-sq}
    \tilde{\omega}(y)+\frac{1}{\tilde{\omega}(y)}=R(y)+\left(\frac{\epsilon}{A}\right)^2 \frac{(1+\tilde{\omega}(y)^2)\tilde{\omega}'(y)^2}{2 \tilde{\omega}(y)(1-\tilde{\omega}(y)^2)^2} +\mathcal{O}(A^{-3}).
\end{equation}
Let us get rid of the derivative in the right-hand side. It appears with an $A^{-2}$ factor, so we need only its leading order. 
Differentiating (\ref{eqn:Rstructure-mod-sq}) and lowering the precision, we get
\begin{equation}
    -\tilde{\omega}'(y)\frac{1-\tilde{\omega}(y)^2}{\tilde{\omega}(y)^2}=R'(y)+\mathcal{O}(A^{-1}).
\end{equation}
Solving this for $\tilde{\omega}'(y)$ and putting it back to (\ref{eqn:Rstructure-mod-sq}), we arrive to an algebraic equation
\begin{equation}\label{eqn:Rstructure-mod-sq-alg}
    \tilde{\omega}(y)+\frac{1}{\tilde{\omega}(y)}=R(y)+\left(\frac{\epsilon}{A}\right)^2 \frac{(1+\tilde{\omega}(y)^2)\tilde{\omega}(y)^3 R'(y)^2}{2 (1-\tilde{\omega}(y)^2)^4} +\mathcal{O}(A^{-3}).
\end{equation}
Let us observe that if we were dealing with the original function $\omega$, we would still be able to get an algebraic equation (see Appendix \ref{app-pertalg} for a general procedure), but the computations would be much more involved, and we would need two subleading of the first  and leading order of the second derivative. 
\par Now we need to expand also $R(y)$. Recall that
\begin{equation}
    R(y)=A^N \kappa \frac{P_N(y)}{Q(Ay)}.
\end{equation}
By (\ref{symQoly}), we have
\begin{equation}
    Q(Ay)=A^N\left( y^N-\frac{T_0}{A y}+  \frac{T_1}{ A^2 y^2}+\mathcal{O}(A^{-3})\right).
\end{equation}

Using the symmetry property (\ref{eqn:Atrans-omega}), which is also shared by the polynomial $Q(Ay)$, we deduce that the coefficients of $P_N(y)$ are of order $p_n=\mathcal{O}(A^{-n-1})$ for $n=0,\ldots,N-2$, so 


\begin{equation}\label{eqn:P-Ainf}
    P_N(y)=y^N-y^{N-1}p_0+y^{N-2}p_1-U^N+\mathcal{O}(A^{-3}).
\end{equation}
Then
\begin{eqnarray}\label{eqn:R-upto3}
    R(y)&=&\kappa \left(1-\left(\frac{U}{y}\right)^N\right)-\frac{T_0}{\sqrt{q} A y}\left(\left(\frac{U}{y}\right)^N (1+q)+q-1\right)+
\\
   && \frac{1}{A^2 y^2\sqrt{q}}\left(B-C \left(\frac{U}{y}\right)^N \right)+\mathcal{O}(A^{-3}), \nonumber
\end{eqnarray}
where
\begin{eqnarray}
    B&=&T_0^2-T_1+q (T_1-T_0 \epsilon)+(1-q)(w_1+k_{0}\epsilon_1\epsilon_2), \\ C&=&(1+q)(T_0^2-T_1). \nonumber
\end{eqnarray}
We also have
\begin{equation}\label{eqn:Rprime}
    R'(y)=\kappa N U^N y^{-N-1}+\mathcal{O}(A^{-1}).
\end{equation}
Recall that we actually need not the function  $\tilde{\omega}$, but its inverse. In other words, we need to perturbatively solve
\begin{equation}\label{eqn:omega-yu}
    \tilde{\omega}(y_u(z))=z+\mathcal{O}(A^{-3}),\ \forall z: |z|\in (r_0,r_0'),
\end{equation}
where $0<r_0<r_0'<1$ are chose so that $|z|\in (r_0,r_0')$ the functions $y_u$ are holomorphic and close to $y_u^{(0)}$.

Combining (\ref{eqn:omega-yu}) with (\ref{eqn:Rstructure-mod-sq-alg}) and substituting (\ref{eqn:R-upto3}) and (\ref{eqn:Rprime}), we get

\begin{eqnarray}\label{eqn:ystructure}
  &&  \kappa \left(\left(\frac{U}{y_u(z)}\right)^N-\left(\frac{U}{y_u^{(0)}(z)}\right)^N\right)=
    -\frac{T_0}{\sqrt{q} A {y_u(z)}}\left(\left(\frac{U}{y_u(z)}\right)^N (1+q)+q-1\right)
\\
   &&  +\frac{1}{A^2 y_u(z)^2\sqrt{q}}\left(B-C \left(\frac{U}{y_u(z)}\right)^N \right)+
  \kappa^2 N^2 \left(\frac{\epsilon}{A}\right)^2 \frac{\left(z+\frac{1}{z}\right)z^4 U^{2N} y_u(z)^{-2N-2} }{2 (1-z^2)^3} +\mathcal{O}(A^{-3}). \nonumber
\end{eqnarray}
It is convenient to look for $y_u(z)$ in the form
\begin{equation}
    y_u(z)=y_u^{(0)}(z)+ \frac{\xi(z)}{A}+\frac{\zeta(z)}{A^2} \frac{1}{y_u^{(0)}(z)} + \mathcal{O}(A^{-3}).
\end{equation}
 Note that by (\ref{eqn:Atrans-omega}), $y_u$ is multiplied by a factor of $e^{-\frac{2\pi \ii d}{N}}$ under the shift of $u$ by $d$. The same transformation rule applies to the products $(y_{u}^{(0)})^k A^{k-1}$ for any $k$. This implies that the coefficients $\xi(z)$, $\zeta(z)$ are independent of $u$.
\par Before proceeding, we observe that for every $m$
\begin{equation}
    y_u(z)^m=y_u^{(0)}(z)^{m-1}\left(y_u^{(0)}(z)+ m\frac{\xi(z)}{A}+\frac{m\zeta(z)+\frac{m(m-1)}{2}\xi(z)^2}{A^2} \frac{1}{y_u^{(0)}(z)} + \mathcal{O}(A^{-3}))\right).
\end{equation}
Substituting this into (\ref{eqn:ystructure}), we get
\begin{eqnarray}
  &&  -N \kappa\rho(z)\frac{\xi(z)}{A y_u^{(0)}(z)}-\kappa\frac{N\rho(z)}{A^2 y_u^{(0)}(z)^2 }\left(\zeta(z)-\frac{N+1}{2}\xi(z)^2\right)=
 \\
   &&        - \frac{T_0}{\sqrt{q} A {y^{(0)}_u(z)}}\left(\rho(z) (1+q)+q-1\right)+ \frac{T_0}{\sqrt{q} A^2 {y^{(0)}_u(z)}^2}\xi(z)\left(\left(N+1\right) \rho(z) (1+q)+q-1\right)
\nonumber \\
 &&  +\frac{1}{A^2\sqrt{q} y_u^{(0)}(z)^2}\left(B-C\rho(z)\right)+
    \kappa^2 N^2 \frac{\epsilon^2}{A^2  y^{(0)}_u(z)^{2}} \frac{\left(z+\frac{1}{z}\right)\rho(z)^{2} }{2 \left(z-\frac{1}{z}\right)^4} \mathcal{O}(A^{-3})).
\nonumber
\end{eqnarray}
There are two types of terms, the ones  proportional to $\frac{1}{A y_u^{(0)}(z)}$, and the ones proportional to $\frac{1}{A^2 y_u^{(0)}(z)^2}$, with coefficients independent of the branch number $u$. Solving this first at the leading order (in terms of $A$) for $\xi$, and then in the subleading order for $\zeta$ we get,as expected, $u$-independent solutions
\begin{equation} \label{eqn:xi}
    \xi(z)=\frac{T_0}{N}\left(1-\lambda\rho(z)^{-1}\right),
\end{equation}
\begin{eqnarray*}
  \zeta(z)&=&\frac{N+1}{2}\xi(z)^2-\frac{T_0\xi(z)}{N}\left(N+1-\frac{1-q}{1+q}\rho(z)^{-1}\right) \\ &&- 
   \frac{1}{N\sqrt{q}\kappa\rho(z)}\left(B-C\rho(z)\right)-     \kappa N \epsilon^2 \frac{\left(z+\frac{1}{z}\right)\rho(z) }{\left(z-\frac{1}{z}\right)^4} . 
\end{eqnarray*}
 With (\ref{eqn:xi}) we can write the latter as 
\begin{equation} \label{eqn:zeta}
    \zeta(z)=\sum_{m=0}^2 \frac{L_m}{\rho(z)^m}-\kappa N \epsilon^2 \frac{\left(z+\frac{1}{z}\right)\rho(z) }{2 \left(z-\frac{1}{z}\right)^4} ,
\end{equation}
where the coefficients are
\begin{eqnarray} \label{eqn:L2}
    L_0&=&\frac{(N-1)T_0^2-2N T_1}{2N^2} ,\\
    L_1&=&-\frac{T_0(T_0-\epsilon)}{2N}-\lambda\frac{T_0^2(N-2)+N T_0 \epsilon +2N(k_0\epsilon_1\epsilon_2+w_1-T_1)}{2N^2},
\nonumber \\
    L_2&=&\left(1-\frac{4}{\kappa^2}\right)T_0^2 \frac{N-1}{2N^2}, \nonumber \\
    \lambda&=&\frac{1-q}{1+q}. \nonumber
\end{eqnarray}

\par A surprising consequence of the analysis above is that we ended up with an algebraic equation (\ref{eqn:Rstructure-mod-sq-alg}). In other words, the quantum Seiberg-Witten curve can be perturbatively approximated by an algebraic one. In fact, this holds for all orders of perturbation theory, and, moreover, the defining equation can be always reduced to a quadratic one in terms of $\omega(y)$.  Although we do not use it, we believe it to be an interesting observation on its own, so we demonstrate it in Appendix \ref{app-pertalg}.

\subsection{Effective number of instantons}\label{ssec:keff}
Now we have $N$ curves defined as 
\begin{eqnarray}
    \label{eqn:yu}
    y_u(z)&=&U \rho(z)^{-1/N}e^{\frac{2 \pi u \ii}{ N}}+ \frac{1}{A}\frac{T_0}{N}\left(1-\lambda\rho(z)^{-1}\right) \\
    && +\frac{1}{ A^2} U^{-1}\rho(z)^{1/N}e^{\frac{-2 \pi u \ii}{ N}}\left(\sum_{m=0}^2 \frac{L_m}{\rho(z)^m}-\kappa N \epsilon^2 \frac{\left(z+\frac{1}{z}\right)\rho(z) }{2 \left(z-\frac{1}{z}\right)^4}\right)  +\mathcal{O}(A^{-3})\nonumber
\end{eqnarray}
with
\begin{equation}
    \rho(z)=\left(1-\frac{\left(z+\frac{1}{z}\right)}{\kappa}\right) .
\end{equation}
\par We can proceed with the integration order by order over the contour as defined in (\ref{eqn:Au-lead}).
\begin{equation} \label{eqn:intonzplane}
    \oint_{|z|=r}y_u(z)\frac{dz}{z}=-2\pi \ii \frac{a_u}{A} =-2 \pi \ii \left(  e^{\frac{2 \pi \ii u}{N}}  -\frac{w_1}{N A^2}  e^{-\frac{2 \pi \ii u}{N}} +\mathcal{O}\left( A^{-3}\right) \right).
\end{equation}
\par We observe the perfect matching of the orders in terms of $A$ and dependence on $u$ in the left- and the right-hand sides: the leading orders always come with the exponent $ e^{\frac{2 \pi \ii u}{N}}$, the subleading order is present only in the left-hand side and is independent of $u$ and, finally, the subsubleading orders appear together with the factor  $e^{-\frac{2 \pi \ii u}{N}}$. 

\par Then, using notation and results of Appendix \ref{app-comp}, from the leading order of (\ref{eqn:intonzplane}) we get 
\begin{equation*}
    Ue^{\frac{2 \pi \ii u}{ N}} \mathcal{I}_{-\frac{1}{N}}(\kappa)=-e^{\frac{2\pi \ii u}{N}}+\mathcal{O}(A^{-1}),
\end{equation*}
or
\begin{equation}\label{eqn:U-solvec}
    U=-\frac{1}{{}_2 F_1(\frac{1}{2N},\frac{N+1}{2N},1;\frac{4}{\kappa^2})}+\mathcal{O}(A^{-1}).
\end{equation}
\par The constant $U$ computed in the leading order is enough for our goals (in fact,  due to the symmetry constraints $U$ can contain only terms proportional to $A^{-lN}$, $l\in \mathbb{Z}$, so (\ref{eqn:U-solvec}) actually holds up to $\mathcal{O}(A^{-N})$).
\par We can verify that the subleading order of (\ref{eqn:intonzplane}), present only in the left-hand side, vanishes. To this end, we write 
\begin{eqnarray}
&&\oint_{|z|=r}\rho(z)^{-1}\frac{dz}{z}=    \oint_{|z|=r}\frac{dz}{z\left(1-\kappa^{-1}\left(z+\frac{1}{z}\right)\right)}=\kappa   \oint_{|z|=r}\frac{dz}{z(\sqrt{q}+\frac{1}{\sqrt{q}}-z-\frac{1}{z})}=
\\
&&-\kappa   \oint_{|z|=r}\frac{dz}{(z-\sqrt{q})\left(z-\frac{1}{\sqrt{q}}\right)}=-2\pi \ii \frac{\sqrt{q}+\frac{1}{\sqrt{q}}}{\sqrt{q}-\frac{1}{\sqrt{q}}}
    =
    2\pi\ii\lambda^{-1}. \nonumber
\end{eqnarray}
Then 
\begin{equation}
    \oint_{|z|=r_0}\frac{dz}{z}=\frac{T_0}{N}    \oint_{|z|=r_0}\frac{dz}{z}\left(1-\lambda \rho(z)^{-1}\right)=0.
\end{equation}
\par Finally, the subsubleading order of (\ref{eqn:intonzplane}) gives us
\begin{equation}
   - \frac{1}{2\pi\ii }\oint_{|z|=r} \left(\sum_{m=0}^2 \frac{L_m}{\rho(z)^m}-\kappa N \epsilon^2 \frac{\left(z+\frac{1}{z}\right)\rho(z) }{2 \left(z-\frac{1}{z}\right)^4}\right)\rho^{\frac{1}{N}}(z)\frac{dz}{z U}=-\frac{w_1}{N} +\mathcal{O}(A^{-1}),
\end{equation}
or
\begin{eqnarray}
    w_1 U&=&N\sum_{m=0}^2 \mathcal{I}_{\frac{1}{N}-m}L_m-\frac{N^2\kappa \epsilon^2}{2} \mathcal{J}_{\frac{1}{N}+1}(\kappa) + \mathcal{O}(A^{-1}),
\end{eqnarray}
where $\mathcal{I_{\alpha}}$, $\mathcal{J_{\alpha}}(\kappa)$ are found in Appendix \ref{app-comp} and given by
   \begin{equation}
        \mathcal{I_{\alpha}}(\kappa)=\sum_{k=0}^{\infty}\frac{(-\alpha)_{2k}}{(k!)^2}\kappa^{-2k}={}_2 F_1\left(-\frac{\alpha}{2},\frac{1-\alpha}{2};1;\frac{4}{\kappa^2}\right) ,
    \end{equation}
    \begin{equation}
        \mathcal{J_{\alpha}}(\kappa)=
   -\frac{1}{6}\frac{(\alpha-2)_3}{\kappa^3} {}_2 F_1\left(\frac{3-\alpha}{2},\frac{4-\alpha}{2},2,\frac{4}{\kappa^2} \right).
    \end{equation}
Now we can extract $k_0$ hidden in $L_1$.
\begin{eqnarray} \label{eqn:k0}
        &&\lambda \epsilon_1 \epsilon_2 k_0= \left(-\lambda + \frac{1}{\mathcal{I}_{\frac{1}{N}-1}\mathcal{I}_{-\frac{1}{N}}} \right)w_1+\frac{1}{2}\frac{(N-1)}{N}\frac{\mathcal{I}_{\frac{1}{N}}+(1-\frac{4}{\kappa^2})\mathcal{I}_{\frac{1}{N}-2}}{\mathcal{I}_{\frac{1}{N}-1}}T_0^2\\
        &&-\frac{(N-2\lambda+N \lambda)}{2 N}T_0^2+\frac{(1-\lambda)}{2}T_0 \epsilon+\left(\lambda-\frac{\mathcal{I}_{\frac{1}{N}}}{\mathcal{I}_{\frac{1}{N}-1}} \right)T_1 - \frac{N^2 \kappa}{2}\frac{ \mathcal{J}_{\frac{1}{N}+1} }{\mathcal{I}_{\frac{1}{N}-1}}\epsilon^2 + \mathcal{O}(A^{-1}). \nonumber
\end{eqnarray}

\par We note that except for the $N=2$ case, when $w_1 \sim A^2$, there is no reason for $k_0$ to be large in the $A \to \infty$ limit for general $\epsilon_1$, $\epsilon_2$, $\lambda$.

\subsection{Renormalisation} \label{subsec:renormalisation}
It will be convenient for further derivations to introduce the effective coupling constant.
\par We will see later that due to the effective number of instantons (\ref{eqn:k0}) not being large in general, the asymptotic $\ln(Z^{(\infty)})$ given by the saddle point method requires a correction, and this correction depends solely on the bare coupling constant $q$, so the asymptotic can be written as
\par 
\begin{equation} \label{eqn:int}
\ln(Z^{(\infty)})=Z_c(q)\int \frac{k_{0}}{q}\dd q,
\end{equation}
choosing the integration constant so that $\ln(Z^{(\infty)})_{q=0}=0$.

\par From (\ref{eqn:k0}), (\ref{eqn:int}), it follows immediately that  $Z^{(\infty)}$ has the form
\begin{equation}
    Z^{(\infty)}=B(q)^{\frac{w_1}{\epsilon_1 \epsilon_2}} Z_0(q),
\end{equation}
where all the dependence on $\bm w$ appears in the exponent of the first factor, and $B(q)$, $Z_0(q)$ are $\bm w$-independent functions of $q$.
This form of the dependence on $\bm w$ coincides with the one we see in the classical part of the partition function
\begin{equation}  \label{eqn:zclass}
  Z_{\rm class}  = q_0^{-   \frac{1}{2N \epsilon_1 \epsilon_2}\sum_{u<v}(a_u-a_v)^2 }
  = q_0^{ \frac{w_1}{\epsilon_1 \epsilon_2} },
\end{equation}
where we have used
\begin{equation}
    0=\left(\sum_{u}a_u\right)^2=\sum_{u}a_u^2+2\sum_{u<v}a_u a_v.
\end{equation}
\par Moreover, for the one-loop contribution, starting from \cite{Nekrasov}
\begin{equation}
        \mathcal{F}_{\rm 1loop}=\frac{1}{2}\sum_{\substack{u,v=1 \\  u \neq v}}^N (a_u-a_v)^2 \ln \frac{a_u-a_v}{\Lambda} - N \sum_{u=1}^N a_u^2 \ln \frac{a_u}{\Lambda} ,
    \end{equation}
taking the variations and choosing branches of logarithms explicitly preserving residual Weyl symmetry, one can find the same dependence
\begin{equation}
    Z_{\rm 1-loop}^{(\infty)}=D_0^{-\frac{w_1}{\epsilon_1 \epsilon_2}} ,
\end{equation}
where
\begin{equation} \label{eqn:logD}
    D_0=(-1)^N D \, , \quad   \ln(D)=2 \ln 2 N -  \sum_{k=1}^{N-1} \cos\left(\frac{2 \pi   k}{N}\right)\ln\left(\sin^2\frac{\pi k}{N}\right)+\pi \ii.
\end{equation}

\par Let us define a new constant
    \begin{equation} \label{eqn:qir}
        q_{\rm IR}=q_0 \, D_0^{-1} \, B(q) 
    \end{equation}
and renormalise the contributions to the partition function as follows:
\begin{eqnarray}
        &&\overline{Z}({\bm w})=\frac{{Z}({\bm w})}{{Z}^{(\infty)}({\bm w})} , \label{eqn:Zrenorm} \\
        &&  \overline{Z}_{\rm 1-loop}({\bm w})=\frac{{Z}_{\rm 1-loop}({\bm w})}  {{Z}_{\rm 1-loop}^{(\infty)}({\bm w})}, \\
       && \overline{Z}_{\rm class}(\bm{w})=Z_{\rm class}(\bm{w}) Z^{(\infty)}_{\rm 1-loop}(\bm{w}) Z^{(\infty)}(\bm{w}). \label{eqn:Zclrenorm}
\end{eqnarray}
\par The full partition function does not change under this transformation
    \begin{equation}
        Z_{\mathrm{full}}={Z}_{\mathrm{class}}Z_{\mathrm{1-loop}}{Z}=\overline{Z}_{\mathrm{class}}\overline{Z}_{\mathrm{1-loop}}\overline{Z}.
    \end{equation}
 The classical part of the partition function now is a function of $q_{\mathrm{IR}}$
\begin{equation} \label{eqn:zclasbar}
    \overline{Z}_{\mathrm{class}}=q_{\mathrm{IR}}^{\frac{w_1}{\epsilon_1\epsilon_2}}Z_0 .
\end{equation}
 As we show in Subsection \ref{sec:recrel}, the renormalised instanton partition function $\overline{Z}({\bm w})$ can be naturally resummed into a series in terms of $q_{\mathrm{IR}}$, while the one loop part in the conformal case does not depend on $q$ at all.
\par We conclude that $q_{\mathrm{IR}}$ can be viewed as the effective coupling parameter of the theory. It is  referred to as the infrared coupling constant in \cite{Poghossian2,Poghossian3}. It might look controversial, since in no region in the space of expectation values $\bm{a}$ the full quantum prepotential is expected to be proportional to the classical action \cite{MinahanNemeschansky}. We resolve this apparent contradiction in Subsection \ref{subsec:effcouplN>=4}.
\par Besides the dependence on the new coupling $q_{\mathrm{IR}}$, the classical part $\overline{Z}_{\mathrm{class}}$ (\ref{eqn:zclasbar}) includes also a factor $Z_0$, which we understand as a constant shift of energy $\ln Z_0$ and call the vacuum energy.

\subsection{Modular properties of effective coupling and vacuum energy}


\subsubsection{Effective coupling constant}  \label{subsubsec:wdependasympt}

\par From (\ref{eqn:k0}),  (\ref{eqn:qir}) follows that a coupling constant
$\tau_{\mathrm{IR}}$, defined as $q_{\rm IR}=e^{2\pi \ii \tau_{\rm IR}}$, satisfies the equation
\begin{equation}\label{eqn:tauIRder}
    2\pi \ii q  \frac{ \partial \tau_{\mathrm{IR}}}{\partial q}
    =\frac{1}{\lambda \mathcal{I}_{-\frac{1}{N}}(\kappa)\mathcal{I}_{\frac{1}{N}-1}(\kappa)}=\frac{1+q}{1-q} \,{}_2F_1\left(\frac{1}{2N},\frac{N+1}{2N},1;\frac{4}{\kappa^2}\right)^{-2}\left(1-\frac{4}{\kappa^2}\right)^{\frac{N+2}{2N}},
\end{equation}
where we used Euler's transformation formula for hypergeometric functions.
\par Let us keep in mind that $\tau$ can be viewed as a function of a new variable
\begin{eqnarray}
    t=\frac{4}{\kappa^2}=\frac{4}{q+2+\frac{1}{q}}.
\end{eqnarray}
In terms of $t$, equation (\ref{eqn:tauIRder}) turns into
\begin{equation} \label{eqn:tauIRdert}
    t \frac{\dd \tau}{\dd t}={}_2F_1\left(\frac{1}{2N},\frac{N+1}{2N},1;t\right)^{-2}\left(1-t\right)^{\frac{2-N}{2N}}.
\end{equation}
We impose a boundary condition
\begin{equation}\label{eqn:tauD}
2\pi \ii \tau_{\mathrm{IR}}=\ln(q)-\ln(D)+\mathcal{O}(q)=\ln(t)-2\ln 2 -\ln(D)+\mathcal{O}(t),\quad q\to 0 \, (t\to 0).
\end{equation}
Here we used $D=(-1)^N D_0$ to remove the sign factor in $q_0$.

We expect the coupling constant $\tau_{\mathrm{IR}}$ to be an inverse modular function of $q(t)$ with respect to certain triangle groups \cite{Lerda}. Such functions can often be represented as ratios of two solutions of a hypergeometric equation \cite{Doran}. Thus, we can try to find  a solution in the form
\begin{equation}\label{eqn:tau-ans}
    2\pi \ii \tau_{\mathrm{IR}}=\frac{F_1}{F_2},
\end{equation}
where $F_{1,2}$ are two different solutions of the same unknown hypergeometric differential equation.
\par To justify the ansatz (\ref{eqn:tau-ans}), we recall that the hypergeometric equation
\begin{equation} \label{eqn:hyp-eqn}
    z(1-z)F''(z)+(c-(a+b+1)z)F'(z)-abF(z)=0
\end{equation}
is equivalent to 
\begin{equation}\label{eqn:hypeqn-G}
    G''(z)+Q(a,b,c;z)G(z)=0,
\end{equation}
where
\begin{equation}
    G(z)=z^{c/2}(1-z)^{\frac{c-a-b-1}{2}}F(z)
\end{equation}
and
\begin{equation}
    Q(a,b,c;z)=\frac{q^2(1-(a-b)^2)+q(2c(a+b-1)-4 a b)+c(2-c)}{4q^2(1-q)^2}.
\end{equation}
\par The ansatz (\ref{eqn:tau-ans}) is equivalent to
\begin{equation}\label{eqn:tau-ans-G}
    2\pi \ii \tau_{\mathrm{IR}}=\frac{G_1}{G_2},
\end{equation}
where $G_1$, $G_2$ are two solutions of (\ref{eqn:hypeqn-G}).
\par It is known \cite{Hille} that a function can be represented in the form (\ref{eqn:tau-ans-G}) if and only if it satisfies
\begin{equation}
    S\tau_{\mathrm{IR}}=2Q(a,b,c;z),
\end{equation}
where the Schwarzian derivative operator $S$ is defined by
\begin{equation}
S\tau_{\mathrm{IR}}=\frac{\tau_{\mathrm{IR}}{}'''}{\tau_{\mathrm{IR}}{}'}-\frac{3}{2}\left(\frac{\tau_{\mathrm{IR}}{}''}{\tau_{\mathrm{IR}}{}'}\right)^2.
\end{equation}
\par Note that the Schwarzian derivative depends only on the derivatives of the function, so we can find it using (\ref{eqn:tauIRder}) or (\ref{eqn:tauIRdert}) without  solving them. It is convenient to rewrite the hypergeometric function in (\ref{eqn:tauIRder})
\begin{equation}
    {}_2F_1\left(\frac{1}{2N},\frac{N+1}{2N},1;t\right)=G_0\left(t\right)\left(t\right)^{-\frac{1}{2}}\left(1-t\right)^{-\frac{1}{2}+\frac{N+2}{4N}}, 
\end{equation}
where $G_0$ satisfies the equation
\begin{equation}
    G_0''(t)+Q\left(\frac{1}{2N},\frac{N+1}{2N},1;t\right)G_0(t)=0.
\end{equation}
With this, (\ref{eqn:tauIRdert}) turns into
\begin{equation}
    \frac{\dd \tau}{\dd t}=G_0(t)^{-2}
\end{equation}
and it is easy to show that the Schwarzian of $\tau$ with respect to $t$ is
\begin{equation} \label{eqn:schwartzt}
    S_t\tau_{\mathrm{IR}}=2Q\left(\frac{1}{2N},\frac{N+1}{2N},1;t\right), 
\end{equation}
so a presentation in the form (\ref{eqn:tau-ans}) actually exists and we need only to choose $F_1$, $F_2$ appropriately.
\par On the other hand, the Schwarzian with respect to $q$ can be found as 
\begin{equation}
    S_q\tau_{\mathrm{IR}}=\left(S_t\tau_{\mathrm{IR}}-S_q t\right)\left(\frac{\dd t}{\dd q}\right)^{-2}=2Q\left(\frac{1}{N},\frac{1}{N},1;q\right),
\end{equation}
which leads us to yet another presentation of the infrared coupling constant $\tau$. Both forms will be useful for us in the future.
\par The hypergeometric functions should be chosen such that (\ref{eqn:tauD}) is satisfied.
\par Note that 
\begin{equation}
{}_2F_1\left(\frac{1}{N},\frac{1}{N},1,q\right)=1+\mathcal{O}(q),\, q\to 0  ,  
\end{equation}
\begin{equation}
  {{}_2F_1\left(\frac{1}{N},\frac{1}{N},\frac{2}{N},1-q\right)}=-\left(\ln(q)+2\gamma_E+2\psi\left(\frac{1}{N}\right)\right) \frac{\Gamma\left(\frac{2}{N}\right)}{\Gamma\left(\frac{1}{N}\right)^2}+\mathcal{O}(1), \, q\to 0 ,
\end{equation}
where $\psi(z)=\frac{\Gamma'(z)}{\Gamma(z)}$ is the digamma function and $\gamma_E$ is the  Euler–Mascheroni constant
and using
the Gauss's diagamma theorem
\begin{equation}
    \psi\left(\frac{1}{N}\right)=
-\gamma_E-\ln(2N)-\frac{\pi}{2} \cot\left(\frac{\pi}{N}\right)+2\sum_{n=1}^{\lfloor\frac{N-1}{2}\rfloor} \cos\left(\frac{2\pi n}{N}\right)\ln\left(\sin\left(\frac{2\pi n}{N}\right)\right)+\pi \ii.
\end{equation}
together with (\ref{eqn:logD})
we construct an appropriate solution
    \begin{equation}\label{eqn:tau-final}
        2\pi \ii \tau_{\mathrm{IR}}=-\frac{\Gamma\left(\frac{1}{N}\right)^2}{\Gamma\left(\frac{2}{N}\right)}\frac{{}_2F_1\left(\frac{1}{N},\frac{1}{N},\frac{2}{N},1-q\right)}{{}_2F_1\left(\frac{1}{N},\frac{1}{N},1,q\right)}+\pi\left(\cot\left(\frac{\pi}{N}\right) +\ii\right).
    \end{equation}
Another form, following from (\ref{eqn:schwartzt}), which we will use in the future is
\begin{equation}\label{eqn:tau-t}
    2\pi \ii \tau_{\mathrm{IR}}=-\frac{\Gamma\left(\frac{1}{N}\right)^2}{\Gamma\left(\frac{2}{N}\right)}2^{\frac{1}{N}} \frac{{}_2F_1\left(\frac{1}{2N},\frac{N+1}{2N},\frac{N+2}{2N};1-t\right)}{{}_2F_1\left(\frac{1}{2N},\frac{N+1}{2N},1;t\right)} + \pi\left(\cot\left(\frac{\pi}{N}\right) +\ii\right).
\end{equation}

    \subsubsection{Modular properties} \label{subsubsec:modularprop}

    The form (\ref{eqn:tau-ans}) was motivated by expected modularity\footnote{We use the word `modular' to indicate invariance (or particular law of transformation) with respect to any subgroup of $SL(2,\mathbb{R})$, not necessary a subgroup of the modular group $SL(2,\mathbb{Z})$. In mathematical literature, e.g. \cite{Doran}, the term `automorphic' is more common in this context.} of the inverse function $\tau_{\rm{IR}}$ with respect to triangle groups. Let us now see if this is actually the case, assuming for simplicity that $N>2$. Let us assume for now that there exists a single-valued inverse function $q$ such that
    \begin{equation}
        q_{\mathrm{UV}}(\tau_{\mathrm{IR}}(q))=q.
    \end{equation}
    The transformations of $\tau_{\mathrm{IR}}$ which leave $q_{\mathrm{UV}}$ invariant are related to the monodromy group of the hypergeometric equation. It has three singular points: $q=0$, $q=1$, and $q=\infty$. 
    \par 
    We already know that at $q=0$ the function $\tau_{\mathrm{IR}}(q)$ behaves as $\frac{1}{2\pi\ii}\ln(q_0)$. Thus $\tau_{\mathrm{IR}}(q)$ and $\tau_{\mathrm{IR}}(q)+k$, $k\in\mathbb{Z}$ can be understood as different branches of the same holomorphic function. Any of these branches should be inverse of $q_{\mathrm{UV}}$ by analytical continuation. Hence
    \begin{equation}\label{eqn:modularT}
        q_{\mathrm{UV}}(\tau)=q_{\mathrm{UV}}(\tau+k),\ k\in \mathbb{Z}.
    \end{equation}
    \par
    Another type of symmetry comes from the point $q=1$. To see this, it is convenient to change the basis of solutions of the hypergeometric equation. Instead of ${}_2F_1\left(\frac{1}{N},\frac{1}{N},1,q\right)$ we take $$(1-q)^{1-\frac{2}{N}}{}_2F_1\left(1-\frac{1}{N},1-\frac{1}{N},2-\frac{2}{N},q\right).$$ We have
    \begin{eqnarray}
       && {}_2F_1\left(\frac{1}{N},\frac{1}{N},1,q\right)=\frac{\Gamma\left(1-\frac{2}{N}\right)}{\Gamma\left(1-\frac{1}{N}\right)^2}{}_2F_1\left(\frac{1}{N},\frac{1}{N},\frac{2}{N},1-q\right) \nonumber \\
       &&+\frac{\Gamma\left(\frac{2}{N}-1\right)}{\Gamma\left(\frac{1}{N}\right)^2}(1-q)^{1-\frac{2}{N}}{}_2F_1\left(1-\frac{1}{N},1-\frac{1}{N},2-\frac{2}{N},1-q\right).
    \end{eqnarray}
    Note that the formula above fails for $N=2$. Using
    \begin{eqnarray*}
        \Gamma\left(1-\frac{2}{N}\right)=\frac{\pi}{\sin\left(\frac{2\pi}{N}\right)\Gamma\left(\frac{2}{N}\right)}, \quad \Gamma\left(1-\frac{1}{N}\right)=\frac{\pi}{\sin\left(\frac{\pi}{N}\right)\Gamma\left(\frac{1}{N}\right)},
    \end{eqnarray*}
    we present (\ref{eqn:tau-final}) as
\begin{equation}
2\pi\ii \tau_{\mathrm{IR}}=
-\frac{\frac{\Gamma\left(\frac{1}{N}\right)^2}{2\Gamma\left(\frac{2}{N}\right)}{{}_2F_1\left(\frac{1}{N},\frac{1}{N},\frac{2}{N},1-q\right)}-\pi\cot\left(\frac{\pi}{N}\right) \frac{\Gamma\left(\frac{2}{N}-1\right)}{\Gamma\left(\frac{1}{N}\right)^2}(1-q)^{1-\frac{2}{N}}{}_2F_1\left(1-\frac{1}{N},1-\frac{1}{N},2-\frac{2}{N},q\right)}
{\frac{\Gamma\left(\frac{1}{N}\right)^2}{2\pi\Gamma\left(\frac{2}{N}\right)}{{}_2F_1\left(\frac{1}{N},\frac{1}{N},\frac{2}{N},1-q\right)}\tan\left(\frac{\pi}{N}\right)+ \frac{\Gamma\left(\frac{2}{N}-1\right)}{\Gamma\left(\frac{1}{N}\right)^2}(1-q)^{1-\frac{2}{N}}{}_2F_1\left(1-\frac{1}{N},1-\frac{1}{N},2-\frac{2}{N},q\right)}+\pi \ii.
\end{equation}
    
    In this case, the source of the non-trivial monodromy is the multivalued function $(1-q)^{1-\frac{2}{N}}$ which gets a phase $e^{2\pi\ii \frac{N-2}{N}}$ if we go around the singular point $q=1$.  Thus,  
    \begin{eqnarray}
&&2\pi \ii \tau_{\mathrm{IR}}^{(k)}= \\ \nonumber&&-
\frac{\frac{\Gamma\left(\frac{1}{N}\right)^2}{2\Gamma\left(\frac{2}{N}\right)}{{}_2F_1\left(\frac{1}{N},\frac{1}{N},\frac{2}{N},1-q\right)}-e^{\frac{4\pi\ii k}{N}}\cot\left(\frac{\pi}{N}\right) \frac{\Gamma\left(\frac{2}{N}-1\right)}{\Gamma\left(\frac{1}{N}\right)^2}(1-q)^{1-\frac{2}{N}}{}_2F_1\left(1-\frac{1}{N},1-\frac{1}{N},2-\frac{2}{N},q\right)}
{\frac{\Gamma\left(\frac{1}{N}\right)^2}{2\pi\Gamma\left(\frac{2}{N}\right)}{{}_2F_1\left(\frac{1}{N},\frac{1}{N},\frac{2}{N},1-q\right)}\tan\left(\frac{\pi}{N}\right)+ e^{\frac{4\pi\ii k}{N}}\frac{\Gamma\left(\frac{2}{N}-1\right)}{\Gamma\left(\frac{1}{N}\right)^2}(1-q)^{1-\frac{2}{N}}{}_2F_1\left(1-\frac{1}{N},1-\frac{1}{N},2-\frac{2}{N},q\right)}+\pi\ii
\end{eqnarray}
    with various $k\in\mathbb{Z}$ are all different branches of the same multivalued function. As long as we are interested in the generators of the  monodromy group of $\tau_{\mathrm{IR}}$ (or, equivalently, the group of automorphisms of the inverse function $q_{\mathrm{UV}}$), it is enough to consider $k=1$. 
    \par 
    We find
    \begin{equation}
        \tau_{\mathrm{IR}}^{(1)}=1+\frac{\tau_{\mathrm{IR}}}{1-\tau_{\mathrm{IR}}\sin^2\left(\frac{\pi}{N}\right)}.
    \end{equation}
    Thus, taking (\ref{eqn:modularT}) into account,  for the inverse function we have
    \begin{equation}
        q_{\mathrm{UV}}(\tau)=q_{\mathrm{UV}}\left(\frac{\tau}{1-\tau\sin^2\left(\frac{\pi}{N}\right)}\right).
    \end{equation}
    \par
    We do not have to consider the third point $q=\infty$ because the contour around $\infty$ is equivalent to the sum of contours encircling $q=0$ and $q=1$, so no new monodromy generators will appear.
    \par Writing the fractional linear transforms as the corresponding elements of $PSL(2,\mathbb{C})$, we get that $q$ is a modular function with respect to the group generated by
    \begin{equation}
        \gamma_2=\left(
        \begin{matrix}
            1 & 0 \\
            -\sin^2\left(\frac{\pi}{N}\right) & 1
        \end{matrix}
        \right),\ \gamma_3=\left(
        \begin{matrix}
            1 & 1 \\
            0 & 1
        \end{matrix}
        \right).
    \end{equation}
    Let us compare this result with the results of the analysis of the $S$-group in \cite{Lerda}. In the transformation $\gamma_3$, we easily recognize the $T$-duality (for $N>2$). The other generator, the $S$-duality, is expected to be
    \begin{equation}
        \gamma_S^{(k)}= \left( \begin{matrix}
            0 & \frac{1}{2\sin^2\left(\frac{\pi k}{N}\right) } \\
            -2\sin^2\left(\frac{\pi k}{N}\right)  & 0
        \end{matrix}
        \right),
    \end{equation}
    where $k=1,\ldots,N-1$.
    \par 
    The second generator of the group preserving $q_{\mathrm{UV}}$ can then be expressed as
    \begin{equation}
        \gamma_2= -\gamma_S \gamma_3^{-1}  \gamma_S. 
    \end{equation}
    Here the minus sign can be dropped because it does not affect the $PSL(2,\mathbb{C})$ class.
    It is easy to see then that the group generated by $\gamma_2$, $\gamma_3$ is an index two subgroup of the $S$-group presented in \cite{Lerda,ArgyresBuchel}. It consists of products of $T^{\pm 1}$ and $S$ transformations such that the number of the latter is always even.    This is consistent with the results of \cite{Lerda}, since according to that paper the $S$-duality acts on the bare coupling constant as $q\to q^{-1}$, so the transformations containing an even number of $S$-dualities when presented as a product of the generators are exactly the ones which preserve $q$. In fact, using the relations
    \begin{equation}
        q^{-\frac{1}{N}}{{}_2F_1\left(\frac{1}{N},\frac{1}{N},\frac{2}{N},1-q^{-1}\right)}={{}_2F_1\left(\frac{1}{N},\frac{1}{N},\frac{2}{N},1-q\right)},
    \end{equation}
    \begin{equation}
        q^{-\frac{1}{N}}{{}_2F_1\left(\frac{1}{N},\frac{1}{N},1,q^{-1}\right)}={{}_2F_1\left(\frac{1}{N},\frac{1}{N},\frac{2}{N},1-q\right)\frac{e^{-\frac{\pi\ii}{N}}\sin\left(\frac{\pi}{N}\right)\Gamma\left(\frac{1}{N}\right)}{\Gamma\left(\frac{2}{N}\right)^2\pi}+{{}_2F_1\left(\frac{1}{N},\frac{1}{N},1,q\right)}}e^{-\frac{2\pi\ii}{N}}
    \end{equation}
    one finds that
    \begin{equation}
        \tau_{\mathrm{IR}}(q^{-1})=-\frac{1}{4\sin^2\left(\frac{\pi}{N}\right) \left(\tau_{\mathrm{IR}}(q^{-1})-1\right)}.
    \end{equation}
    Together with the $\gamma_3$ transformation it follows that
    \begin{equation}
        q_{\mathrm{UV}}\left(\frac{1}{4\sin^2\left(\frac{\pi}{N}\right)\tau}\right)=q_{\mathrm{UV}}(\tau)^{-1}.
    \end{equation}
    The transformation of the argument above is exactly the one given by $\gamma_S$. 
    \par 
    To finish this analysis, let us classify the group generated by $\gamma_2$ and $\gamma_3$. To do that we consider an isomorphic group, generated by
    \begin{equation}
        \tilde{\gamma}_2=\gamma_0^{-1}\gamma_2\gamma_0,\ \tilde{\gamma}_3=\gamma_0^{-1}\gamma_3\gamma_0,
    \end{equation}
    where 
    \begin{equation}
        \gamma_0=\begin{pmatrix}
            \frac{1}{2\sin\left(\frac{\pi}{N}\right)} & - \frac{1}{2\sin\left(\frac{\pi}{N}\right)}\\
            0 & 2 \sin\left(\frac{\pi}{N}\right)
        \end{pmatrix}
    \end{equation}
    Note that conjugation with $\gamma_0$ is equivalent to rescaling and shift of $\tau$. It is also convenient to introduce $\tilde{\gamma}_1=\tilde{\gamma}_3^{-1}\tilde{\gamma}_2^{-1}$. Then we have
    \begin{equation}
        \tilde{\gamma}_1=-\begin{pmatrix}
            2\cos(\pi \frac{N-2}{N}) & 1\\
            -1 & 0
        \end{pmatrix},\ \tilde{\gamma}_2=\begin{pmatrix}
            0 & 1\\
            -1 & 2
        \end{pmatrix},
        \tilde{\gamma}_3=\begin{pmatrix}
            1 & 2+ 2\cos(\pi \frac{N-2}{N}) \\
            0 & 1
        \end{pmatrix}.
    \end{equation}
    As before, the extra sign of $\tilde{\gamma}_2$ can be dropped as it does not affect the class in $PSL(\mathbb{C},2)$.
    Comparing the above with \cite[eqn. (10)]{Doran} we conclude that the group of interest is isomorphic to the triangle group $\Gamma_{(\frac{N}{N-2},\infty,\infty)}$. On the other hand, the same conjugation transforms the $S$-group to the Hecke group  $\Gamma_{(\frac{2N}{N-2},2,\infty)}$ in agreement with \cite{Lerda} (for $k=1$). Comparing (\ref{eqn:tau-final}) with \cite[Section 7.1]{Doran} we conclude that up to appropriate shift and renormalisation of $\tau$, $q_{\mathrm{UV}}=J_{(\frac{N}{N-2},\infty,\infty)}^{-1}$, where $J_{(\frac{N}{N-2},\infty,\infty)}$ is the Hauptmodul\footnote{We recall that the Hauptmodul of a group is such a modular function of the group that any other modular function can be represented as a rational function of the Hauptmodul.} of $\Gamma_{(\frac{N}{N-2},\infty,\infty)}$. 
    \par 
    We note that if  $J_{(\frac{N}{N-2},\infty,\infty)}$ is a Hauptmodul, then so is  $\pm J_{(\frac{N}{N-2},\infty,\infty)}^{-1}$. So, (\ref{eqn:tau-final}) can be formulated as follows: the bare coupling constant $q$ (or $q_0$) is the Hauptmodul of the triangle group $\Gamma_{(\frac{N}{N-2},\infty,\infty)}$ with the argument shifted and rescaled by the matrix $\gamma_0$ and specified by the normalisation conditions (\ref{eqn:tauD}). 
    This characterisation of $q_{\mathrm{UV}}$ is more direct than the one via the irrational functions of the Hauptmodul of the Hecke group $\Gamma_{(\frac{2N}{N-2},2,\infty)}$ suggested in \cite{Lerda}. 
    \par 
    Still, a  proof of the conjecture made in \cite{Lerda} (for the constant $\tau_1$ in their notation) can be extracted from our results. Indeed, we just have shown that $q$ transforms exactly as it is predicted in  \cite{Lerda}, and satisfies the same normalisation conditions. Alternatively, one can repeat the same analysis for (\ref{eqn:tau-t}) instead of (\ref{eqn:tau-final}) to derive that the $S$-invariant combination
    \begin{equation}\label{eqn:t-q}
        t=\frac{4}{\kappa^2}=\frac{4}{q+\frac{1}{q}+2}
    \end{equation}
    is a modular function, and, in fact, a Hauptmodul for the group $\Gamma_{(\frac{2N}{N-2},2,\infty)}$. Solving (\ref{eqn:t-q}) for $q$, one can get the conjecture of \cite{Lerda} up to a rational transform of the Hauptmodul. 
    
    \paragraph{Remark.} Strictly speaking, unless $N-2\vert N$, the groups we constructed here are not the triangle groups $\Gamma_{(m_1,m_2,m_3)}$, where all $m_i$ should be either positive integers, or infinities. In the case of general rational parameters it is not clear what the actual modular properties of the formally constructed Hauptmodul are. The problem in the argument above lies in the assumption that the function of $q_{\rm{UV}}$ is single-valued. One can verify that this is the case when all $m_i$ are integer (or infinite), but may fail in general. Then the modular properties should be considered as equality between properly chosen branches of $q_{\rm{UV}}$. We will not focus our attention on this problem.
    \par
    Interestingly enough, the cases $N-2\vert N$, \textit{i.e.} $N=3$ and $N=4$, are exactly the ones when $\Gamma_{(\frac{N}{N-2},\infty,\infty)}$ is isomorphic to an arithmetic group (\textit{i.e.} a subgroup of $PSL(2,\mathbb{Z})$). Namely, we have $\Gamma_{(3,\infty,\infty)}=\Gamma_0(3)$, $\Gamma_{(2,\infty,\infty)}=\Gamma_0(2)$. The corresponding Hauptmoduln $J_{(2,\infty,\infty)}$, $J_{(3,\infty,\infty)}$ are known to coincide (with appropriate scaling and shift of the argument) with the famous modular functions of $\Gamma_0(2)$, $\Gamma_0(3)$ expressed via the $\eta$-quotients \cite{Doran}. This allows a simple comparison with the known results for $N=3,4$ in the next section. 
    \par 
    The group $\Gamma_{(2,\frac{2N}{N-2},\infty)}$, besides the cases $N=2,3$, is also a Hecke triangle group in the strict sense for $N=6$. Again, these are precisely all the cases when a Hecke group is isomorphic to an arithmetic group \cite{Doran}. The group $\Gamma_{(\frac{3}{2},\infty,\infty)}$, which according to our analysis corresponds to $N=6$, is also arithmetic, and one may expect that at least in this special case the modular properties we found are exact. 
    \par 
    Finally, we note that although we excluded the case $N=2$ from our analysis, the results formally apply to it. Indeed, the modular $\lambda$-function is a Hauptmodule of $\Gamma_{(\infty,\infty,\infty)}=\Gamma(2)$.

    \subsubsection{Vacuum energy}
    
    Let us now find a more explicit form of $Z_0$. We recall that in the $N=2,3$ cases \cite{Poghossian2,Poghossian3} one of the factors was presented as a power of a modular form of low weight of the relevant modular group. Combining this with the analysis from the previous subsection, it is natural to expect that the $\Gamma_{(\frac{N}{N-2},\infty,\infty)}$ modular forms of low weight should play some role. From \cite{Doran} we can find that all such modular forms have even weights, and in the lowest possible weight of two we have exactly one,  
\begin{equation}   \label{eqn:f2}
f_2(q_{\mathrm{IR}})=-q_{\mathrm{IR}}\frac{\dd q_{\mathrm{UV}}^{-1}}{\dd q_{\mathrm{IR}}}q_{\mathrm{UV}}=\frac{q_{\mathrm{IR}}}{q_{\mathrm{UV}}}\frac{\dd q_{\mathrm{UV}}}{\dd q_{\mathrm{IR}}}=\frac{\dd \tau_{\mathrm{UV}}}{\dd \tau_{\mathrm{IR}}}.
\end{equation}
We added here an extra minus sign with respect to \cite{Doran} for convenience.
\par In fact, the modular forms of higher weights are proportional to the same expression (\ref{eqn:f2}), together with powers of $q_{\mathrm{UV}}$ and $1-q_{\mathrm{UV}}$. The powers of $q_{\mathrm{UV}}$ cannot appear in $Z_0$, because it would mean that $\ln(Z_0)\sim C \ln(q)$ in the $q\to 0$ limit, in contradiction with (\ref{eqn:k0}). In contrast, $\ln(1-q)$ is regular in a neighbourhood of $q=0$. Moreover, factors of this term already appeared in \cite{Poghossian3}. It is thus natural to look for an expression of the form
\begin{equation}
    Z_0=Z_c(q)f_2(q_{\mathrm{IR}})^{\nu_1}(1-q_{\mathrm{UV}}(q_{\mathrm{IR}}))^{\nu_2}.
\end{equation}
The coefficients then can be found from the power series of (\ref{eqn:k0}) to be
\begin{equation}
    \nu_1={N\frac{12T_0^2-24 T_1+(N+1)\epsilon^2}{24 \epsilon_1 \epsilon_2}}
\end{equation}
\begin{equation}
    \nu_2=-\frac{T_0^2(N-2)+N T_0 \epsilon-2N T_1}{N \epsilon_1\epsilon_2}.
\end{equation}

\subsection{Case of arbitrary masses} \label{subsec:arbmass}

\par The derivation above was performed for the case of a special mass spectrum
\begin{equation} 
    {m}_f={m}_{f+N}-\epsilon, \, f=1,\ldots, N.
\end{equation}
All the computations can be repeated in full generality. The main technical complication is that in the absence of factorisation (\ref{eqn:Qfactor}) of the polynomial $\tilde{Q}$, we cannot factorise the exponentiated energy cost of one instanton as in (\ref{eqn:omegaomega}). So, instead of $\omega$, we define $\hat{\omega}$ as
\begin{equation}\label{eqn:omegadef-Q}
    \hat{\omega}(y, \, {\bm \phi})= \sqrt{q}\frac{1}{P_0(Ay)}\prod_j\frac{\left(y-\frac{\phi_j}{A}\right)\left(y-\frac{\phi_j-\epsilon}{A}\right)}{\left(y-\frac{\phi_j-\epsilon_1}{A}\right)\left(y-\frac{\phi_j-\epsilon_2}{A}\right)}.
\end{equation}
The derivation in Subsection \ref{subsec:sadptapp} can be repeated with minimal changes. The mass term, which is removed from $\hat{\omega}$, appears explicitly in the saddle point equation, turning (\ref{onacut}) into
\begin{equation} \label{onacut-Q}
     \hat{\omega}(y \pm \ii 0)\hat{\omega}\left(y-\frac{\epsilon}{A}\mp \ii 0\right) \tilde{Q}(Ay) = 1,
\end{equation}
and thus the Quantum Seiberg-Witten equation takes a form similar to the one presented in \cite{PoghossianQSW,Fucitoetal}
\begin{equation}\label{eqn:genmassQSW}
   \tilde{Q}(Ay) \hat{\omega}\left(y-\frac{\epsilon}{A}\right) +\frac{1}{\hat{\omega}(y)} = \left(\sqrt{q}+\frac{1}{\sqrt{q}}\right)P_{N}(y).
\end{equation}
To repeat Subsections \ref{ssec:omegapers}-\ref{ssec:keff} we would need to adjust Appendix \ref{app-rec} so that it can be applied to (\ref{eqn:genmassQSW}). We omit these details. Instead, we note that in the end only the expansion of (\ref{eqn:genmassQSW}) up to $\mathcal{O}(A^{-3})$ matters. In particular, only the three highest  powers of the polynomial $\tilde{Q}$ determine the asymptotic behaviour. As we show in this subsection, any polynomial can be factorised with such precision. 
\par We define
\begin{equation}
    \tilde{T}_0= \sum_{i=1}^{2N} m_i, \quad \tilde{T}_1= \sum_{i_1=1}^{2N}\sum_{i_2=1}^{i_1-1} m_{i_1} m_{i_2}
\end{equation}
and observe that
\begin{eqnarray}  \label{eqn:newT0T1} 
    &&\tilde{Q}(x) =x^{2N}- x^{2N-1}\tilde{T}_0+x^{2N-2} \tilde{T}_1+\mathcal{O}\left(x^{-3} \right)   = (x^{N}-x^{N-1} {T}_0+x^{N-2} {T}_1+\mathcal{O}\left(x^{-3} \right) )  \\
    && \cdot ((x-\epsilon)^{N}-(x-\epsilon)^{N-1} {T}_0+(x-\epsilon)^{N-2} {T}_1+\mathcal{O}\left(x^{-3} \right) ), \nonumber
\end{eqnarray}
where
\begin{eqnarray}    \label{eqn:TviatildeT}
    &&T_0=\frac{\tilde{T}_0-N \epsilon}{2} , \\
    &&T_1=\frac{1}{2}\left(\tilde{T}_1-\frac{\tilde{T}_0^2}{4}-\frac{(N-1)\tilde{T}_0 \epsilon}{2}+\frac{N^2 \epsilon^2}{4} \right) . \nonumber
\end{eqnarray}
  \par To adjust the results of previous subsections to the case of arbitrary mass distribution it is enough to replace $T_0$, $T_1$ everywhere with their new definitions (\ref{eqn:TviatildeT}).

\subsection{Asymptotic behaviour provided by the saddle point}

We conclude that the saddle point approximation leads to the following asymptotic form of the instanton partition function

    \begin{eqnarray}  \label{eqn:asympt2Nfund}
        Z=&&\left(\frac{q}{D q_{\mathrm{IR}}
        } \right)^{-\frac{w_1}{\epsilon_1\epsilon_2}}  f_2(q_{\mathrm{IR}})^{\frac{N}{4 \epsilon_1 \epsilon_2}\left(\tilde{T}_0^2-2\tilde{T}_1-\tilde{T}_0 \epsilon+\frac{\epsilon^2(N+1)}{6} \right)} \\ && \cdot (1- q)^{-\frac{1}{\epsilon_1 \epsilon_2}\left( \frac{N-1}{2N} \tilde{T}_0^2+\tilde{T}_0\epsilon-\tilde{T}_1-N \epsilon^2\right)} \left(1+\mathcal{O}\left(w_{N-1}^{-1}\right) \right). \nonumber
    \end{eqnarray}

Let us also write the asymptotic for $2N-k$ fundamental multiplets and $k$ antifundamental ones. To do that, we have to replace $q\to (-1)^k q$ and shift the masses according to (\ref{eqn:goodMass}), which means shifting $\tilde{T}_0$, $\tilde{T}_1$ as
    \begin{eqnarray}
        &&\tilde{T}_0 \to \tilde{T}_0+k \epsilon ,\\
        &&\tilde{T}_1 \to \tilde{T}_1 +  \epsilon (k \tilde{T}_0 - \tilde{T}_0^{af}) + \frac{k (k-1)}{2} \epsilon^2 ,  \nonumber 
    \end{eqnarray}
 where $\tilde{T}_0^{af}$ is the sum of masses of antifundamental multiplets, $\tilde{T}_0=\tilde{T}_0^{f}+\tilde{T}_0^{af}$.
This yields
    \begin{eqnarray}  \label{eqn:asympt2Nmenokandk}
        Z^{(2N-k,k)}&=&\left(\frac{q}{D q_{\mathrm{IR}}
        } \right)^{-\frac{w_1}{\epsilon_1\epsilon_2}} f_2(q_{\mathrm{IR}})^{\frac{N}{4 \epsilon_1 \epsilon_2}\left(\tilde{T}_0^2-2\tilde{T}_1- \epsilon(\tilde{T}_0^f-\tilde{T}_0^{af})+\epsilon^2\frac{N+1}{6} \right)} \\ &&\cdot(1- q
        )^{-\frac{1}{\epsilon_1 \epsilon_2}\left( -\tilde{T}_1+\frac{N-1}{2N}\tilde{T}_0^2 +\tilde{T}_0 \epsilon \frac{3N-2k}{2N}-\frac{1}{2}\epsilon(\tilde{T}_0^f-\tilde{T}_0^{af})+\epsilon^2(\frac{k}{2}(3-\frac{k}{N})-N)\right)} \left(1+\mathcal{O}\left(w_{N-1}^{-1}\right) \right) .\nonumber
    \end{eqnarray}
In particular, for $2N$ antifundamental multiplets the saddle point method gives
    \begin{eqnarray} \label{eqn:asympt2Nantifund} 
        Z^{(0,2N)}&=&\left(\frac{q}{D q_{\mathrm{IR}}
        } \right)^{-\frac{w_1}{\epsilon_1\epsilon_2}} f_2(q_{\mathrm{IR}})^{\frac{N}{4 \epsilon_1 \epsilon_2}\left(\tilde{T}_0^2+ \epsilon \tilde{T}_0+\epsilon^2\frac{N+1}{6} \right)} \\ && \cdot  (1-q)^{-\frac{1}{\epsilon_1 \epsilon_2}\left( -\tilde{T}_1+\frac{N-1}{2N}\tilde{T}_0^2 \right)} \left(1+\mathcal{O}\left(w_{N-1}^{-1}\right) \right) . \nonumber
    \end{eqnarray}

\section{Tests and comparisons} \label{sec:comparison}

\subsection{\texorpdfstring{$N=2$}{N=2}}

In \cite{Poghossian2}, based on the AGT conjecture, the author found the asymptotic behaviour of the instanton partition function in the $N=2$ theory with four antifundamental multiplets\footnote{Here we changed the sign of $q_{\rm IR}$  with respect to  \cite{Poghossian2} due to different conventions. }. 
\begin{equation}\label{eqn:pogh2}
    Z^{(0,2N)} =\left(-\frac{q}{16 q_{\rm IR}} \right)^{-\frac{w_1}{\epsilon_1 \epsilon_2}} (1-q)^{\frac{1}{4\epsilon_1 \epsilon_2}\left(\epsilon-\sum_i m_i\right)^2} \theta_3(-q_{\rm IR})^{\frac{1}{\epsilon_1 \epsilon_2}\sum_i (m_i^2+(\epsilon-m_i)^2 -3 \epsilon^2/4)} \left( 1+\mathcal{O}(w_1^{-1}) \right) .
\end{equation}
We have written the asymptotic in terms of $w_1=-a_1^2=-A^2$ to make it more consistent with our general result.
According to \cite{Poghossian2, Billoetal}, in the $N=2$ case the IR and UV couplings are connected as
\begin{equation} \label{qUVN2}
    q=\left( \frac{\theta_2(-q_{\rm IR})}{\theta_3(-q_{\rm IR})}\right)^4=-16\left(\frac{\eta(q_{\rm IR}^4)}{\eta(q_{\rm IR})}\right)^8.
\end{equation}
\par Note that in the $N=2$ case there is only one way to approach infinity. 
\par
Let us compare (\ref{eqn:pogh2}) with (\ref{eqn:asympt2Nantifund}). First of all, the right-hand side of (\ref{qUVN2}), up to a sign in the argument, is known as the modular $\lambda$-function which is defined by $\exp(\pi \ii \tilde{\tau}(\lambda(q)))=q$, where
\begin{equation}
\tilde{\tau}=\ii \, \frac{{{}_2F_1\left(\frac{1}{2},\frac{1}{2},1,1-q\right)}}{{}_2F_1\left(\frac{1}{2},\frac{1}{2},1,q\right)}.
\end{equation}
It follows that (\ref{qUVN2}) agrees with (\ref{eqn:tau-final}).
\par The one-loop contribution for $N=2$ is
\begin{equation}
    D= -16 ,
\end{equation}
and thus the effective couplings in (\ref{eqn:pogh2}) and (\ref{eqn:asympt2Nantifund}) coincide. We also note that (\ref{qUVN2}) is equal to the Hauptmodul $J_{(\infty,\infty,\infty)}^{-1}$ constructed in \cite{Doran} up to an appropriate shift and rescaling of $\tau$, and $\lambda$ is the famous Hauptmodul of $\Gamma(2)$ that happens to coincide with $\Gamma_{(\infty,\infty,\infty)}$ \cite{Doran}.
\par 
As for the vacuum energy contribution, let us note that
\begin{equation}
    \frac{\dd}{\dd q_{\rm{IR}}}\frac{\theta_2(-q_{\rm{IR}})}{\theta_3(-q_{\rm{IR}})}=\frac{\theta_2(-q_{\rm{IR}})\theta_4(-q_{\rm{IR}})^4}{4q_{\rm{IR}}\theta_3(-q_{\rm{IR}})},
\end{equation}
so
\begin{equation}
f_2=\frac{q_{\mathrm{IR}}}{q_{\mathrm{UV}}}\frac{\dd q_{\mathrm{UV}}}{\dd q_{\mathrm{IR}}}=\theta_4(-q_{\rm{IR}})^4=\theta_3(-q_{\rm{IR}})^4-\theta_2(-q_{\rm{IR}})^4=\theta_3(-q_{\rm{IR}})^4\left(1-q_{\mathrm{UV}}\right).
\end{equation}
Thus, (\ref{eqn:pogh2}) can be rewritten as
\begin{eqnarray}\label{eqn:pogh2-r}
    Z =\left( -\frac{q}{16 q_{\rm IR}} \right)^{-\frac{4 w_1}{\epsilon_1 \epsilon_2}} & (1-q)^{\frac{1}{4\epsilon_1 \epsilon_2}(\epsilon-\sum_i m_i)^2-\frac{1}{4\epsilon_1 \epsilon_2}\sum_i (m_i^2+(\epsilon-m_i)^2 -3 \epsilon^2/4)} \nonumber\\ & \cdot f_2(q_{\rm IR})^{\frac{1}{4\epsilon_1 \epsilon_2}\sum_i (m_i^2+(\epsilon-m_i)^2 -3 \epsilon^2/4)} \left( 1+\mathcal{O}(w_1^{-1}) \right) \nonumber  \\
    =\left( -\frac{q}{16 q_{\rm IR}} \right)^{-\frac{4 w_1}{\epsilon_1 \epsilon_2}} & (1-q)^{-\frac{1}{4\epsilon_1 \epsilon_2} (\tilde{T}_0^2-4 \tilde{T}_1)}f_2(q_{\rm IR})^{\frac{1}{4\epsilon_1 \epsilon_2}(2\tilde{T}_0^2-4\tilde{T}_1+\epsilon^2-2\epsilon\tilde{T}_0)}\left( 1+\mathcal{O}(w_1^{-1}) \right) .
\end{eqnarray}
which coincides with (\ref{eqn:asympt2Nantifund}) up to the sign of the masses of the antifundamental multiplets, which is a question of convention.

\subsection{\texorpdfstring{$N=3$}{N=3}}

 In the $SU(3)$ theory one can approach infinity in multiple ways, but since \cite{Poghossian3} is tied to the AGT duality, the chosen large vev limit is the same as the one we consider in the present paper.
 \par The conjectured asymptotic in the $N=3$ theory with six fundamental multiplets is\footnote{In \cite{Poghossian3} the variables are $u=-3w_1$ and $v=-27w_2$. Note that here we did not change the sign of $q_{\rm IR}$  with respect to  \cite{Poghossian3}.  }
 \begin{eqnarray}\label{eqn:pogh3}
     &&Z^{(0,2N)}=\left(\frac{q}{27 q_{\rm IR}} \right)^{-\frac{w_1}{\epsilon_1 \epsilon_2}}\left(\frac{\eta(q_{\rm IR}^3)}{\eta^3(q_{\rm IR} )} \right)^{\frac{3T_2-T_1^2}{\epsilon_1 \epsilon_2}} \\
     &&\cdot \left( \left( \frac{\eta^3(q_{\rm IR})}{\eta(q_{\rm IR}^3)}\right)^3+27\left(\frac{\eta^3(q_{\rm IR}^3)}{\eta(q_{\rm IR})}\right)^3 \right)^{\frac{T_1^2-3T_1\epsilon+2\epsilon^2}{6\epsilon_1 \epsilon_2}}\left(1+\mathcal{O}(w_2^{-1})\right) , \nonumber
 \end{eqnarray}
where the UV coupling $q$ and the effective coupling $q_{\rm IR}$ are related by \cite{Poghossian3,Billoetal}
\begin{equation} \label{qUVN3}
    q=-27 \left(\frac{\eta(q_{\rm IR}^3)}{\eta(q_{\rm IR})}\right)^{12} .
\end{equation}
{
Here we change the sign of $q_{\rm{IR}}$ relative to \cite{Poghossian3} to take the difference in conventions into account. We note that this does not affect the asymptotic behaviour of the instanton partition function.
}

\par It is known that the fraction in (\ref{qUVN3}) is proportional to $J_{(3,\infty,\infty)}^{-1}$, the Hauptmodul of $\Gamma_{(3,\infty,\infty)}=\Gamma_0(3)$ constructed in \cite{Doran}, up to a rescaling and a shift of the argument.  So, to ensure that it agrees with (\ref{eqn:tau-final}) it is enough to compare finitely many terms and observe a perfect agreement.
\par The one-loop contribution for $N=3$ is
\begin{equation}
    D=-27,
\end{equation}
so we again see that the effective couplings in 
(\ref{eqn:asympt2Nantifund}) and (\ref{eqn:pogh3}) coincide.
\par In \cite{Poghossian3} it was noted that
\begin{equation}
    f_1(q_{\rm{IR}})=\left( \left( \frac{\eta^3(q_{\rm IR})}{\eta(q_{\rm IR}^3)}\right)^3+27\left(\frac{\eta^3(q_{\rm IR}^3)}{\eta(q_{\rm IR})}\right)^3 \right)^{1/3}
\end{equation}
is a weight-one modular form of $\Gamma_1(3)$. The larger group $\Gamma_{(3,\infty,\infty)}=\Gamma_0(3)$ does not admit such a modular form,  but both $f_2$ defined by (\ref{eqn:f2}) and $f_1^2$ are weight-two modular forms for the smaller group $\Gamma_1(3)$. Comparing, say, the first twenty terms, one may guess that $f_1^2=f_2$.
Since the vector space formed by such forms is known to be low-dimensional, this is enough to conclude that they indeed coincide.
\par 
Now note that
\begin{equation}
    f_1(q_{\rm IR})= \frac{\eta^3(q_{\rm IR})}{\eta(q_{\rm IR}^3)}\left(1-q \right)^{1/3},
\end{equation}
so (\ref{eqn:pogh3}) can be written as
\begin{equation}
     Z=\left(-\frac{q}{27 q_{\rm IR}} \right)^{-\frac{w_1}{\epsilon_1 \epsilon_2}}\left(1-q \right)^{\frac{3\tilde{T}_1-\tilde{T}_0^2}{3\epsilon_1 \epsilon_2}} f_2(q_{\rm IR})^{\frac{3\tilde{T}_0^2-6\tilde{T}_1-3\tilde{T}_0\epsilon+2\epsilon^2}{4\epsilon_1 \epsilon_2}}\left(1+\mathcal{O}(w_2^{-1})\right).
 \end{equation}
It coincides with (\ref{eqn:asympt2Nfund}), again, up to the sign of $\tilde{T}_0$.

\subsection{Smaller number of multiplets}

We can decouple some of the multiplets and write the asymptotic behaviour of the partition function also in these cases. To do that, we have to send the masses $M$ of the multiplets to be decoupled to infinity and to expand $\ln Z$ in a series in large $M$. Note that it is crucial to perform the expansion of $\ln Z$ (or effective number of instantons) rather than of the asymptotic (\ref{eqn:asympt2Nmenokandk}) itself. 
\par From (\ref{Zkfaf}) we see that to correctly redefine the partition function when masses of $n$ fundamental multiplets tend to infinitely large mass $M$, we need to renormalise the parameter $q$ as $q \rightarrow q/(-M)^n$, and when masses of $l$ antifundamental multiplets are large a suitable renormalisation is $q \rightarrow q/M^l$.
\par We immediately note that
\begin{equation}
    B(q)=1+\mathcal{O}(q) ,
\end{equation}
so, from (\ref{eqn:qir}), after decoupling of any number of multiplets,
\begin{equation}
    q_{\rm IR}=q/D +\mathcal{O}(M^{-1}) ,
\end{equation}
and the first factor in the asymptotic disappears.
\par Moreover,
\begin{equation}
    \ln f_2  (q_{\rm IR})=-\frac{2(N-1)}{N^2}q- \frac{\left(2N^3-N^2-2N+1\right)}{2N^4}q^2+\mathcal{O}(q^{3}),
\end{equation}
\begin{equation}
    \ln (1-q)=-q-\frac{q^2}{2}+\mathcal{O}(q^{3}).
\end{equation}

\subsubsection{\texorpdfstring{$2N-1$}{2N-1} multiplets}

With the indications above, we easily find the asymptotics

 \begin{eqnarray} \label{eqn:menounofund}
        Z^{(2N-k-1,k)}&=& e^{ \ln(f_2(q_{\mathrm{IR}}))\frac{N}{4 \epsilon_1 \epsilon_2}\left(M^2- \epsilon M \right)} \\ &&\cdot e^{ -\frac{1}{\epsilon_1 \epsilon_2} \ln (1- q
        )\left(\frac{N-1}{2N}M^2-M\frac{ \tilde{T}_0}{N} +M \epsilon \frac{3N-2k}{2N}-\frac{1}{2}\epsilon M+\epsilon^2(\frac{k}{2}(3-\frac{k}{N})-N)\right) }\left(1+\mathcal{O}\left(M^{-1},w_{N-1}^{-1}\right) \right) \nonumber  \\
        &&=e^{\frac{q}{8N^3 \epsilon_1 \epsilon_2}\left( 8N^2 \tilde{T}_0-q(N-1)^2-4N^2 \epsilon(3 N-2k-1 )\right)}\left(1+\mathcal{O}\left(M^{-1},w_{N-1}^{-1}\right) . \right) \nonumber
\end{eqnarray}
 \begin{eqnarray} \label{eqn:menounoantifund}
        Z^{(2N-k,k-1)}&=& e^{ \ln(f_2(q_{\mathrm{IR}}))\frac{N}{4 \epsilon_1 \epsilon_2}\left(M^2+ \epsilon M \right)} \\ &&\cdot e^{ -\frac{1}{\epsilon_1 \epsilon_2} \ln (1- q
        )\left(\frac{N-1}{2N}M^2-M\frac{ \tilde{T}_0}{N} +M \epsilon \frac{3N-2k}{2N}+\frac{1}{2}\epsilon M+\epsilon^2(\frac{k}{2}(3-\frac{k}{N})-N)\right) }\left(1+\mathcal{O}\left(M^{-1},w_{N-1}^{-1}\right) \right)  \nonumber \\
        &&=e^{\frac{q}{8N^3 \epsilon_1 \epsilon_2}\left(- 8N^2 \tilde{T}_0-q(N-1)^2+4N^2 \epsilon(3 N-2k+1 )\right)}\left(1+\mathcal{O}\left(M^{-1},w_{N-1}^{-1}\right) \right). \nonumber
    \end{eqnarray}
One can see that (\ref{eqn:menounoantifund}) turns into  (\ref{eqn:menounofund}) by a mass shift $\tilde{T}_0\to \tilde{T}_0 +\epsilon$ and redefinition $q \to -q$, as it should.
\par For $k=2N$ we get
 \begin{eqnarray} \label{eqn:2Nmenounoantifund} 
        Z^{(0,2N-1)}Z^{(2N-k,k-2)}=e^{\frac{q}{8N^3\epsilon_1\epsilon_2}\left(8 N^2 \tilde{T}_0-q(N-1)^2+4N^2\epsilon(N-1) \right)}\left(1+\mathcal{O}\left(M^{-1},w_{N-1}^{-1}\right) \right) .
    \end{eqnarray}
    The asymptotic (\ref{eqn:2Nmenounoantifund}) coincides with the results reported in \cite{Poghossian2, Poghossian3}, again, up to the sign of $\tilde{T}_0$ and $q$. It is difficult to trace the root of this discrepancy due to more complicated introduction of masses and partition function in the cited papers.

\subsubsection{\texorpdfstring{$2N-2$}{2N-2} multiplets}

We have already found the asymptotic in this case in \cite{SysoevaBykov}, so it serves as another check of (\ref{eqn:asympt2Nmenokandk}).
\par Decoupling one more fundamental multiplet in the same way as before, we get

 \begin{eqnarray} \label{eqn:menodue}
        Z^{(2N-k-2,k)}=Z^{(2N-k,k-2)}=e^{-\frac{q}{N \epsilon_1 \epsilon_2}}\left(1+\mathcal{O}\left(M^{-1},w_{N-1}^{-1}\right)  \right) , 
\end{eqnarray}
which is in perfect agreement with \cite{SysoevaBykov}. Also,
 \begin{eqnarray} 
        Z^{(2N-k-1,k-1)}=e^{\frac{q}{N \epsilon_1 \epsilon_2}}\left(1+\mathcal{O}\left(M^{-1},w_{N-1}^{-1}\right)  \right) .\nonumber
\end{eqnarray}

\subsection{Effective coupling for \texorpdfstring{$N \geq 4$}{N>=4} theories} \label{subsec:effcouplN>=4}

For $N \geq 4$, the asymptotic of the  partition functions has not been found anywhere before, but the non-equivariant limit of the prepotential
\begin{eqnarray}
    \mathcal{F}=\epsilon_1 \epsilon_2 \ln(Z)
\end{eqnarray}
and the coupling matrix
\begin{equation} \label{tauirmat}
   2\pi \ii \tau^{\rm{IR}}_{uv}=\frac{\partial^2 \mathcal{F}}{\partial a_u \partial a_v}
\end{equation}
were widely studied in the literature. Here $\mathcal{F}$ is considered as a function of the independent variables $a_u$, $u=1,\ldots,N-1$.
\par One could assume that, since the main tool of those studies was the (non deformed) Seiberg-Witten theory, our result should automatically reproduce their result as soon as we put $\epsilon=0$. However, from (\ref{eqn:asympt2Nmenokandk}) it seems that the coupling matrix, up to a scalar factor, is the classical coupling matrix
\begin{equation}\label{eqn:tauclass}
    -\frac{\partial^2 w_1}{\partial a_u \partial a_v}= 
    \left(
    \begin{matrix}
        2 & 1 & \ldots & 1 \\
        1 & 2 & 1 & \ldots \\
        \ldots &\ldots & \ldots & \ldots \\
        1 & \ldots & 1 & 2
    \end{matrix}
    \right).
\end{equation}
This gives an immediate contradiction with the known fact that for $N\geq 4$ in no region of the space of expectation values $\bm{a}$ the coupling matrix can be proportional to the classical coupling matrix (\ref{eqn:tauclass}) \cite{MinahanNemeschansky}. In other words, the prepotential is not expected to be proportional to the classical action.
\par To resolve the problem, it is enough to point out that the asymptotic (\ref{eqn:asympt2Nmenokandk}) written in terms of the \textit{symmetric variables} is not adequate for finding the derivatives with respect to $a_u$. Indeed, for any $n=1,\ldots, N-2$, $w_n$ is an $(n+1)$-degree polynomial of $\bm{a}$. Thus, according to (\ref{au}) its expected behaviour in the large $A$ limit should be $A^{n+1}$. In reality, due to the symmetry of $w_n$, first $n$ orders cancel out, leaving a tautological $A$-independent identity $w_n=w_n$ (here we assume that the expansion (\ref{au}) is continued to the next $N$ orders, which can be easily done). But when we take derivatives with respect to $a_u$, the symmetry is broken. For example, we have
\begin{equation}
    \frac{\partial w_1}{\partial a_u}=-\sum_{v=1}^{N-1}a_v-a_u=a_N-a_u=A(1-e^{\frac{2\pi\ii u}{N}}).
\end{equation}
This means that when we compute the coupling matrix $\tau^{\rm{IR}}_{uv}$, the terms  hidden in $\mathcal{O}(w_{N-1}^{-1})$ in (\ref{eqn:asympt2Nmenokandk}) are no longer suppressed and can contribute significantly. 
\par
From the analysis above, we see that the coupling matrix cannot be read from the asymptotic (\ref{eqn:asympt2Nmenokandk}). In particular, what we call the effective infrared coupling in Subsection \ref{subsec:renormalisation} is not a priori a component of the coupling matrix (\ref{tauirmat}) in a suitable basis. Instead, it can be identified with the dual variable
\begin{equation}
    w_1^D=\frac{\partial\mathcal{F}}{\partial w_1}=2 \pi \tau^{\rm IR}.
\end{equation}
To find  $\frac{\partial F}{\partial w_1}$ in the leading order with respect to $A$  it is enough to compute ${\tau}^{\mathrm{IR}}(A)$ precisely at the point 
\begin{equation}
    a_u=A e^{\frac{2\pi\ii u}{N}}.
\end{equation}
Such a configuration is known as the special vacuum \cite{ArgyresPelland} and was studied in the literature before and after the introduction of the name `special vacuum' \cite{MinahanNemeschansky, Billoetal,Lerda}.  We can now proceed to establish the connection between $w_1^D$ and $\tau^{\rm{IR}}_{uv}$, and to compare our results with those available in the literature.
\par In the special vacuum (as well as in the leading order of $A\to \infty$ expansion) holds
\begin{equation}\label{eqn:ddw1}
    \frac{\partial}{\partial w_1}=-\frac{1}{NA}\sum_{u=1}^{N-1}e^{-\frac{2\pi \ii u}{N}}\frac{\partial}{\partial a_{u}},
\end{equation}
\begin{equation}\label{eqn:ddA}
    \frac{\partial}{\partial A}=\sum_{u=1}^{N-1}e^{\frac{2\pi \ii u}{N}}\frac{\partial}{\partial a_{u}},
\end{equation}
where in the left  hand side the set of the independent variables is assumed to be $w_1,\ldots,w_{N-2},A$.
So,
\begin{equation} \label{taumattotau}
    \frac{\partial }{\partial A}A\frac{\partial F}{\partial w_1}=-\frac{2\pi\ii}{N}\sum_{u,v=1}^{N-1}\tau^{\rm IR}_{uv}(A)e^{\frac{2\pi \ii u}{N}}e^{-\frac{2\pi \ii v}{N}} .
\end{equation}
Here we underline the dependence of the coupling matrix on $A$, because it is the only parameter which we are varying at a point.
This implies
\begin{equation}\label{eqn:F-via-tau}
   w_1^D=-\frac{2\pi\ii}{N}\sum_{u,v=1}^{N-1} \frac{1}{A}\int_{A_0}^A \tau^{\rm IR}_{uv}(A)e^{\frac{2\pi \ii (u-v)}{N}}+\frac{\partial F}{\partial w_1}\big \vert_{A=A_0}\frac{A_0}{A},
\end{equation}
where $A_0$ is some constant.
\par In the limit of large $A$, the second term in (\ref{eqn:F-via-tau}) becomes irrelevant, and the mass-dependent terms in $\tau^{IR}(A)$ are suppressed due to dimensional reasons (there are no equivariant parameters and there cannot be singularities with
respect to the masses), hence
\begin{equation} \label{eqn:tauInf}
   \lim_{A\to \infty} w_1^D= 2\pi\ii \tau^{\rm IR}= -\frac{2\pi\ii}{N}\sum_{u,v=1}^{N-1}\tau^{\rm IR}_{uv} e^{\frac{2\pi \ii (u-v)}{N}}.
\end{equation}
Therefore, we expect that $\tau^{\rm IR}$ given by (\ref{eqn:tau-final}) is connected with the coupling matrix $\tau^{\rm IR}_{uv}$, which can be found in the literature, by the contraction procedure (\ref{eqn:tauInf}).

\par As an additional check of the found asymptotic behaviour (\ref{eqn:asympt2Nmenokandk}), we should also verify that for $N>2$
\begin{equation}\label{eqn:sanityA}
    \frac{\partial^2 \mathcal{F}}{\partial A^2}=\frac{2\pi\ii}{N}\sum_{u,v=1}^{N-1}\tau^{\rm IR}_{uv}e^{\frac{2\pi \ii u}{N}}e^{\frac{2\pi \ii v}{N}} \xrightarrow[A\to \infty]{} 0 ,
\end{equation}
because, unless $N=2$, the asymptotic does not depend on $A$ \footnote{In fact, we have to ask that the convergence in (\ref{eqn:sanityA}) is fast enough, e.g. $\mathcal{O}(A^{-2})$.}.

\paragraph{Comparison for the $N=4$ case.}
For $N=4$ matrix $\tau^{\rm IR}_{uv}$ is written explicitly in \cite{Billoetal}.
\begin{equation} \label{eqn:tauuvBillo}
    2 \pi \ii \, \tau ^{\rm IR}_{uv} =
    \pi \ii \, \tau \left( \begin{matrix}
        2 & 1 & 1 \\
        1 & 2 & 1 \\
        1 & 1 & 2
    \end{matrix} \right)_{uv}
     +
    \pi \ii \, \tau'\left(\begin{matrix}
        0 & -1 & 1 \\
        -1 & -2 & -1 \\
        1 & -1 & 0
    \end{matrix} \right)_{uv}
     ,
\end{equation}
where the two couplings $\tau$ and $\tau'$ are given by
\begin{eqnarray}
    \pi \ii \tau_+ =  \pi \ii (\tau + \tau') =  \log q + \pi \ii - \log 16 + \frac{1}{2} q + \frac{13}{64}q^2 + \frac{23}{192} q^3 + \ldots  \nonumber\\
    \pi \ii \tau_- =  \pi \ii (\tau - \tau') = \log q + \pi \ii - \log 64 + \frac{3}{8} q + \frac{141}{1024}q^2 + \frac{311}{4096} q^3 + \ldots 
\end{eqnarray}
Putting (\ref{eqn:tauuvBillo}) into (\ref{eqn:tauInf}) we get\footnote{Note that in \cite{Billoetal} the coupling constants are connected as $q=\exp{\pi \ii \tau}$, hence the factor $1/2$.}
\begin{equation} \label{eqn:tauminus}
    \tau^{\rm IR}=-\frac{\tau_-}{2}.
\end{equation}
Comparing (\ref{eqn:tauminus}) with the first terms of the expansion of (\ref{eqn:tau-final}) we observe a perfect agreement. Moreover, it is known (\cite{Billoetal,Lerda}) that (\ref{eqn:tauminus}) can be inverted as
\begin{equation}
    q_0=-64\left(\frac{\eta(q_{\mathrm{IR}}^2)}{\eta(q_{\mathrm{IR}})}\right)^{24}=J_{\left(2,\infty,\infty\right)}^{-1},
\end{equation}
where the last equality is due to \cite{Doran}. This is in agreement with Subsubsection \ref{subsubsec:modularprop}.
\par  The second independent coupling constant appearing in \cite{Billoetal} is related to the derivative of the prepotential with respect to the second independent symmetrical variable $w_2$
\begin{equation}
    \frac{\partial^2 \mathcal{F}}{\partial w_2^2}=\frac{1}{16 A^4} \sum_{u,v=1}^{3}\tau^{\rm IR}_{uv}e^{\frac{4\pi \ii (u-v)}{4}}=\frac{\tau_+}{8A^4}
\end{equation}
and contributes to terms of order $\mathcal{O}(w_3)=\mathcal{O}(A^{-4})$ suppressed in the asymptotic, as expected.
\par We also see that
\begin{equation}
    \frac{\pi\ii}{2}\sum_{u,v=1}^{3}\tau^{\rm IR}_{uv}e^{\frac{2\pi \ii (u+v)}{3}}= \frac{\partial^2 \mathcal{F}}{\partial A^2}=0 ,
\end{equation}
as we wanted.


\subsection{Comparison with direct computations for \texorpdfstring{$N=4, \, 5$}{N=4, 5}}
So far the asymptotic (\ref{eqn:asympt2Nmenokandk}) has passed all the tests to which it was subjected, and the last check we perform is a comparison with direct computations of the partition function given by (\ref{eqn:ZkNak}) with $a_u$ defined as (\ref{au}) for $N \geq 4$ up to several instantons, and here we encounter a divergence.
\par For $N=4$, the saddle point approach gives
\begin{equation}
   \ln( Z^{(\infty)})=\frac{1}{\epsilon_1\epsilon_2}\left(\frac{6 w_1-4\tilde{T}_1+22\tilde{T}_0\epsilon -69 \epsilon^2}{16}q+\frac{141 w_1-92\tilde{T}_1-18\tilde{T}_0^2+722\tilde{T}_0\epsilon -2223 \epsilon^2}{1024}q^2+\ldots\right),
\end{equation}
while the expansion of the exact expression  (\ref{eqn:ZkNak}) gives
\begin{equation}\label{eqn:ZN4nakaj}
    \ln(Z)= \ln( Z^{(\infty)})+\frac{q^2}{512}+\ldots+\mathcal{O}(w_{N-1}^{-1}). 
\end{equation}
Analogously, for $N=5$, from the saddle point we have
\begin{equation}
   \ln( Z^{(\infty)})=\frac{1}{\epsilon_1\epsilon_2}\left(\frac{8 w_1-5\tilde{T}_1+35\tilde{T}_0\epsilon -135 \epsilon^2}{25}q+\frac{28 w_1-17\tilde{T}_1-4\tilde{T}_0^2+179\tilde{T}_0\epsilon -679 \epsilon^2}{250}q^2+\ldots\right),
\end{equation}
and
(\ref{eqn:ZkNak}) gives
\begin{equation}\label{eqn:ZN5nakaj}
    \ln(Z)= \ln( Z^{(\infty)})+\frac{3q^2}{625}+\ldots+\mathcal{O}(w_{N-1}^{-1}). 
\end{equation}
We see that the actual asymptotic behaviour of the instanton partition function does not match $Z^{(\infty)}$ found by the saddle point method perfectly. This is not a big surprise for us, because, as we already noted, the number of instantons in the saddle point configuration is not large. This means that the leading order of the saddle point approximation actually vanishes and what we are computing is of the same order as the first corrections.
\par 
The coefficients of the corrections to the saddle point asymptotic are small, so, informally, one could say that the saddle point approximation works quite well.

\section{Discrete saddle point method and \texorpdfstring{$qq$}{qq}-characters} \label{sec:q-character-corr}
In this section, we explain why the saddle point method works and compute the missing correction.
\par 
First of all, let us see how the discrete saddle point method derived in Subsection \ref{subsec:sadptapp} is related to the $qq$-characters, since the similarity between the results is clear. To this end, we use the expression  (\ref{eqn:eff-energ-r}) for the effective energy of an instanton defined by (\ref{eqn:eff-enery-r-def}) to get
\begin{equation}
    q^{|\vec{Y}|} Z_{\vec{Y}}=-  q^{|\vec{Y}'|}Z_{\vec{Y}'} \frac{\mathrm{Res}_{y=\overline{\phi_k}}\omega_{\vec{Y}'}\left(y-\frac{\epsilon}{A}\right)}{\mathrm{Res}_{y=\overline{\phi_k}}\omega_{\vec{Y}}(y)^{-1}}.
\end{equation}
Here we use the fact that $\omega_{\vec{Y}}$ has poles at convex corners and zeroes at the concave corners of $\vec{Y}$. 
This implies that
\begin{equation}
    q^{|\vec{Y}'|} Z_{\vec{Y}'}\omega_{\vec{Y}'}(y-\frac{\epsilon}{A})+q^{|\vec{Y}|}Z_{\vec{Y}}\omega_{\vec{Y}'}(y)^{-1}
\end{equation}
has no poles except for the ones coming from the polynomial $Q$ in denominator of $\omega_{\vec{Y}}$. Finally, one notes that the first term has poles at the points where an instanton can be added to $\vec{Y}'$, while the second term has poles exactly at the points where an instanton can be removed from $\vec{Y}$. This means that if we consider a sum
over $\vec{Y}$ and $\vec{Y'}$

\begin{equation}\label{eqn:pre-qq-char}
    \sum_{\vec{Y}}\left(q^{|\vec{Y}|} Z_{\vec{Y}}\omega_{\vec{Y}}(y-\frac{\epsilon}{A})+q^{|\vec{Y}|}Z_{\vec{Y}}\omega_{\vec{Y}'}(y)^{-1}\right),
\end{equation}
all its poles can be grouped into pairs of the form (\ref{eqn:pre-qq-char}), and thus cancel. Dividing by the partition function, we get 
\begin{equation}
\label{eqn:qq1}\left<\omega_{\vec{Y}}\left(y-\frac{\epsilon}{A}\right) +\frac{1}{\omega_{\vec{Y}}(y)}\right>=A^N \kappa \frac{P_N(y)}{Q(y)},
\end{equation}
where $P_N$ is a polynomial of degree $N$, with the leading term $y^N$. Up to notation, this is the first $qq$-character \cite{qqchar}, or double quantised Seiberg-Witten curve \cite{dqSWgeom}. 
It is easy to see that the asymptotic analysis in the limit $y\to \infty$ which we performed assuming the saddle point approximation is still valid, and the periods equation now takes the form
\begin{equation}\label{eqn:qqperiods}
    \left<\oint \frac{\dd \omega_{\vec{Y}}(y)}{\omega_{\vec{Y}}(y)}\right>.
\end{equation}
\par
We note that the equation (\ref{eqn:qq1}) is exact, but not particularly useful. Indeed, in general we do not know how $\left<\omega_{\vec{Y}}(y)\right>$ is related to $\left<\omega_{\vec{Y}}(y)^{-1}\right>$ or $\left<\ln \omega_{\vec{Y}}(y)\right>$, so (\ref{eqn:qq1}) and (\ref{eqn:qqperiods}) do not impose any useful restrictions on the coefficients on the polynomial $P$. In order to arrive at the `semiclassical'  equations (\ref{structure}) and (\ref{eqn:Au-and-au}) we still have to assume that all the sums over Young diagrams are contributed by one `saddle point' contribution, and thus
\begin{equation}\label{eqn:sadpnt-av}
    \left<\frac{1}{\omega(y)}\right>\approx \left<{\omega(y)}\right>^{-1}, \ \left<\ln \omega(y)\right>\approx \ln \left<\omega(y)\right>,
\end{equation}
and all the analysis of Section \ref{sec:asympt} would apply. We conclude that the discrete saddle point equation is equivalent to the first $qq$-character accompanied by the assumptions (\ref{eqn:sadpnt-av}). What is still missing is the motivation for (\ref{eqn:sadpnt-av}) in the case when the average number of instantons is finite. Before addressing this question, let us see how the higher $qq$-characters allow us to find the missing factor in the asymptotic, assuming that (\ref{eqn:sadpnt-av}) holds in the leading order.

\subsection{Second \texorpdfstring{$qq$}{qq}-character and the missing factor} \label{subsec:extrafact}
Analogously to (\ref{eqn:qq1}), one can derive
\begin{align}\label{eqn:qq2}
    \left<\omega_{\vec{Y}}\left(y_1-\frac{\epsilon}{A}\right)\omega_{\vec{Y}}\left(y_2-\frac{\epsilon}{A}\right)+\frac{\omega_{\vec{Y}}\left(y_1-\frac{\epsilon}{A}\right)}{\omega_{\vec{Y}}(y_2)}\mu(y_1-y_2)+\frac{\omega_{\vec{Y}}\left(y_1-\frac{\epsilon}{A}\right)}{\omega_{\vec{Y}}(y_2)}\mu(y_2-y_1)
    +\frac{1}{\omega_{\vec{Y}}(y_1)\omega_{\vec{Y}}(y_2)} \right> \nonumber\\ - \frac{2\epsilon_1\epsilon_2}{A^2(y_1-y_2)^2-\epsilon^2}=
    \kappa^2\frac{P_{2N}(y_1,y_2)}{Q(y_1)Q(y_2)},
\end{align}
where 
\begin{equation}
    \mu(y)=A\frac{(t+\epsilon_1/A)(t+\epsilon_2/A)}{t(t+\epsilon/A)},
\end{equation}
and $P_{2N}$ is a polynomial of total degree $2N$. With the same asymptotic analysis as before, one can show that the leading order of $P_{2N}$ in the limit $y_2\to \infty$ is $y_2^{N}P_1(y_1)$, and similarly for $y_1\to \infty$.  
\par 
To see this, let us first compute the residues at the poles of $\mu$, which are $y_1=y_2$ and $y_1=y_2\pm \epsilon$. 
\par
The residue of the left-hand side at $y_2=y_1=y$ is equal to
\begin{equation}\label{eqn:symmureg}
    \frac{\omega(y-\frac{\epsilon}{A})}{\omega(y)}{\rm Res}_{y'=0}\left(\mu(y')-\mu(-y')\right)=0.
\end{equation}
At $y_2=y_1+\epsilon/A$  the residue of the  left-hand side is
\begin{equation}
   {\mathrm{Res}}_{y'=-\epsilon/A}\left
   (\mu(y')-\frac{2\epsilon_1\epsilon_2}{A^2y'^2-\epsilon^2}\right)=0,
\end{equation}
and similarly for $y_2=y_1-\epsilon/A$. Finally, all the poles of $\omega_{\vec{Y}}$
cancel similarly to the first order case (we refer to \cite{qqchar} for the details) except for those coming from $Q$. This implies that $P_{2N}$ is an entire function, and the asymptotic analysis implies that it is a polynomial of maximal degree $N$ in $x$ and $y$. The average in the left-hand side is nothing but the second $qq$-character \cite{qqchar}.
\par 
Let us now assume that all the averages are dominated by the contribution of a 'saddle point' diagram, but now we will take small fluctuations around it. It is convenient to define\footnote{At this point, we ignore the issue of choosing the branch of the logarithm. In the saddle point one can argue that $\omega(y)\approx \left<\omega_{\vec{Y}}(y)\right>$ selects the right one. } 
\begin{equation}\label{eqn:omega-av}
    \omega(y)=\exp\left(\left< \ln \omega_{\vec{Y}}(y)\right>\right).
\end{equation}
Then the period equation (\ref{eqn:Au-and-au}) is exact. The fluctuations around the saddle point can then be characterised by 
\begin{equation}\label{eqn:delta-def}
    \delta_{\vec{Y}}(y)=\ln\left(\frac{\omega_{\vec{Y}}(y)}{\omega(y)}
    \right).
\end{equation}
It satisfies 
\begin{equation}\label{eqn:delta-per}
    \oint   \delta_{\vec{Y}}(y)\dd y=-\oint y \dd \delta_{\vec{Y}}(y)=0.
\end{equation}
By construction,
\begin{equation}
    \left<\delta_{\vec{Y}}(y)\right>=0,
\end{equation}
but the higher correlators 
\begin{equation}\label{eqn:Delta-n-def}
   \Delta_n(x_1,\ldots,x_n)= \left<\delta_{\vec{Y}}(x_1)\cdots \delta_{\vec{Y}}(x_n) \right>
\end{equation}
can be non-zero. If we believe that this is valid in the leading order (\ref{eqn:sadpnt-av}), we should expect $\delta$ to be small, and thus assume that the higher correlators are negligible. Then the first correction to the saddle point approximation would be given by taking only $\Delta_2$ into account. So, we have
\begin{equation}\label{eqn:sadpnt-av-cor}
    \left<\omega_{\vec{Y}}(y)^{\pm 1}\right>\approx \omega(y)\left(1+\frac{\Delta(y,y)}{2}\right),
\end{equation}
\begin{equation}
    \left<\omega_{\vec{Y}}(x)^{s_1}\omega_{\vec{Y}}(y)^{s_2 1}\right>\approx \left<\omega_{\vec{Y}}(x)^{s_1 }\right>\left<\omega_{\vec{Y}}(y)^{s_2}\right>(1+s_1 s_2 \Delta_2(x,y)),
\end{equation}
which can be understood as the subleading corrections to  (\ref{eqn:sadpnt-av}).
Substituting this into (\ref{eqn:qq2}), and assuming that $\Delta_2/A^2$ is of the next order of smallness, we get
\begin{equation} \label{eqn:deltaxy}
    \Delta_2(x,y)=\frac{\frac{\epsilon_1 \epsilon_2}{A^2}\frac{(\omega(x)-\omega(y))^2}{(x-y)^2}+\kappa^2\frac{P^c_{2(N-1)}(x,y)}{x^N y^N}\omega(x)\omega(y)}{(\omega(x)^2-1)(\omega(y)^2-1)},
\end{equation}
where 
\begin{equation}
    P^c_{2(N-1)}(x,y)=P_{2N}(x,y)-P_{N}(x)P_{N}(y)
\end{equation}
is a polynomial that has total degree $2(N-1)$, and with maximal degree in each of its variables being $N-1$.
Combining (\ref{eqn:delta-per}) with (\ref{eqn:Delta-n-def}), we get
\begin{equation} \label{eqn:IntDeltaZero}
    \oint_{\mathcal{C}_u} \Delta_2(x,y) \dd y =0,
\end{equation}
where $u$ runs from one to $N$. The variable $x$ in the equation above can be treated as a fixed parameter. Then we have $N$ linear equations for $N$ coefficients of $P^c_{n(N-1)}(x,y)$ as a polynomial in $y$ of degree $N-1$. It is not evident at this point why these coefficients would be $N-1$-degree polynomials in the parameter $x$, but we shall see that they are.
\par As in Section \ref{sec:asympt}, we will integrate over the variable $z=\omega(y)$, rather than $y$.
\begin{equation}
    \oint \Delta_2(x,y) \dd y =\oint_{|z|=r} \Delta_2(x,y(z)) \frac{1-\frac{1}{z^2}}{N\kappa U^N}y(z)^{N+1}\dd z,
\end{equation}
where, as in Section \ref{sec:asympt} and Appendix \ref{app-comp}, $0<r<1$ is chosen so that the integrand is holomorphic on the contour.
\par For shortness, we set $\omega(x)=\omega$.
\par Let us first compute the contribution of the first term in the numerator of (\ref{eqn:deltaxy}). Clearly, we can ignore the common factor $\frac{1}{\omega(x)^2-1}$.
\begin{equation} \label{eqn:lhs}
    \frac{\epsilon_1 \epsilon^2}{A^2}\frac{1}{\omega^2-1}\frac{1}{\kappa N U^N}\oint_{|z|=r} \frac{(\omega-z)^2}{(x-y(z))^2}y(z)^{N+1} \frac{\dd z}{z^2}
\end{equation}
Note that
\begin{eqnarray}
    \frac{1}{(x-y)^2}=\frac{1}{x^2}\sum_{k=0}^\infty (k+1) \left(\frac{y}{x}\right)^k=\frac{1}{x^2}\sum_{m=0}^\infty \sum_{l=0}^{N-1}(l+mN+1)\left(\frac{y}{x}\right)^l\left(\frac{y}{x} \right)^{mN} \nonumber \\
    = \frac{1}{x^2}\sum_{l=0}^{N-1}\left(\frac{y}{x}\right)^l\frac{x^{2N}(l+1)+x^N y^N (N-l-1)}{(x^N-y^N)^2}. \label{eqn:resolvesing1}
\end{eqnarray}
\par Moreover, since we are looking for the corrections of order $1/A^2$, we can use the classical SW curve instead of the quantum one. Then we can write
\begin{equation} \label{eqn:resolvesing2}
    x^N-y^N=\frac{x^N y^N}{\kappa U^N}(\omega-z)\frac{\omega z-1}{\omega z}.
\end{equation}
With this, (\ref{eqn:lhs}) turns into
\begin{eqnarray}
    \frac{1}{\kappa N U^N}\oint_{|z|=r} \frac{(\omega-z)^2}{(x-y(z))^2}y(z)^{N+1} \frac{\dd z}{z^2} = \nonumber \\
    \sum_{l=0}^{N-1}\frac{U^N \kappa}{x^{2+l}} \left(\frac{l+1}{N}\oint_{|z|=r} \dd z \frac{y^{1+l-N}}{\left(z-\frac{1}{\omega}\right)^2}+\frac{N-l-1}{N}\frac{1}{x^N}\oint_{|z|=r} \dd z \frac{y^{1+l}}{\left(z-\frac{1}{\omega}\right)^2} \right).
\end{eqnarray}
Taking into account that the leading order of $y(z)$ suffices and the Appendix \ref{app-comp} we get
\begin{eqnarray}
    \frac{1}{\kappa N U^N}\oint_{|z|=r} \frac{(\omega-z)^2}{(x-y(z))^2}y(z)^{N+1} \frac{\dd z}{z^2} = \nonumber \\
    \sum_{l=0}^{N-1}\frac{U^{1+l} \kappa}{x^{2+l}}e^{\frac{2\pi\ii u (l+1)}{N}} \sum_{k=0}^\infty \frac{\kappa^{-k}}{k!}\left(\frac{l+1}{N}\left(\frac{l+1}{N}-1\right)_k+\frac{N-l-1}{N}\left(\frac{l+1}{N}\right)_k \frac{U^N}{x^N}\right) t_k,
\end{eqnarray}
where
\begin{equation}
    t_k=\oint_{|z|=r} \dd z \frac{\left(z+\frac{1}{z} \right)^k}{\left(z-\frac{1}{\omega} \right)^2}. 
\end{equation}
By an analytical continuation argument, it is enough to consider the case when $x$ stays away from $\mathcal{C}_u$, so we may assume that $|\omega|<1$, thus the pole $z=\omega^{-1}$ stays out of the contour of integration.
\par

Note that $t_0=0$. Then, using the classical SW again, we get 
\begin{eqnarray}\label{eqn:lhs-res}
    \frac{1}{\kappa N U^N}\oint_{|z|=r} \frac{(\omega-z)^2}{(x-y(z))^2}y(z)^{N+1} \frac{\dd z}{z^2} = \nonumber \\
     \sum_{l=0}^{N-1}\frac{U^{1+l} \kappa}{x^{2+l}}e^{\frac{2\pi\ii u (l+1)}{N}} \sum_{k=1}^\infty \frac{\left(\frac{1+l}{N}\right)_k}{k! \kappa^k}\frac{1}{\frac{l+1}{N}-1+k}((1-k)t_k+(\omega+\frac{1}{\omega})t_{k-1}) = \\
     2\pi\ii \sum_{l=0}^{N-1}\frac{U^{1+l} }{x^{2+l}} e^{\frac{2\pi\ii u (l+1)}{N}} \omega \left(\frac{l+1}{N}-1\right) \frac{1+l}{N \kappa} {}_2F_1\left(\frac{N+l+1}{2N},\frac{N+l+1}{2N},2,\frac{4}{\kappa^2}\right). \nonumber
\end{eqnarray}
To get the last line, we used
\begin{equation}
    (1-k)\frac{\left(z+\frac{1}{z}\right)^k}{\left(z-\frac{1}{\omega}\right)^2}+\left(\omega+\frac{1}{\omega}\right)\frac{\left(z+\frac{1}{z}\right)^{k-1}}{\left(z-\frac{1}{\omega}\right)^2}=k \omega \frac{\left(z+\frac{1}{z}\right)^{k-1}}{z^2}-\frac{\dd}{\dd z}\left(\frac{\left(z+\frac{1}{z}\right)^k}{z-\frac{1}{w}}\right),
\end{equation}
which implies
\begin{equation}
    (1-k)t_{k}+\left(\omega+\frac{1}{\omega}\right)t_{k-1}=k\omega \oint_{|z=r|}\frac{\left(z+\frac{1}{z}\right)^{k-1}}{z^2}.
\end{equation}
Note that the original integrals $t_k$ have very non-trivial dependence on $\omega$, but the relevant combination is just linear in it. As a result, the expression in the last line of (\ref{eqn:lhs-res}) has the form $\omega=\omega(x)$ times some polynomial of $x$. This is exactly the form that can be cancelled by the contribution of the second term of (\ref{eqn:deltaxy}). In fact, it is easy to see that we should look for the polynomial of the form
\begin{equation}\label{eqn:Pc2}
\kappa^2 P_{2(N-1)}^c(x,y)=\sum_{l=0}^{N-2}\tilde{p}_l y^l x^{N-l-2},    
\end{equation}
 and the coefficients are
\begin{equation} \label{eqn:pl}
    \tilde{p}_l=-\frac{\epsilon_1 \epsilon_2}{A^2}\frac{(1+l)(1+l-N)U^N}{N}\frac{{}_2F_1\left(\frac{N+l+1}{2N},\frac{N+l+1}{2N},2,\frac{4}{\kappa^2}\right)}{{}_2F_1\left(\frac{l+1}{2N},\frac{N+l+1}{2N},1,\frac{4}{\kappa^2}\right)}.
\end{equation}
Note that by  (\ref{eqn:deltaxy}) and (\ref{eqn:Pc2}-\ref{eqn:pl}), $\Delta_2(x,y)=\mathcal{O}(A^{-2})$, so our assumption that the correlators of $\delta_{\vec{Y}}$ is at least self-consistent at this level.
\par
We now return to the first $qq$-character (\ref{eqn:qq1}), but take into account the subleading corrections (\ref{eqn:sadpnt-av-cor}). This leads to a new contibution to the function $y_u$,
\begin{equation}
    \delta y_{u}(z) = \frac{\left(z+\frac{1}{z} \right) \Delta(y_u(z),y_u(z))}{2 \kappa N U^N} \left(y^{(0)}_u\right)^{N+1} .
\end{equation}
This correction will enter the left-hand sides of (\ref{eqn:intonzplane}), that should be compensated by a correction of the effective instanton number
\begin{equation}
    \delta k_{\mathrm{eff}}=-\frac{1+q}{1-q}\frac{U N}{\mathcal{I}_{\frac{1}{N}-1}} \frac{A^2}{\epsilon_1 \epsilon_2}e^{-\frac{2\pi\ii u}{N}}\oint_{|z|=r} \frac{\dd z}{z} \delta y_u(z) ,
\end{equation}
Resolving the apparent singularity in $\Delta_2(x,x)$ with (\ref{eqn:resolvesing1}), (\ref{eqn:resolvesing2}), we find
\begin{equation}
    \Delta_2(y_u(z),y_u(z))=\frac{z^2}{(z^2-1)^2}\frac{1}{\left(y_u(z)\right)^{2N+2}} \left(\frac{\epsilon_1 \epsilon_2 }{A^2}\frac{N^2 z^2 \kappa^2 U^{2N}}{ (z^2-1)^2} + \left(y_u(z)\right)^N \sum_{l=0}^{N-2}\tilde{p}_l \right),
\end{equation}
which leads to
\begin{equation}
    \delta k_{\mathrm{eff}}=-\frac{1+q}{1-q} \frac{1}{2 U \mathcal{I}_{\frac{1}{N}-1} } \left( N^2 \kappa^2 \oint_{|z|=r} \frac{\left(z+\frac{1}{z}\right)}{\left(z-\frac{1}{z}\right)^4}\rho^{1+\frac{1}{N}} \frac{\dd z}{z}+N^2\frac{A^2}{\epsilon_1\epsilon_2}\frac{\sum_{l=0}^{N-2}\tilde{p}_l}{U^N}\oint_{|z|=r} \frac{\left(z+\frac{1}{z}\right)}{\left(z-\frac{1}{z}\right)^2}\rho^{\frac{1}{N}} \frac{\dd z}{z} \right)
\end{equation}
By the same methods as in Appendix \ref{app-comp}, we finally arrive to
\begin{eqnarray}
    \delta k_{\mathrm{eff}}=\frac{1+q}{1-q} \frac{1}{U \mathcal{I}_{\frac{1}{N}-1} }\pi \ii  \sum_{k=0}^{\infty}\kappa^{-(2k+1)}\left(\frac{\epsilon_1\epsilon_2}{A^2}\frac{\sum_{l=0}^{N-2} \tilde{p}_l}{U^N}\frac{\left(-\frac{1}{N}\right)_{2k+1}}{k! k!} +  N^2\frac{\left(-\frac{1}{N}-1\right)_{2k+3}}{6 k! (k+1)!} \right) = \nonumber \\
    =\frac{1+q}{1-q} \frac{1}{U \mathcal{I}_{\frac{1}{N}-1} }\pi \ii \left(-\frac{\epsilon_1\epsilon_2}{A^2}\frac{\sum_{l=0}^{N-2} \tilde{p}_l}{U^N}\frac{\mathcal{I}_{\frac{1}{N}-1}}{N \kappa} +\kappa^2 N^2\mathcal{J}_{\frac{1}{N}+1} \right).
\end{eqnarray}
To find the correction for the partition function $Z_c$, we have to solve
\begin{equation}
    q\frac{\dd}{\dd q}\ln(Z_c)=\delta k_{\rm eff},
\end{equation}
with boundary condition $Z_c\vert_{q=0}=0$. Then one can verify that the solution is
\begin{equation} \label{eqn:Zc}
    Z_c=\frac{f_2^{\frac{N(N+1)}{24}}}{\prod_{l=1}^{N-1}(f_2^{(l)})^{\frac{1}{4}}}.
\end{equation}
Here $f_2^{(l)}=\left(\frac{\dd \tau^{(l)}_{\rm IR}}{\dd \tau_{\rm UV}}\right)$, and

    \begin{equation}\label{eqn:tau-l}
        2\pi \ii \tau^{(l)}_{\mathrm{IR}}=-\frac{\Gamma\left(\frac{l}{N}\right)^2}{\Gamma\left(\frac{2l}{N}\right)}\frac{{}_2F_1\left(\frac{l}{N},\frac{l}{N},\frac{2l}{N},1-q\right)}{{}_2F_1\left(\frac{l}{N},\frac{l}{N},1,q\right)}+\pi\left(\cot\left(\frac{\pi l}{N}\right) +\ii\right)
    \end{equation}
    is just formally obtained from (\ref{eqn:tau-final}) by a substitution $N\to N/l$. Clearly, $\tau^{(1)}_{\mathrm{IR}}=\tau_{\mathrm{IR}}$, and one can verify that $\tau^{(l)}_{\mathrm{IR}}=\tau^{(N-l)}_{\mathrm{IR}}$. So, one may guess that $\tau^{(l)}_{\mathrm{IR}}$ with $l=2,\ldots,\left\lfloor \frac{N}{2}\right\rfloor$ are the rest of the $\left\lfloor \frac{N}{2}\right\rfloor$ coupling constants introduced in \cite{Lerda}. In fact, they are, as we show in \cite{OneConst}. Independently, by minimal modification of the analysis of Subsubsection \ref{subsubsec:modularprop}, $f^{(l)}_2$, as functions of $\tau^{(l)}_{\mathrm{IR}}$, can be understood as the modular forms for the group $\Gamma_{(\frac{N}{N-2l},\infty,\infty)}$.
    \par 
As one can see, for $N=2$ and $N=3$, when there exists only one coupling constant, the correction indeed disappears and $Z_c=1$. For $N=4$, one has
\begin{equation}
    Z_c=\frac{q^2}{512}+\ldots,
\end{equation}
and for $N=5$
\begin{equation}
    Z_c=\frac{3q^2}{625}+\ldots,
\end{equation}
in agreement respectively with (\ref{eqn:ZN4nakaj}) and (\ref{eqn:ZN5nakaj}). We underline that this result is correct even in the case of $w_1=m_f=0$, when $Z_c$ is the only non-trivial contribution, and the formulas above predict that the effective number of instantons is, in fact, much smaller than one. Numerical evaluations show that this is not an artifact of the small $q$ expansion. For example, if $N=4$, and $w_1=m_f=0$,
\begin{equation}
    k_{\rm eff}(q=0.9)\approx 0.1.
\end{equation}
This is a clear indication that the validity of our result goes far beyond the applicability range of the original derivation by Nekrasov and Okounkov. Before giving an explanation why it is so, let us write down the final result.
\begin{eqnarray} \label{eqn:finalanswer} 
        Z^{(0,2N)}&=&\left(\frac{q}{D q_{\mathrm{IR}}
        } \right)^{-\frac{w_1}{\epsilon_1\epsilon_2}} f_2(q_{\mathrm{IR}})^{\frac{N}{4 \epsilon_1 \epsilon_2}\left(\tilde{T}_0^2+ \epsilon \tilde{T}_0+\epsilon^2\frac{N+1}{6} \right)} \\ && \cdot  (1-q)^{-\frac{1}{\epsilon_1 \epsilon_2}\left( -\tilde{T}_1+\frac{N-1}{2N}\tilde{T}_0^2 \right)}  \frac{f_2^{\frac{N(N+1)}{24}}}{\prod_{l=1}^{N-1}(f_2^{(l)})^{\frac{1}{4}}}\left(1+\mathcal{O}\left(w_{N-1}^{-1}\right) \right) . \nonumber
    \end{eqnarray}

\subsection{Higher corrections and justification of the saddle point approximation}

Let us now briefly explain how we can justify (\ref{eqn:sadpnt-av}) without relying on the saddle point method. 
\par 
The $qq$-characters can be written to any order. Following \cite{ABCD}, we introduce
\begin{equation}\label{eqn:qq-high}
    X_{n}(y_1,\ldots,y_n)=\sum_{\substack{\{1,\ldots,n\}= \\
    \{i_1,\ldots,i_l\}\cup \{j_1,\ldots,j_{n-l}\}}}
    \left<\frac{\prod_{s=1}^{l}\omega_{\vec{Y}}\left(y_{i_s}-\frac{\epsilon}{A}\right)}{\prod_{r=1}^{n-l}\omega_{\vec{Y}}\left(y_{j_r}\right)}\right>\prod_{r=1}^l\prod_{s=1}^{n-l}\mu(y_{i_r}-y_{j_s}).
\end{equation}
For the same reasons as in the cases $n=1,2$ these functions have no poles, except for the ones caused by the polynomials $Q$, and the ones at $y_i-y_j=\pm \frac{\epsilon}{A}$. Both kinds of singularities can be removed if we consider $P_{nN}$, defined by
\begin{eqnarray} \label{eqn:qq-reg-high}
\sum_{0\leq 2r \leq n}(-1)^r \sum_{ {\bf k}_r}X_{n-2r}(y_1,\ldots,\hat{y}_{{\bf k}_r},\ldots,y_n)\prod_{s=1}^r \frac{2\epsilon_1\epsilon_2}{A^2\left(\Delta y_{{\bf k}_r(s)}^2-\frac{\epsilon^2}{A^2}\right)}=\kappa^n
     \frac{P_{nN}(y_1,\ldots,y_N)}{\prod_{i=1}^n Q(Ay_i)},
\end{eqnarray}
where ${\bf k}_r$ is a set of $r$ pairs ${\bf k}_r=\{(i_1,j_1), \ldots, (i_r,j_r) \}$ such that $1\leq i_1<i_2 \ldots< i_r\leq n$, $ 1\leq j_1<j_2< \ldots < j_r\leq n$ and $i_s<j_s$ for all $s=1,\ldots, r$ ; $\hat{y}_{{\bf k}_r}$ denotes that all arguments with an index appearing in ${\bf k}_r$ are omitted; $\Delta y_{{\bf k}_r(s)}$ stands for $y_{i_s}-y_{j_s}$. 
\par
The function $P_{nN}$ is an entire function, and by the asymptotic analysis it is a symmetric polynomial of total degree $nN$, and in the $y_n\to \infty$ limit it behaves as
\begin{equation}
    P_{nN}(y_1,\ldots,y_n)\sim y_n^N P_{(n-1)N}(y_1,\ldots,y_{n-1}).
\end{equation}
Analogously to (\ref{eqn:sadpnt-av-cor}), we can express the averages
\begin{equation}
    \left<\omega_{\vec{Y}}(y_1)^{s_1}\cdots \omega_{\vec{Y}}(y_n)^{s_n}\right>
\end{equation}
via the logarithmic averages $\omega(y_i)$ and the correlators (\ref{eqn:Delta-n-def}). The main problem is that without any a priori assumptions we would get an infinite series, including correlators of all possible orders. As a result, (\ref{eqn:qq-reg-high}) is a system of infinitely many equations on infinitely many unknown functions $\Delta_n$, which is not easy to operate with. So, the first step would be to find some way to drop the higher correlators without the saddle point argument. 
\par 
To do that, one observes that for generic $y$, $\mu(y)=1+\mathcal{O}(A^{-2})$. Thus, by (\ref{eqn:omegadef}) and (\ref{eqn:delta-def}), for fixed $\vec{Y}$ and generic $y$,
\begin{equation}
    \delta_{\vec{Y}}(y)=\mathcal{O}(A^{-2}).
\end{equation}
Taking into account (\ref{Zksim}), we conclude that
\begin{equation}
    \Delta_n=\sum_{k=1}^{\infty}q^k \Delta_n^{(k)}(y_1,\ldots,y_n),
\end{equation}
where 
\begin{equation}\label{eqn:Delta-first-est}
\Delta_n^{(k)}(y_1,\ldots,y_n)=\mathcal{O}(A^{2k-2n})    
\end{equation}
 This estimation is very rough, in particular it allows unlimited growth with $A\to \infty$ of the high instanton number contributions of the correlators, but it is good enough as a first step.
\par 
Another important ingredient is the connected $qq$-character. To save space, we shall not give all the details here, but only explain the idea. Instead of the correlators (\ref{eqn:Delta-n-def}), it is convenient to introduce the connected correlators $\Delta_n^c$, defined in the usual way. Then (\ref{eqn:pre-qq-char}) and (\ref{eqn:qq-reg-high}) can be provided with a natural diagrammatic interpretation. The connected $qq$-character by definition sums over the connected diagrams only. By simple induction, one can rewrite (\ref{eqn:qq-reg-high}) as
\begin{equation}\label{eqn:qq-c-high}
    X_n^c(y_1,\ldots,y_n)=\kappa^n 
     \frac{P^c_{n(N-1)}(y_1,\ldots,y_N)}{\prod_{i=1}^n Q(Ay_i)},
\end{equation}
where $P^c_{n(N-1)}$ is a symmetric polynomial, that does not have terms of degree higher than $N-1$ in either of its variables. In the case of $n=2$, it is precisely $P^c_{2(N-1)}$ we have seen in (\ref{eqn:deltaxy}). 
\par 
We claim that this construction allows one to prove that
\begin{equation}\label{eqn:Deltac-est}
    \Delta^c_n(y_1,\ldots,y_n)=\mathcal{O}(A^{2-2n}). 
\end{equation}
We note that this is enough to justify our assumption that as long as we are not interested in the corrections of order higher than $A^{-2}$, it is enough to consider $\Delta_2=\Delta_2^c$.
\par 
It is actually easier to prove that for the coefficients of $q$-expansion of the connected correlators holds
\begin{equation}\label{eqn:Deltac-est-ind}
    \Delta^{c(k)}_n(y_1,\ldots,y_n)=\mathcal{O}(A^{2(k-l)-2n}), \ l=0,\ldots,k-1, 
\end{equation}
which can be shown by induction in $k$ and $l$. Then (\ref{eqn:Deltac-est}) follows from $l=n-1$.
\par
In the case $k=1$ or $l=0$, (\ref{eqn:Deltac-est-ind}) follows from (\ref{eqn:Delta-first-est}), giving the base of the induction. As for the induction step, assuming that (\ref{eqn:Deltac-est-ind}) holds for $k<k_0$, and for $k=k_0$ but $l<l_0$, by diagrammatic analysis, one can show that 
\begin{equation}
    X_n^c=\frac{q^k \Delta^{c(k)}_n(y_1,\ldots,y_n)+\mathcal{O}(q^{k+1},A^{2(k-l_0)-2n})}{\omega(y_1)\cdots\omega(y_n)}.
\end{equation}
This and (\ref{eqn:qq-c-high}) allow one to express $\Delta^{c(k)}_n$ via the coefficients of $P^c_{n(N-1)}$ and the terms of desired smallness. Moreover, from the periods constraint
\begin{equation}
   \oint_{\mathcal{C}_u} \Delta_n^c(y_1,\ldots,y_n)=0,
\end{equation}
one derives that the coefficients of the polynomial have the same order. This proves (\ref{eqn:Deltac-est-ind}) for $k=k_0$ and $l=l_0$, and thus for all allowed values of $k$ and $l$.
\par 
Note that in this derivation we used only the fact that
\begin{equation} \label{eqn:unitmu}
    \mu(y)=1+\mathcal{O}(A^{-2}).
\end{equation}
We have already seen that the condition $\mu(y)\to 1$ was crucial in the 
 discrete saddle point approximation of Subsection \ref{subsec:sadptapp}. 
\par We conclude that (\ref{eqn:sadpnt-av}) is relatively accurate if (\ref{eqn:unitmu}) holds, and that to find the effective number of instantons in the leading order with respect to large $A$, the second $qq$-character is enough.

\section{Recurrence relation} \label{sec:recrel}

\subsection{Residue formula}

In \cite{SysoevaBykov} we established a residue formula for the instanton partition function in an $SU(N)$ theory with any number of colours $N$ and with different types of matter hypermultiplets, in particular, with $2N$ fundamental multiplets. This formula connects a residue of the partition function with its value at another, less singular, point.
\par In terms of the variables $\bm a$, the connected points differ by the partial Weyl permutation. Namely, let us take two points, $\bm a$ and $\hat{\bm a}$, such that  $a_w=\hat{a}_w$ for every $w \neq u, \, v$ and
\begin{eqnarray}
     \begin{aligned}[c]
        a_u = \alpha_u+m_u \epsilon_1 + n_u \epsilon_2  \\
        a_v = \alpha_v+m_v \epsilon_1 + n_v \epsilon_2 
        \end{aligned}
            \quad  &\rightarrow& \quad
        \begin{aligned}[c]
           \hat{a}_u^{(uv)}= \alpha_u+m_u \epsilon_1 + n_v \epsilon_2  \\
          \hat{a}_v^{(uv)} = \alpha_v+m_v \epsilon_1 + n_u \epsilon_2  
        \end{aligned} 
    \end{eqnarray}
Here $m_u,n_u \in \mathbb{Z}$, $\alpha_i\in\mathbb{C}$. 
Then 
\begin{equation} \label{residuea}
{\rm Res}_{\alpha_{uv}=0} Z( {\bf a})=q_0^{m n}\frac{\mathcal{P}^{(uv)}_{N,{\rm R}}(m,n|{\bf a})}{\mathcal{P}^{(uv)}_{N}(m,n|{\bf a}) }Z(\hat{{\bf a}}^{(uv)}),
\end{equation}
where $\alpha_{uv}=\alpha_u-\alpha_v$, and
$\mathcal{P}^{(uv)}_{N}$, $\mathcal{P}^{(uv)}_{N,{\rm R}}$ are some polynomials defined in \cite{SysoevaBykov}. The subscript $R$ indicates dependence on the representation of the matter hypermultiplet.
\par The equation (\ref{residuea}) can be rewritten in terms of the symmetric variables $\bm w$ as
\begin{equation} \label{residuesym}
    {\rm Res}_{w_{N-1}=\bar{w}_{N-1}^{(k|m,n)}} Z({\bf w})=q_0^{m n}J^{(m,n)} \frac{\mathcal{P}^{(mn)}_{N, {\rm R}}}{\mathcal{P}^{(mn)}_{N}}Z({\bf \hat{w}}^{(k|m,n)}),
 \end{equation}
where $\bar{w}_{N-1}^{(k|m,n)}$ is the $k$-th root (in an arbitrarily chosen enumeration) of the equation
\begin{equation} \label{determinator}
    \Delta^{(m,n)}(\bm{w})=\Delta^{(m,n)}({\bf a})=\prod_{u\neq v}((a_{u}-a_v)^2-(m_1\epsilon_1+n_2\epsilon_2)^2) ,
\end{equation} 
$\hat{w}^{(k|m,n)}$ is a set of symmetric variables corresponding to $\hat{\bm{a}}$ and containing the transformed root $\bar{w}_{N-1}^{(k|m,n)}$ as the last element and $J^{(m,n)}$ is the Jacobian compensating the change of variables from $a_{uv}$ to $w_{N-1}$, symbolically defined as
\begin{equation}
   J^{(m,n)}{\rm Res}_{a_{12}=\bar{a}_{12}}
   ={\rm  Res}_{w_{N-1}=\bar{w}_{N-1}^{(k|m,n)}}
\end{equation}

 In \cite{SysoevaBykov} we show that all poles of $Z$ with respect to $w_{N-1}$ are roots of $\Delta^{(m,n)}$ for some $m,n\in\mathbb{N}$, and provide the exact forms of $\Delta^{(m,n)}$,  $J^{(m,n)}$, $\mathcal{P}^{(mn)}_{N, {\rm R}}$, $\mathcal{P}^{(mn)}_{N}$ in terms of the symmetric variables. We also explain how $\hat{w}$ can be found without finding all $a_u$. Note that, unlike the original variables $\bm{a}$, most of which remain unchanged, 
 all $N-1$ independent variables change in general.



\subsection{Zamolodchikov-like recurrence relation}
The formula (\ref{residuesym}) gives us all the residues of $Z$ with respect to $w_{N-1}$, while the expansion (\ref{eqn:finalanswer}) provides its asymptotic behaviour at $w_{N-1}\to \infty$. This is enough to reconstruct a meromorphic function completely, leading to the recurrence relation 
\begin{equation} \label{rec}
    Z({\bf w})=Z^{(\infty)}({\bf w})+\sum_{k=1}^{N-1}\sum_{m,n=1}^{\infty} \frac{q_0^{mn} J^{(m,n)} }{(w_{N-1}-\bar{w}^{(k|m,n)}_{N-1})}\frac{\mathcal{P}^{(m,n)}_{N, {\rm fund}}}{\mathcal{P}^{(m,n)}_{N}}{Z}({\bf \hat{w}}^{(k|m,n)}) .
\end{equation}
\par Now note that
\begin{equation}\label{eqn:q-renorm-inrel-1}
    \frac{Z^{(\infty)}({\bf \hat{w}}^{(k|m,n)})}{ Z^{(\infty)}({\bf w})} = \left(\frac{q_0}{Dq_{\mathrm{IR}}
} \right)^{\frac{w_1-\hat{w}_1^{(m,n)}}{\epsilon_1\epsilon_2}}=\left(\frac{D q_{\mathrm{IR}}}{q_0
} \right)^{m n } ,
\end{equation}
where the latter was shown in \cite{SysoevaBykov}.
\par The residue formula (\ref{residuesym}) can be written for the renormalised partition function as
   \begin{equation}
            {\rm Res}_{w_{N-1}=\bar{w}_{N-1}^{(k|m,n)}} \overline{Z}({\bf w})=(D q_{\rm IR})^{m n}J^{(m,n)} \frac{\mathcal{P}^{(mn)}_{N, {\rm R}}}{\mathcal{P}^{(mn)}_{N}}\overline{Z}({\bf \hat{w}}^{(k|m,n)})
    \end{equation}
and the series (\ref{rec}) can be resummed as
\begin{equation} \label{recnorm}
    \overline{Z}({\bf w})= 1 +\sum_{k=1}^{N-1}\sum_{m,n=1}^{\infty} \frac{(Dq_{\rm IR})^{mn} J^{(m,n)} }{(w_{N-1}-\bar{w}^{(k|m,n)}_{N-1})}\frac{\mathcal{P}^{(m,n)}_{N, {\rm fund}}}{\mathcal{P}^{(m,n)}_{N}} \overline{Z}({\bf \hat{w}}^{(k|m,n)}) .
\end{equation}
\par To decouple several multiplets, we simply note from \cite{SysoevaBykov} that
\begin{equation}
    \mathcal{P}^{(m,n)}_{N, {\rm fund}}\big|_{N_f, N_{af}}= (-M)^{p m n }M^{r m n}\mathcal{P}^{(m,n)}_{N, {\rm fund}}\big|_{N_f-p, N_{af}-r} ,
\end{equation}
and the extra factors are exactly absorbed by the transformation of $q_0^{mn}$, hence the relation (\ref{rec}) stays the same. The effective constant $D q_{\rm IR}$ transforms in the same way as $q_0$, so (\ref{recnorm}) also stays the same. 
\paragraph{Remark.} An interesting result follows from the formal observation (\ref{eqn:q-renorm-inrel-1}): the bare coupling constant is replaced by the renormalised one in the final recurrence relation. Let us explain the appearance of the renormalised constant from a different point of view. 
As shown in \cite{SysoevaBykov}, geometrically, a more fundamental form of (\ref{residuea}) is
\begin{equation}
    \label{eqn:zfull-lead}Z_{\mathrm{full}}(\bm{a})=-{\rm Sign}(\epsilon_1)Z_{\mathrm{full}}(\hat{\bm{a}})(1+\mathcal{O}(\alpha_{uv})).
\end{equation}
In particular, the $q^{nm}$ factor in (\ref{residuea}) arises from the ratio $Z_{\mathrm{class}}({\bm{a}})/Z_{\mathrm{class}}(\hat{\bm{a}})$ hidden in (\ref{eqn:zfull-lead}). With the renormalisation performed in Subsection \ref{subsec:renormalisation}, it is natural to expect the replacement of the bare coupling constant with the effective one in the recurrence relation for the renormalised instanton partition function.
\par Note also that the equation (\ref{eqn:zfull-lead}) keeps the same form when written in terms of the symmetric variables.

\section{Summary and discussion} \label{sec:concl}

The main purpose of this paper was to identify the asymptotic behaviour of the instanton partition function in $SU(N)$ theory with $2N$ matter multiplets in the fundamental representation, in the limit of large vacuum expectation values of the Higgs field $\bf a$, and to present the recurrence relation for the instanton partition function of this theory. However, in the process, we made a few observations that we find interesting enough to include in this Section.
\par The behaviour we were investigating essentially depends on how infinity is approached. We consider the symmetric way to infinity, required for the recurrence relation, in which in the leading order, all the $a_u$ are placed at the vertices of a regular $N$-polygon, and its diameter tends to infinity.
\par We found that, contrary to intuition, in this regime the typical contributions to the instanton partition function come from the configurations with a finite, and not large, number of instantons. However, the discrete saddle point approach catches all the essential dependence of the asymptotic behaviour on vevs of the Higgs field and masses of the multiplets, and requires only a minor correction depending solely on the bare ultraviolet coupling constant. The saddle point equation turns out to be the Quantum Seiberg-Witten equation, precisely the same as the one describing the Nekrasov-Shalikashvili limit. We discovered that the reason behind the saddle point method being relatively accurate is that in the regime of interest the potential energy of interaction of instantons is almost always absent, except for several singular points, exactly as it is in the non-equivariant and NS limits. Due to this suppression of instanton interaction, the first $qq$-character equation reduces to the  saddle point equation  up to a correction, which can be found from the second $qq$-characters, and  the rest of the infinite chain of the $qq$-character equations does not play a role. The correction appears for the theories with $N \geq 4$.
\par We concluded that the same QSW equation is valid for any regime that can be described with the saddle point method.
\par In our case, the QSW equation can be solved perturbatively, reducing in each order to an algebraic equation. Moreover, thanks to the conformality of the theory and the chosen approach to infinity, the constraints on the (quantum) periods, which are usually transcendental, are just linear equations. These significant simplifications with respect to the general case allow us to find explicit solutions at the saddle point.
\par The asymptotic form of the partition function possesses some interesting features. 
\par Firstly, the dependence of the asymptotic partition function on the vevs has the same exponential form as the classical part of the partition function. This naturally allows the introduction of a unique effective coupling constant. This effective coupling does not allow us to determine the usual coupling matrix, but can be computed from the latter by a contraction procedure.
\par Another feature is revealed if we change the perspective and consider the renormalised coupling constant as a parameter of the theory, similarly to particle physics. Then the bare ultraviolet coupling turns out to be a modular function of the renormalised constant (in fact, a Hauptmodul) with respect to a certain triangle group. Moreover, the other two factors of the asymptotic are a modular form and a modular function with respect to the same triangle group.

\par The found asymptotic expression withstood multiple checks, such as reproducing the results found in the non-equivariant and Nekrasov-Shatashvili limits, comparison with the known asymptotics for $N=2$ and $N=3$ theories, reproducing the correct asymptotic for general $N$ after decoupling of two multiplets, recovering the relevant part of the coupling matrix for $N=4$ theory, and possessing the modular properties with respect to a group in reasonable agreement with the modular group predicted before. Furthermore, for general $N$, we have shown that the modular properties of the effective coupling coincide with the modular properties of one of the $\lfloor \frac{N}{2}\rfloor$ coupling constants found in previous works (it would be interesting to understand why for any rank of the theory exactly one coupling constant is relevant for the asymptotic behaviour).
\par As the final check, we compared the found asymptotic with the partition function computed up to several instantons for the $N=4,5$ theories and observed the necessity of the correction to the asymptotic found with the saddle point method. The extra factor found from the second $qq$-character equation reproduced exactly the difference between the results of the saddle point method and the direct symbolic computations.
\par 
It is interesting that in the recurrence relation for the instanton partition function, valid for whatever values of $a_u$ and $m_f$, appears one of the massless special vacuum infrared coupling constants (we discuss it further in \cite{OneConst}). This hints that this constant may have some meaning also for general vacuum of the theory. Since this coupling constant is characterised by the way the $S$ duality group acts on it, it is natural to ask if this action has any meaning for general vacuum. In particular, one may try to see if the recurrence relation implies any modular properties of the instanton partition function as function of the effective coupling. 
\par
We also believe that there should be an underlying physical reason for the dependence of the asymptotic of the instanton partition function on vevs to be of the precise form which is required for introduction of the effective coupling constant. At the very least we think that this fact should have an interesting interpretation on the conformal field theory side of the AGT duality.

\section*{Acknowledgements}

E.S. would like to thank Marialuisa Frau and Marco Billò for discussions. Research of E.S. is partially supported by the MUR PRIN contract 2020KR4KN2 "String Theory as a bridge between Gauge Theories and Quantum Gravity" and by the INFN project ST\&FI "String Theory \& Fundamental Interactions".


\appendix

\section{Convergence of generalised continued fractions perturbed by noise}\label{app-rec}

\par The purpose of this Appendix is to study the relation between the solution $\omega(y)$ of (\ref{eqn:Rstructure}) and $\omega_0(y)$ of (\ref{eqn:Rstructure-lead}). Investigation of this relation led us to an apparently unrelated problem of classical (or even numerical) analysis. Namely, we had to consider the convergence of a recurrent scheme solving a quadratic equation in the presence of noise slightly perturbing the scheme at each step. The results of this analysis allow us to find the perturbative solution of the quantum Seiberg-Witten equations (\ref{eqn:Rstructure}) in the limit of small ratio $\epsilon/A$. To apply this approach, we change the perspective: instead of proving that $\omega_0(y)$ is a good enough approximation for $\omega(y)$, we show the opposite, namely that $\omega(y)$ is an approximation for $\omega_0(y)$. In fact, it is a modified version of a well-known scheme for numerical solutions of a quadratic equation. 
\par Since this Appendix is rather disconnected from the main text, we allow occasional overlaps of notation which should not cause any confusion.
\subsection{Recurrent schemes for quadratic equations and their convergence}
First, let us recall how the classical recurrent schemes for quadratic equations work.
\par We consider a function 
\begin{equation} \label{eqn:fdef}
    f(z)=\frac{a z+b}{c z+d}.
\end{equation}
In the current Appendix we work on the extended complex plane, so the point $z=-d/c$ is not special in any sense, and the function $f$ is invertible, with inverse function given by
\begin{equation}
    f^{-1}(z)=\frac{d z-b}{-c z+a}.
\end{equation}
\par Let us determine the stable points of the function (\ref{eqn:fdef}), \textit{i.e.} such $z$ that
\begin{equation}\label{eqn:cfrac-fp}
    f(z)=z \, .
\end{equation}
Up to multiplying both sides by the denominator (recall that we work on the extended complex plane, so this operation is completely legitimate), this is a quadratic\footnote{It is actually a linear equation if $c=0$. But one can note that exactly in this case $z=\infty$ is also a fixed point, so in the generic case we still have two fixed points, as if (\ref{eqn:cfrac-fp}) still were quadratic. The degenerate case $c=0$, $a=d$ but $b\neq 0$ (so we exclude the trivial case $b=c=0$ and $a=d$, $f(z)\equiv z$), when (\ref{eqn:cfrac-fp}) has no solutions at all, can be interpreted as a double fixed point at infinity. These special cases can be treated uniformly together with the general case if we identify the extended complex plane with $\mathbb{CP}^1$.} equation for $z$, and in general has two solutions, say $z_1$ and $z_2$. 
\par We can apply function $f$ repeatedly, defining $f^n=f\circ f\circ \cdots \circ f$ ($n$ times) for $n\in\mathbb{N}$. We also define  $f^{-n}=f^{-1}\circ f^{-1}\circ \cdots \circ f^{-1}$ ($n$ times), and $f^{0}(z)=z$. Clearly, for any $n\in \mathbb{Z}$, $f^n$ is a fractional linear function, and for every $n,m\in\mathbb{Z}$
\begin{equation}
    f^{n}\circ f^m=f^{n+m}.
\end{equation}
\par We are interested in the behaviour of $f^n(z)$ in the limit $n\to \pm \infty$. If these limits exist, they must coincide with the stable points of $f(z)$. In other words, we can see the construction of $f^n$ as an algorithm for solving a quadratic equation  (\ref{eqn:cfrac-fp}) via the recurrent scheme
\begin{equation}
    f^{n+1}(z)=\frac{a f^n(z)+b}{c f^n(z)+d} \, .
\end{equation}
\par In this subsection we answer the question of whether this scheme converges and, if it does, how fast. 
\par We start with rewriting $f(z)$ as
\begin{equation}\label{eqn:f-rec}
    f(z)=\frac{a+\frac{b}{z}}{c+\frac{d}{z}}=\frac{b}{d}+\frac{a-\frac{bc}{d}}{c+\frac{d}{z}},
\end{equation}
and, schematically, the supposed solution $f^{+\infty}(z)$ as a generalised continued fraction\footnote{The fact that the roots of a generic quadratic equation can be approximated by continued fractions is of course well known and the convergence of such approximations was extensively studied \cite{Hubert}. It is, however, easier to reproduce the elementary analysis than to reduce the general formulation we consider here to the standard approximation scheme described in the literature.} 
\begin{equation}
    f^{+\infty}(z)=\frac{b}{d}+\frac{a-\frac{bc}{d}}{c+\frac{b}{d}+\frac{a-\frac{bc}{d}}{c+\frac{b}{d}+\frac{a-\frac{bc}{d}}{c+\ldots}}}.
\end{equation}

\par Let us assume first that $z_1\neq z_2$ (we consider the case of a double fixed point later). Then it is convenient to perform a conformal transformation of the extended complex plane mapping the fixed points to $0$ and $\infty$. So, we introduce
\begin{equation}
    \xi_{z_1,z_2}(z)=\frac{z-z_1}{z-z_2}.
\end{equation}
In the main part of the paper, the most important case is $z_1=z_2^{-1}=\omega$, so we introduce
\begin{equation}
    \xi_{\omega}=\xi_{\omega,\omega^{-1}}.
\end{equation}
Then we have (remember that $f(z_{1,2})=z_{1,2}$)
\begin{equation}
    \xi_{z_1,z_2}(f(z))=\frac{\frac{a z+b}{c z+d}-\frac{a z_1+b}{c z_1+d}}{\frac{a z+b}{c z+d}-\frac{a z_2+b}{c z_2+d}}=\frac{c z_2+d}{c z_1+d}\frac{P(z,z_1)-P(z_1,z)}{P(z,z_2)-P(z_2,z)},
\end{equation}
where
\begin{equation}
    P(z,w)=(a z+b)(c w+d).
\end{equation}
Since both numerator and denominator are of degree one in terms of 
$z$ and vanish respectively at $z=z_1$ and $z=z_2$, we conclude that
\begin{equation}\label{eqn:structure-xi}
    \xi_{z_1,z_2}(f(z))=\lambda \xi_{z_1,z_2}(z).
\end{equation}
It is convenient for us to express the constant $\lambda$ as 
\begin{equation}\label{eqn:lambda}
    \lambda= \xi_{z_1,z_2}(f(\infty))=\frac{1}{\xi_{z_1,z_2}(f^{-1}(\infty))}.
\end{equation}
Here we have used the fact that $\xi_{z_1,z_2}(\infty)=1$.

\par It is clear (for example, from  (\ref{eqn:lambda}) and $\xi_{z_1,z_2}(z)=\xi_{z_2,z_1}(z)^{-1}$) that swapping $z_1$ with $z_2$ replaces $\lambda$ with $\lambda^{-1}$. So, without any of generality, we can assume that $|\lambda|\leq 1$. 

\par Let us start with $|\lambda|<1$. From (\ref{eqn:structure-xi}) we see that $\lim_{n\to +\infty} \xi_{z_1,z_2}(f^n(z))=0$ unless $z=z_2$. In physical terms, we can say that $z=z_1$ is an attractor, or a stable fixed point, while $z=z_2$ is a repulsive, or an unstable fixed point of the recurrent scheme. Similarly, $\lim_{n\to -\infty} \xi_{z_1,z_2}(f^n(z))=z_2$ unless $z=z_1$. Therefore, when we change the direction, the roles of the fixed points are swapped.
\par Let us now write some estimates and see explicitly that at $n \to +\infty$ we indeed find $f^n(z) \to z_1$ even if we start very close to $z_2$.
\par Since 
\begin{equation} 
    |z-z_1| = |z_1-z_2| \frac{|\xi_{z_1,z_2}(z)|}{|1-\xi_{z_1,z_2}(z)|}.
\end{equation}
for $|\xi{z_1,z_2}(z)|<1/2$, we have an estimation
\begin{equation}\label{eqn:z-z1-est}
    |z-z_1| < 2 |z_1-z_2| |\xi_{z_1,z_2}(z)|,
\end{equation}
while for $|\xi_{z_1,z_2}(z)|>2$ by swapping $z_1$ and $z_2$ we have
\begin{equation}\label{eqn:z-z2-est}
    |z-z_2|< 2 |z_1-z_2| |\xi_{z_1,z_2}(z)^{-1}|.
\end{equation}

\par For every $z\neq z_2$ we fix $\varepsilon_2>0$ so that $|z-z_2| > \varepsilon_2$ for some $\varepsilon_2< |z_1-z_2|$. Observe that  $|\xi_{z_1,z_2}(z)|< \frac{2|z_1-z_2|}{\varepsilon_2}$, because we either have $|\xi_{z_1,z_2}(z)|\leq 2 < \frac{2|z_1-z_2|}{\varepsilon_2}$ or can apply (\ref{eqn:z-z2-est}).
\par Let $n$ be so large that $|\lambda|^{n}|\xi_{z_1,z_2}(z)|<|\lambda|^{n}\frac{2|z_1-z_2|}{\varepsilon_2}<\frac{1}{2}$, then with (\ref{eqn:z-z1-est}) we have
\begin{eqnarray}
    && |f^n(z)-z_1|<2|z_1-z_2||\xi_{z_1,z_2}(f^n(z))| \\
    &&=2|z_1-z_2||\lambda|^n|\xi_{z_1,z_2}(z)|<  \varepsilon_1= |\lambda|^n \frac{ 4|z_1-z_2|^2}{r_2}, \ \forall z: |z-z_2|>\varepsilon_2.
\end{eqnarray}
Therefore for any initial point $z$ separated from $z_2$, however small the separation length $\varepsilon_2$ was, after many enough steps $n$ we can guarantee that $\varepsilon_1$ is arbitrary small, \textit{i.e.} that $f^n(z)$ has indeed converged to $z_1$.
\par Note that the product
\begin{equation}\label{eqn:prod-radii}
    \varepsilon_1 \varepsilon_2= 4|z_1-z_2|^2|\lambda|^n.
\end{equation}
does not depend on $\varepsilon_2$ and decays exponentially fast to $0$ with $n\to \infty$.

\par Qualitative, we can describe $f^{n}$ as a map, sending the whole Riemann sphere but a small neighbourhood of $z_2$ of radius $\varepsilon_2$ to a small neighbourhood of $z_1$ of radius $\varepsilon_1$, where $\varepsilon_{1,2}$ satisfy (\ref{eqn:prod-radii}).

\par The case of the inverse function can be treated similarly. For sufficiently large $n \to \infty$ we have 
\begin{equation}
     |f^{-n}(z)-z_2|< \varepsilon_2, \ \forall z:\  |z-z_1|>\varepsilon_1,
\end{equation}
where again $\varepsilon_1$, $\varepsilon_2$ obey (\ref{eqn:prod-radii}).

\par Now let us consider the case $|\lambda|=1$. From (\ref{eqn:structure-xi}) we see that $\xi_{z_1,z_2}(f^{n}(z))$, multiplied by $\lambda$ on each step, just moves on circles. Since $\xi_{z_1,z_2}(z)$ is a linear fractional function, it means that $f^{n}(z)$ also moves on circles, so we can say that both fixed points are equally attractive (or equally repulsive). 

\par For completeness, let us consider also the case when (\ref{eqn:cfrac-fp}) has a double root, which we denote with $z_0$. Then 
\begin{equation}
    (cz+d)(z-f(z))=z(cz+d)-az-b=c(z-z_0)^2.
\end{equation}
It is convenient to put the only fixed point to infinity, introducing the function
\begin{equation}
    \chi_{z_0}(z)=\frac{1}{z-z_0}.
\end{equation}
We have
\begin{eqnarray}
    \chi_{z_0}(f(z))&=&\frac{1}{f(z)-z_0}=\frac{1}{f(z)-z+(z-z_0)} \nonumber\\
    &=& \frac{c z+d}{(z-z_0)(c z_0+d)}= \chi_{z_0}(z)+\frac{c}{c z_0+d} \, .
\end{eqnarray}
We see that $\chi_{z_0}(f(z))$ shifted from $ \chi_{z_0}(z)$ by a fixed value independent of the argument. The same happens with $\chi^{-1}_{z_0}(f(z))$. It means that $\chi(f^{n}(z))$ tends to infinity both for $n\to \infty$ and for $n\to -\infty$, and hence $f^{n}(z)$ converges to the unique fixed point $z=z_0$. However, unlike the previous case with the exponential convergence, here the convergence is much slower and there is no typical convergence scale.

\paragraph{Important example}
Let us now apply the analysis developed above to a specific case interesting in light of the main goal of this Appendix. 
\par Consider a recurrent scheme
\begin{equation} \label{eqn:schemewonoise}
    x_{n-1}+\frac{1}{x_{n}}=R,
\end{equation}
where $R=\omega_0+\frac{1}{\omega_0}$, and $|\omega_0|\leq 1$. Then $x_n=f^{n}(x_0)$, and $f$ is determined by
\begin{equation}
    z+\frac{1}{f(z)}=R.
\end{equation}
It is easy to see that the fixed points are 
\begin{equation} \label{eq:fixedpoints}
    z_1=\omega_0, \, z_2=\omega_0^{-1}. 
\end{equation}
The convergence rate is controlled by
\begin{equation}
    \lambda=\frac{f(\infty)-\omega_0}{f(\infty)-\frac{1}{\omega_0}}=\frac{0-\omega_0}{0-\frac{1}{\omega_0}}=\omega_0^2.
\end{equation}
So, if $|\omega_0|<1$,  we have exponential convergence to $\omega_0$ at large positive $n$, and to $\omega_0^{-1}$ at large negative $n$. By (\ref{eqn:prod-radii}) the rate of convergence is characterised by
\begin{equation}
    \varepsilon_1 \varepsilon_2=4 D_{\omega_0} |\omega_0|^{2n}, \ D_{\omega_0}=\left|\omega_0-\frac{1}{\omega_0}\right|.
\end{equation}
If instead $|\omega_0|=1$, we have no convergence at all with a small exception for $\omega_0=\pm 1$. In the latter case, the convergence is present but is very slow. Writing $|\omega_0|=1$ as $\omega_0=e^{i\alpha}$, we have $R=2\cos(\alpha)$, so the unstable zone can also be described as $R\in [-2,2]$.

\subsection{Recurrent scheme in the presence of noise}
The next step is to add some noise to the scheme (\ref{eqn:schemewonoise}). Consider a recurrent scheme
\begin{equation}\label{eqn:noise-rs}
    X_{n-1}+\frac{1}{X_{n}}=\tilde{R}_n,
\end{equation}
where $|\tilde{R}_n-R|<\delta$ for some $\delta$ to be chosen later.
\par It is convenient to think of $\tilde{R}_n-R$ as a small noise perturbing the scheme considered above. The questions are: to which extent the former analysis can be applied to this case? Do we still have the same attractor and repulsive fixed point? And how does the convergence rate change? Before proceeding with long and thorough estimates, let us write the answer: yes, if we stay far enough from the critical line $R\in [-2,2]$ and $\delta$ is small enough, the attractor and the repulsive point remain the same as in the unperturbed scheme. However, unlike the case without the noise, when we could have started with any point separated from the repulsive point, no matter how close the two points were, and ended up arbitrarily close to the attractor if we had made enough steps, in the present case we will find a minimum separation distance, and the convergence will get saturated at a certain radius. In other words, we cannot start too close to the repulsive point and we cannot get as close to the attractor as we want. We will find that both of these radii are set by the noise amplitude $\delta$. The convergence again is exponentially fast up to the point of saturation.
\par Of course we need to make careful estimations to justify these claims, since there is a possibility that a considerable cumulative effect of the small noise shifts the fixed points after asymptotically many steps. Let us now make these estimates.


\par Let $\omega_0$ be a solution of the equation
\begin{equation} \label{eqn:omega0const}
    \omega_0+\frac{1}{\omega_0}=R,
\end{equation}
We assume that $R\notin [-2,2]$, so we can also assume that $|\omega_0|<1$.
\par We introduce
\begin{equation}
    Y_n=\xi_{\omega_0,\omega_0^{-1}}(X_n)=\frac{X_n-\omega_0}{X_n-\omega_0^{-1}}
\end{equation} 
and
\begin{equation}
    Z_n=\frac{|X_n-\omega_0|}{|X_n|\left|X_{n-1}-\frac{1}{\omega_0}\right|}.
\end{equation}
With the triangle inequality, we can estimate $Z_n$ as
\begin{eqnarray}
    Z_n&=&|\omega_0|\frac{\left|\frac{1}{X_n}-\frac{1}{\omega_0}\right|}{\left|X_{n-1}-\frac{1}{\omega_0}\right|}=|\omega_0|\frac{\left|\tilde{R}_n-X_{n-1}-\frac{1}{\omega_0}\right|}{\left|X_{n-1}-\frac{1}{\omega_0}\right|}\leq |\omega_0|\frac{\left|X_{n-1}-\omega_0\right|+\delta}{\left|X_{n-1}-\frac{1}{\omega_0}\right|} \\
    &=&
    |\omega_0|\left(|Y_{n-1}|+\frac{\delta}{\left|X_{n-1}-\frac{1}{\omega_0}\right|}\right).
\end{eqnarray}
Note that
\begin{equation}
    Y_{n-1}=\frac{X_{n-1}-\omega_0}{X_{n-1}-\frac{1}{\omega_0}}=1-\frac{\omega_0-\frac{1}{\omega_0}}{X_{n-1}-\frac{1}{\omega_0}}
\end{equation}
so
\begin{equation}
    \frac{1}{|X_{n-1}-\frac{1}{\omega_0}|}\leq \frac{1}{D_{\omega_0}}\left(1+|Y_{n-1}|\right).
\end{equation}
Therefore
\begin{equation}\label{eqn:magic-above}
   Z_n\leq |\omega_0|\left(|Y_{n-1}|\left(1+ \frac{\delta}{D_{\omega_0}} \right)+\frac{\delta}{D_{\omega_0}}\right).
\end{equation}
On the other hand, we can write
\begin{eqnarray}
    \frac{1}{Z_{n}}= \frac{|X_n|\left|\tilde{R}_n-\frac{1}{X_n}-\frac{1}{\omega_0}\right|}{|X_n-\omega_0|}
    \leq \frac{|X_n|\left|\omega_0-\frac{1}{X_n}\right|+\delta |X_n|}{|X_n-\omega_0|}=\frac{|\omega_0|}{|Y_n|}+\delta \left|\frac{1}{1-\frac{\omega_0}{X_n}}\right|.
\end{eqnarray}
Taking into account that
\begin{equation}
    \frac{1}{Y_{n}}=\frac{X_{n}-\frac{1}{\omega_0}}{X_{n}-\omega_0}=\frac{1-\frac{1}{\omega_0 X_n}}{1-\frac{\omega_0}{X_n}}=\frac{1}{\omega_0^2}\left(\frac{1-\frac{1}{\omega_0^2}}{1-\frac{\omega_0}{X_n}}+\frac{1}{\omega_0^2}\right)=\frac{1}{\omega_0}\left(\frac{\omega_0-\frac{1}{\omega_0}}{1-\frac{\omega_0}{X_n}}+\frac{1}{\omega_0}\right),
\end{equation}
we get
\begin{equation}
    \frac{1}{|1-\frac{\omega_0}{X_n}|}\leq \frac{1}{D_{\omega_0}} \left(\frac{|\omega_0|}{|Y_n|}+\frac{1}{|\omega_0|}\right),
\end{equation}
and hence
\begin{equation}\label{eqn:magic-below}
    \frac{1}{Z_n}\leq \frac{|\omega_0|\left(1+\frac{\delta}{D_{\omega_0}}\right)}{|Y_n|}+\frac{\delta}{|\omega_0|D_{\omega_0}}.
\end{equation}
Multiplying (\ref{eqn:magic-below})
by (\ref{eqn:magic-above}) we arrive at
\begin{equation}\label{eqn:magic-ewhr}
    1\leq \left(|\omega_0||Y_{n-1}|\left(1+ \frac{\delta}{D_{\omega_0}} \right)+|\omega_0|\frac{\delta}{D_{\omega_0}}\right)\left(\frac{|\omega_0|\left(1+\frac{\delta}{D_{\omega_0}}\right)}{|Y_n|}+\frac{\delta}{|\omega_0|D_{\omega_0}}\right).
\end{equation}
\par Now, let $g$ be a function defined by the equation
\begin{equation}\label{eqn:magic-g}
     \left(|\omega_0|r\left(1+ \frac{\delta}{D_{\omega_0}} \right)+|\omega_0|\frac{\delta}{D_{\omega_0}}\right)\left(\frac{|\omega_0|\left(1+\frac{\delta}{D_{\omega_0}}\right)}{g(r)}+\frac{\delta}{|\omega_0|D_{\omega_0}}\right)=1.
\end{equation}
Clearly, $g$ is a fractional linear function, and we can apply the analysis from the previous section to investigate the convergence of $g^n$ when $n \to \pm \infty$. In the next step we will use the estimation of $g^n$ to bound $|Y_n|$ and hence $|X_n|$.
\par First, we find the fixed points of $g$. If $r=g(r)$, then from (\ref{eqn:magic-g}) we have
\begin{equation}\label{eqn:magic-fp}
     \left(|\omega_0|r\left(1+ \frac{\delta}{D_{\omega_0}} \right)+|\omega_0|\frac{\delta}{D_{\omega_0}}\right)\left(\frac{|\omega_0|\left(1+\frac{\delta}{D_{\omega_0}}\right)}{r}+\frac{\delta}{|\omega_0|D_{\omega_0}}\right)=1.
\end{equation}
It is, of course, possible to solve this quadratic equation, but its solutions are rather involved. For our purposes it is enough to understand the qualitative behaviour of the fixed points (namely, we intend to show that there are two distinct positive fixed points) for sufficiently small $\delta$. To do this, it is enough to find asymptotic solutions in the limit $\delta\to 0$. For that we write (\ref{eqn:magic-fp}), leaving only the leading power of $\delta$ in front of each power of $r$:
\begin{equation}\label{ref:eqn:magic-fp-as}
    \frac{|\omega_0|\delta }{D_{\omega_0}}\left(\frac{r}{|\omega_0|}+\frac{|\omega_0|}{r}\right)=1-|\omega_0|^2+\mathcal{O}(\delta,\delta^2 r,\delta^2 r^{-1}).
\end{equation}
By the standard Newton polygon method, or simply guessing the correct power, there are two branches of solutions $r_{\pm}\sim \delta^{\mp 1}$. Here the signs are chosen so that $|r_-|<|r_+|$. Substituting these asymptotics into (\ref{ref:eqn:magic-fp-as}), we get\footnote{We note that the corrections of higher power of $\delta$ cannot become complex, because we are solving a quadratic equation with real coefficients, so passage from real to complex solutions is possible only when the discriminant is close to zero, which contradicts $|r_-|\ll |r_+|$.}
\begin{equation}        
    r_{\pm}=|\omega_0|\left(\frac{D_{\omega_0}D_{|\omega_0|}}{\delta}\right)^{\pm 1}\left(1+\mathcal{O}(\delta)\right).
\end{equation}
Here, we used that $|\omega_0|<1$, so
\begin{equation}
    D_{|\omega_0|}=\frac{1}{|\omega_0|}-|\omega_0|=\frac{1-|\omega_0|^2}{|\omega_0|}
\end{equation}
for brevity.
\par In order to identify the convergence rate $\lambda$, we need to find $g(\infty)$. To compensate $r\to \infty$ in the first bracket in the right-hand side of (\ref{eqn:magic-g}),  the second one should vanish, so
\begin{equation}\label{eqn:magic-ginf}
     \frac{|\omega_0|\left(1+\frac{\delta}{D_{\omega_0}}\right)}{g(\infty)}+\frac{\delta}{|\omega_0|D_{\omega_0}}=0
\end{equation}
and
\begin{equation}
    g(\infty)=-\frac{|\omega_0|^2 D_{\omega_0}}{\delta}(1+\mathcal{O}(\delta)) \, .
\end{equation}
Then
\begin{eqnarray}
    \lambda =\frac{g(\infty)-r_-}{g(\infty)-r_+}= \frac{-\frac{|\omega_0|^2 D_{\omega_0}}{\delta}-|\omega_0|\left(\frac{\delta}{D_{\omega_0}D_{|\omega_0|}}\right)}{-\frac{|\omega_0|^2 D_{\omega_0}}{\delta}-|\omega_0|\left(\frac{D_{\omega_0}D_{|\omega_0|}}{\delta}\right)}+\mathcal{O}(\delta) = |\omega_0|^2 +\mathcal{O}(\delta) .
\end{eqnarray}
Recall that we assume $|\omega_0|<1$, so $\lambda<1$, and, at large $n$, $g^{n}(z)$ converges to $r_-$ for all values of $z$ except for $r_+$, and vice versa for the inverse function $g^{-n}$. More precisely, we have 
\begin{equation} \label{eqn:gconvergence}
    |g^n(r)-r_-|\leq \varepsilon_1, \quad \mathrm{if}   \ |r-r_+|\geq \varepsilon_2,
\end{equation}
\begin{equation} \label{eqn:gconvergence2}
    |g^{-n}(r)-r_+|\leq \varepsilon_2, \quad \mathrm{if} \ |r-r_+|\geq \varepsilon_1,
\end{equation}
  where $\varepsilon_1,\varepsilon_2\in (0,r_{+}-r_{-})$ are such that
  \begin{equation}\label{eqn:eps12-est}
      \varepsilon_1 \varepsilon_2=4|r_{+}-r_{-}||\lambda|^{n}.
  \end{equation}
\par Let us now understand how to use this information to estimate $Y_n$.
\par Since the right-hand side of (\ref{eqn:magic-ewhr}) is an increasing function of $|Y_{n-1}|$ and a decreasing function of $|Y_n|$, we have
\begin{equation} \label{eqn:initialineq}
    |Y_n|\leq g(|Y_{n-1}|),\ |Y_{n-1}|\geq g^{-1}(|Y_n|).
\end{equation}
It is tempting to conclude by induction that
\begin{equation}
    |Y_n|\leq g^n(|Y_{0}|),\ |Y_{n-1}|\geq g^{-n}(|Y_0|).
\end{equation}
However, we should be careful, because only increasing functions preserve inequalities, and function $g$ is increasing only on connected regions of its continuity. Therefore, we need to find out if there are regions of continuity (and hence monotonicity)  of $g^n$ and $g^{-n}$ large enough to make (\ref{eqn:initialineq}) useful.
\par To find the point of discontinuity of $g$, we put $g(r)\to \infty$ in (\ref{eqn:magic-g}):
\begin{equation}
     \left(|\omega_0|r_{\infty}\left(1+ \frac{\delta}{D_{\omega_0}} \right)+|\omega_0|\frac{\delta}{D_{\omega_0}}\right)\frac{\delta}{|\omega_0|D_{\omega_0}}=1,
\end{equation}
or
\begin{equation}\label{eqn:magic-rinf}
    r_{\infty}=\frac{\frac{D_{\omega_0}}{\delta}-|\omega_0|\frac{\delta}{D_{\omega_0}}}{1+ \frac{\delta}{D_{\omega_0}}}=\frac{D_{\omega_0}}{\delta}\left(1+\mathcal{O}(\delta)\right).
\end{equation}
As we are interested in the case of small $\delta$, we may assume $r_{\infty}>0$. 
\par We have already found the fixed points (\ref{ref:eqn:magic-fp-as}) and now we can note that
\begin{equation}\label{eqn:rplus-rinft}
    r_+=r_{\infty}(1-|\omega_0|^2)+\mathcal{O}(\delta).
\end{equation}
We conclude that for small enough $\delta$ 
\begin{equation}
   0<r_-\ll r_+ < r_{\infty}.
\end{equation}
\par From the above, we see that $g$ is continuous on $[0,r_{+}]$, so it is an increasing function there. Moreover, for small enough $\delta$ we have $g(0)>0$ since putting $r=0$ into (\ref{eqn:magic-g}) we get
\begin{equation}\label{eqn:magic-g0}
     |\omega_0|\frac{\delta}{D_{\omega_0}}\left(\frac{|\omega_0|\left(1+\frac{\delta}{D_{\omega_0}}\right)}{g(0)}+\frac{\delta}{|\omega_0|D_{\omega_0}}\right)=1,
\end{equation}
so 
\begin{equation}
    g(0)=\frac{|\omega_0|^2\delta}{D_{\omega_0}}+\mathcal{O}(\delta^2),
\end{equation}
This means that 
\begin{equation}
    g([0,r_{+}])\subset [0,r_{+}]. 
\end{equation}
By induction, for any positive $n$, $g^n$ is a continuous increasing function on $[0,r_{+}]$, mapping this interval to itself, and we can write
\begin{equation}\label{eqn:Yn-est}
    |Y_n|\leq g^{n}(|Y_{0}|),\quad \mathrm{if} \ |Y_0|<r_+.
\end{equation}
    \par Similarly, we have to find the point where the inverse function $g^{-1}$ is discontinuous, \textit{i.e.} $g(\infty)$, which we have already done above. Observe that, for sufficiently small $\delta$,
\begin{equation}
    g(\infty)<r_- \ll r_+ .
\end{equation}
 Function $g^{-1}$ is increasing and continuous on the interval $[r_-,+\infty)$, moreover it maps this interval to itself. Thus, the same is true for $g^{-n}$ with any positive $n$ and we can write
    \begin{equation} \label{eqn:Yn-west}
        |Y_{-n}|\geq g^{-n}(|Y_0|) \, , \quad \mathrm{if} \  |Y_0|>r_{-}. 
    \end{equation}
Note that for small $\delta$ we have $r_+ \to \infty$ and $r_- \to 0$, so the estimations (\ref{eqn:Yn-est}) and (\ref{eqn:Yn-west}) are indeed useful.
\par Finally, with (\ref{eqn:gconvergence}), (\ref{eqn:Yn-est}) and the triangle inequality we arrive to
  \begin{equation}
      |Y_n|\leq r_{-}+\varepsilon_1, \quad  \ \mathrm{if} \ |Y_0|\leq r_{+}-\varepsilon_2 ,
  \end{equation}
   where $\varepsilon_1,\varepsilon_2\in (0,r_{+}-r_{-})$ and (\ref{eqn:eps12-est}) holds.
  \par Moreover, for small enough $\delta$ we have
   \begin{equation}
      r_{-}+\varepsilon_1<\frac{1}{2},\  r_{+}-\varepsilon_2>2 
  \end{equation}
and thus, with the help of (\ref{eqn:z-z1-est}), (\ref{eqn:z-z2-est}), we conclude that the recurrent scheme converges to $\omega_0$ as
  \begin{equation}\label{eqn:Xnest}
      |X_n-\omega_0|\leq 2 D_{\omega_0}(r_{-}+\varepsilon_1)    \,
      , \quad \mathrm{if} \ \left|X_0-\frac{1}{\omega_0}\right|>\frac{2D_{\omega_0}}{r_{+}-\varepsilon_2}.
  \end{equation}
For the inverse function, we use (\ref{eqn:gconvergence2}) and (\ref{eqn:Yn-west}) and, acting in the same way, find that the recurrent scheme converges to $\omega_0^{-1}$. 
    
  \begin{equation}
      \left|X_{-n}-\frac{1}{\omega_0}\right|\leq \frac{2D_{\omega_0}}{r_{+}-\varepsilon_1}, \quad \mathrm{if} \ \left|X_0-{\omega_0}\right|>2D_{\omega_0}(r_{-}+\varepsilon_2).
  \end{equation}
  This justifies the initial claim that if the noise amplitude $\delta$ is small, then just as in the unperturbed case $X_n$ at large positive (respectively, negative) $n$ lies in a small neighbourhood of $\omega_0$ (respectively, $\frac{1}{\omega_0}$) unless the initial condition $X_0$ lies in a small neighbourhood of the other solution. The radius of the neighbourhoods still shrinks exponentially with a base close to $|\omega_0|^2$, but only until it saturates at 
  \begin{equation}\label{eqn:noise-rs-err}
      \frac{2D_{\omega_0}}{r_{+}} , \ 2D_{\omega_0}r_{-}  \sim \frac{2 \delta}{D_{\omega_0}}|\omega_0|. 
  \end{equation}
  The presence of this minimal radius is the cumulative effect of the noise over all steps. 
  The crucial result of this analysis is that its effect is finite even after asymptotically many steps $n\to \pm \infty$.
  \par It is also interesting that the error (\ref{eqn:noise-rs-err}) is proportional to
  \begin{equation}
  \frac{2 \delta}{D_{\omega_0}}|\omega_0|=\frac{2\delta}{|1-\omega_0^2|}. 
  \end{equation}
  In particular, near $\omega_0=\pm 1$, the scheme appears to have additional problems besides the slow divergence of unperturbed recurrent scheme.

  \subsection{Mutual convergence}\label{app:mutconv}
  
  Let $X_n$, $X'_n$ be two different solutions of (\ref{eqn:noise-rs}), evolved from the initial points $X_0$, $X'_0$, which are far enough from the repulsive point $\frac{1}{\omega_0}$, and let us ask how $|X_n-X'_n|$ behaves as $n \to +\infty$.
  \par On the one hand, it is natural to expect that $|X_n-X'_n|\to 0$, since the recurrent equation (\ref{eqn:noise-rs}) tends to 'forget' the initial condition as long as it lies in the stable region. On the other hand, we actually have derived only that both $X_n$ and $X_n'$ lie in a small, but finite neighbourhood of $\omega_0$, so at large $n$, $|X_n-X'_n|$ could be as large as $\frac{2D_{\omega_0}}{r_+}$. In this subsection we show that the na\"ive guess is actually correct if $\delta$ is small enough (but still finite). Our intention is not to find the necessary and sufficient condition for $|X_n-X'_n|\to 0$, but only to show that the sufficient condition can be satisfied by adjusting the free parameters. Such a rough analysis will be enough for our purposes.
  
   \par We assume that the initial conditions are far from the unstable region 
  \begin{equation}\label{eqn:mut-conv-X0}
      \left|X_0-\frac{1}{\omega_0}\right|, \left|X'_0-\frac{1}{\omega_0}\right|>\frac{4D_{\omega_0}}{r_+},
  \end{equation}
  
  Then, by choosing appropriately $\epsilon_1$ and using the analysis of previous subsection, one can find some $N\in\mathbb{N}$ such that 
  \begin{equation}\label{eqn:mut-conv-N}
      \left|X_n-\omega_0\right|, \left|X'_n-\omega_0\right|<4D_{\omega_0}r_-,\  \forall n\geq N
  \end{equation}
  Let us introduce
  \begin{equation}\label{eqn:small-delta-mutconv}
      t=|\omega_0|(\delta+4D_{\omega_0}r_-)
  \end{equation}
   and assume that $t<1$. Then for any $m$ with the triangle inequality we can write
  \begin{eqnarray}
      \left|R_m-X_n \right|&=&\left|\left(R_m-\omega_0-\frac{1}{\omega_0}\right)-(X_n-\omega_0)+\frac{1}{\omega_0} \right| \nonumber \geq  \frac{1}{|\omega_0|}-\left|\left(R_m-\omega_0-\frac{1}{\omega_0}\right)-(X_n-\omega_0)\right| \\ 
      &>& \frac{1}{|\omega_0|}- \delta -4D_{\omega_0}r_-\geq \frac{1-t}{|\omega_0|}.
  \end{eqnarray}
  and similarly for $X'_n$. Note that for small $\delta$ we have
  \begin{equation}
      t=|\omega_0|\delta(1+8D_{\omega_0}^{-1})+\mathcal{O}(\delta), 
  \end{equation}
 so $t<1$ is easy to achieve by the choice of $\delta$. Moreover, by making $\delta$ even smaller if necessary, we can always ensure 
\begin{equation}\label{eqn:mutconv-lambda}
    \frac{|\omega_0|}{1-t}<\Lambda
\end{equation}
 for some $\Lambda\in (0,1)$.
  Now we can write
  \begin{equation}
      \left|X'_n-X_n\right|=     \left|\frac{1}{R_n-X'_{n-1}}-\frac{1}{R_n-X_{n-1}}\right|=\frac{|X_{n-1}-X'_{n-1}|}{|R_n-X_{n-1}||R_n-X'_{n-1}|}\leq \Lambda^2|X_{n-1}-X'_{n-1}|.
  \end{equation}
  As this is true for any $n> N$, we have
  \begin{equation}\label{eqn:mutconv}
      |X_n-X_n'|\leq |X_N-X_N'|\Lambda^{2(n-N)} \leq \frac{2}{|\omega_0|} \Lambda^{2(n-N)},\ \mathrm{for}\ n>N,
  \end{equation}
  which converges to 0 when $n\to+\infty$. Here we used the condition $t<1$ to simplify the right-hand sides. 
  This proves our initial guess under the assumption that $\delta$ is small enough (but finite). 

  \par To conclude this analysis, let us note how the parameters enter into the bound and its applicability conditions. The condition (\ref{eqn:mut-conv-X0}) restricts the initial conditions $X_0$ and $X'_0$, and depends on values of $\omega_0$ and $\delta$. Condition (\ref{eqn:mut-conv-N}) restricts $N$ and depends again on $\omega_0$ and $\delta$. Note that the particular values of $X_0$ and $X'_0$ are not important as long as they satisfy (\ref{eqn:mut-conv-X0}). Condition (\ref{eqn:mutconv-lambda}) restricts the noise $\delta$ and depends on $\omega_0$ only. Finally, the estimation (\ref{eqn:mutconv}) depends only on $\omega_0$ and the bound $\Lambda$, provided that the conditions (\ref{eqn:mut-conv-X0}), (\ref{eqn:mut-conv-N}) and (\ref{eqn:mutconv-lambda}) hold.

    \subsection{Application to quantum Seiberg-Witten curves} \label{apptoSWC}

   Finally, we show how to use the analysis from the previous subsections to study solutions of the functional equation (\ref{eqn:Rstructure})

    \begin{equation}\label{eqn:Rstructure-app}
      \omega\left(y-\frac{\epsilon}{A}\right)+\frac{1}{\omega(y)}=R(y) \ \forall\, y\in\mathbb{C}
  \end{equation}
  with the boundary condition
  \begin{equation} \label{boundarycond}
      \lim_{y\to \infty}\omega(y)=\sqrt{q}.
  \end{equation}
  \par The problem is consistent only if 
  \begin{equation}
  \lim_{y\to \infty}R(y)=\sqrt{q}+\frac{1}{\sqrt{q}
}, \label{eqn:Rasmpt}    
  \end{equation}
   which was already taken into account in the main text of the paper.
    
  \paragraph{Quantum Seiberg-Witten equation as a recurrent scheme}

  Let us define $\omega_0(y)$ so that
  \begin{equation}
      \omega_0(y)+\frac{1}{\omega_0(y)}=R(y).
  \end{equation}
 The difference from (\ref{eqn:omega0const}) is that now everything depends on $y$.

  \par We choose the branch of the square root so that $|\omega_0(y)|\leq 1$. Explicitly, we can write it as
  \begin{equation}
      \omega_0(y)=\frac{R(y)}{2}\left(1- \sqrt{1-\frac{4}{R(y)^2}}\right),
  \end{equation}
  where the square root should be understood as its standard branch (positive on the positive real semiaxis, cut on the negative real semiaxis).
  
  \par Note that (\ref{eqn:Rstructure-app}) is in fact a family of recurrent equations  parametrised by $y$. It is convenient to introduce functions $\rho^n_y$ defined as follows:
  \begin{equation}
    \rho^{0}_y(z)=z,
    \end{equation}
    \begin{equation}\label{eqn:rho-def-rec}
        \rho^{n-1}_{y}(z)+\frac{1}{\rho^{n}_y(z)}=R(y+n\epsilon/A),\quad \forall n\in \mathbb{N}. 
    \end{equation}
    One can see inductively that the introduced function $\rho^n_y(z)$ is subject to a composition rule
    
    \begin{equation} \label{eqn:comprule}
        \rho^n_{y}\circ \rho^{m}_{y-m\epsilon/A}(z) = \rho^{m+n}_{y-m\epsilon/A}(z) .
    \end{equation}

    Clearly, for fixed values of $y$ and $z$, (\ref{eqn:rho-def-rec}) is a recurrent scheme of type (\ref{eqn:noise-rs}). Thus, if only we can control the variation of $R(y)$, it can be considered as an approximation to $\omega_0(y)$ or $\omega_0(y)^{-1}$.
    \par  We postpone the question of the controlled variation of $R(y)$ for later, and for now we note that the solution of (\ref{eqn:Rstructure-app}) satisfies
    \begin{equation} \label{kindofperiodic}
        \omega(y)=\rho^{n}_{y-n\epsilon/A}(\omega(y-n \epsilon/A))\,, \qquad \forall n \in \mathbb{N}.
    \end{equation}
    It means that the function $\omega$ is defined by its values on a strip of the width $\epsilon/A$. Note that (\ref{kindofperiodic}) introduces a gluing condition on the sides of the strip. 
    It still does not completely fix the solution, because infinitely many functions satisfy this condition. To select the right one, we need to use the boundary condition (\ref{boundarycond}). Let us do it carefully.

    \paragraph{Boundary conditions}
   In terms of the sequence $\omega(y+n \epsilon/A)$, (\ref{boundarycond}) leads to two distinct boundary conditions
    \begin{equation} \label{eqn:boundaryintermsofn}
        \lim_{n\to +\infty}\omega(y+n \epsilon/A)=\sqrt{q} \, , \quad       \lim_{n\to -\infty}\omega(y+n \epsilon/A)=\sqrt{q}
    \end{equation}
    These two conditions apparently lead to two different expressions for the solution, namely
    \begin{equation}\label{eqn:QSW-sol-cand}
        \omega(y)= \lim_{n\to +\infty} \rho^{-n}_{y+n\epsilon/A}(\sqrt{q}) \, , \quad  \omega(y)= \lim_{n\to +\infty} \rho^n_{y-n\epsilon/A}(\sqrt{q}) ,
    \end{equation}
    which are incompatible unless very strict conditions are imposed on $R$. The reason behind this apparent contradiction is the peculiar behaviour of function $\rho^n_{y-n\epsilon/A}$ around the point $\sqrt{q}$ when $n$ is large. In fact, in the previous subsection we have already seen that depending on the sign of $n$ a stable point is either repulsive or attracting. From this follows that only one of the conditions (\ref{eqn:boundaryintermsofn}) is actually restrictive, while the second one will be satisfied almost automatically, and only one of them can be continued to the region of finite arguments as in (\ref{eqn:QSW-sol-cand}).
    
    \par In order to show this rigorously, we look at the part of the complex plane where $R$ is close to $\sqrt{q}+\frac{1}{\sqrt{q}}$. Let us fix some $\delta>0$ so that 
    \begin{equation}
        \delta\ll D_{\sqrt{q}}.
    \end{equation}
    Due to (\ref{eqn:Rasmpt}), there is some $d_{\infty}\in (0,\infty)$ such that whenever $|y|>d_{\infty}$ we have
    \begin{equation}\label{eqn:Rinf-var}
        \left|R(y)-\sqrt{q}-\frac{1}{\sqrt{q}}\right|<\delta.
    \end{equation}
    Define
    \begin{equation}
        U^{(\infty)}=\{y\in\mathbb{C}\big| |y|>d_{\infty}\}.
    \end{equation}
    Take $y\in U^{(\infty)}$ such that $y+n\epsilon/A\in U^{(\infty)}$ for any positive $n$ (in other words, we take $y$ to infinity in the right direction. For example, if $\epsilon/A>0$, any $y$ with positive real part and $|y|>d_\infty$ satisfies this condition).
    \par We write
    \begin{equation}
        \omega(y+n\epsilon/A)=\rho_y^{n}(\omega(y))
    \end{equation}
        and note that, assuming $n>0$, $\rho^n_y$ depends only on values of $R$ in the set $U^{(\infty)}$, where (\ref{eqn:Rinf-var}) holds. This means that results of the previous subsection can be applied and, for large enough $n$, $\omega(y+n\epsilon/A)$ lies in a small neighbourhood of $\sqrt{q}$ unless $\omega(y)$ lies in a small neighbourhood of $\sqrt{q}^{-1}$. In fact, taking $n$ really large we make $\varepsilon_{1},\varepsilon_{2}$ in (\ref{eqn:eps12-est}) as small as we desire, so, by (\ref{eqn:Xnest}), for small $\delta$ we have
    \begin{equation}
        |\omega(y+n\epsilon/A)-\sqrt{q}|\lesssim \frac{2\delta}{D_{\sqrt{q}}}\sqrt{q}, \quad \mathrm{unless} \ \left|\omega(y)-\frac{1}{\sqrt{q}}\right|\lesssim \frac{2\delta}{D_{\sqrt{q}}}\sqrt{q}.
    \end{equation}
    Taking into account the fact that $\delta$ can be chosen arbitrarily small (with the cost of moving $d_{\infty}$ and hence $y$ farther to the infinity), this means that  the condition
    \begin{equation}\label{eqn:wrongbc}
        \lim_{n\to +\infty}\omega(y+n\epsilon/A)=\sqrt{q}
    \end{equation}
    gives us almost no information about values of  $\omega(y)$ at finite $y $. It is satisfied almost automatically, excluding only the case of $\omega(y)$ being very close to the other solution $\frac{1}{\sqrt{q}}$. 
    \par We can also understand uselessness of this condition in a different way. Write
    \begin{equation}\label{eqn:wrong-approx}
        \omega(y)=\rho^{-n}_{y+n\epsilon/A}(\omega(y+n\epsilon/A)).
    \end{equation}
    We have $\omega(y+n\epsilon/A)\to \sqrt{q}$ for $n\to +\infty$, which means that we have to evaluate $\rho^{-n}$ in its unstable region. For this reason, despite (\ref{eqn:wrongbc}),
        \begin{equation}
        \omega(y)=\lim_{n\to+\infty}\rho^{-n}_{y+n\epsilon/A}       \left(\omega\left(y+n\frac{\epsilon}{A}\right)\right)\neq \lim_{n\to+\infty}\rho^{-n}_{y+n\epsilon/A}(\sqrt{q}).
    \end{equation}

     \par On the other hand, the second boundary condition
    \begin{equation}\label{eqn:right-bc}
    \lim_{n\to -\infty}\omega(y+n\epsilon/A)=\sqrt{q}    
    =\lim_{n\to +\infty}\omega(y-n\epsilon/A)  
    \end{equation}
    is very restrictive. Indeed, assuming now that $y-n \epsilon/A\in U^{(\infty)}$ for every $n\geq 0$ and acting in the same way as before, we conclude that 
    \begin{equation}
        \left|\omega(y-n\epsilon/A)-\frac{1}{\sqrt{q}}\right|\lesssim \frac{2\delta}{D_{\sqrt{q}}}\sqrt{q}, \quad \mathrm{unless} \ \left|\omega(y)-\sqrt{q}\right|\lesssim \frac{2\delta}{D_{\sqrt{q}}}\sqrt{q}.
    \end{equation}
    We see that the condition (\ref{eqn:right-bc}) is extremely strong. It implies that for finite, but large $y$ (large `in the right direction') $\omega(y)$ has to remain close to $\sqrt{q}$. This gives us a hope that the second equation in (\ref{eqn:QSW-sol-cand}) is correct. In fact we will show much stronger formula, namely
    \begin{equation}\label{eqn:QSW-sol-r}
        \omega(y)=\lim_{n\to +\infty}\rho^{n}_{y-n\epsilon/A}(X_0)
    \end{equation}
    for every $X_0\neq \frac{1}{\sqrt{q}}$. 
    \paragraph{Solution}
    Let us choose $\delta$ such that, assuming $\omega_0=\sqrt{q}$, the bound (\ref{eqn:mutconv-lambda}) holds for some $\Lambda\in (0,1)$  together with the first condition in (\ref{eqn:mut-conv-X0}) and $r_+>4$. It is easy to see that, provided $X_0\neq \frac{1}{\sqrt{q}}$, such a choice is always possible. Now, let $d$ be such that 
    \begin{equation}
        \left| R(y)-\sqrt{q}-\frac{1}{\sqrt{q}}\right|<\delta,
    \end{equation}
    and
    \begin{equation}
        \left|\omega(y)-\frac{1}{\sqrt{q}}\right|>\frac{4D_{\sqrt{q}}}{r_+},
    \end{equation}
    whenever $|y|>d$. Both of the inequalities hold for large enough $y$: the first one due to (\ref{eqn:Rasmpt}), and the second one because in the limit $y\to \infty$, where $\omega_0(y)\to \sqrt{q}$, it becomes just $r_+>4$ which we have assumed already.
    
    Set 
    \begin{equation}
        V_{\infty}=\{y\in\mathbb{C}| \left|y-s \epsilon/A\right|>d,\,\forall s>0\}.
    \end{equation}
    
    Let $N\in\mathbb{N}$ be such that (\ref{eqn:mut-conv-N}) holds for the choice of $\delta$ we have made. Then for $m\in\mathbb{N}$ such that $m>N$ and $y\in V_{\infty}$ consider  \begin{equation}
        X_n=\rho^n_{y-m\epsilon/A}(X_0),
    \end{equation}
    \begin{equation}
        X'_n=\omega(y-(m-n)\epsilon/A).
    \end{equation}
    It is easy to see that these are two solutions of a recurrence relation of type (\ref{eqn:noise-rs}) satisfying all conditions introduced in Subsection \ref{app:mutconv}. Thus, from (\ref{eqn:mutconv}) we have
    \begin{equation}
        |X_{n}-X'_{n}|\leq \frac{2}{|\omega_0|}\Lambda^{2(n-N)}.
    \end{equation}
    Putting $n=m$ we have
    \begin{equation}
        \left|\omega(y)-\rho^{m}_{y-m\epsilon/A}(X_0) \right|\leq \frac{2}{\omega_0}\Lambda^{2(m-N)},\ \forall m>N, \forall y\in V_{\infty},
    \end{equation}
   therefore
    \begin{equation}
        \omega(y)=\lim_{m\to +\infty}\rho^{m}_{y-m\epsilon/A}(X_0), \forall y\in V_{\infty}. 
    \end{equation}
    Finally, for any $y\in\mathbb{C}$ we can find $n\in\mathbb{N}$ such that $y-n\epsilon/A\in V_{\infty}$. Then
    \begin{eqnarray}
        \omega(y)&=&\rho^n_{y-n\epsilon/A}(\omega(y-n\epsilon/A))=\rho^n_{y-n\epsilon/A}\lim_{m\to +\infty}\rho^{m}_{y-(m+n)\epsilon/A}(X_0)=\lim_{m\to +\infty}\rho^n_{y-n\epsilon/A}\circ \rho^{m}_{y-(m+n)\epsilon/A}(X_0) \nonumber \\ &=& \lim_{m\to +\infty}\rho^{m+n}_{y-(m+n)\epsilon/A}(X_0)=\lim_{m\to +\infty}\rho^{m}_{y-m\epsilon/A}(X_0),
    \end{eqnarray}
    where we have used continuity\footnote{The apparent problem caused by the pole of $\rho^m$ can be cured by using the topology of the Riemann sphere.}  of $\rho^n_{y-n\epsilon/A}$, the composition rule (\ref{eqn:comprule}) and the shift invariance of the limit.
    
    So, for general $y$ and arbitrary $X_0\neq \sqrt{q}$ we can ultimately write \footnote{We could have also written the solution in a continued fraction-like form
        \begin{equation} \label{eqn:solutionomega}
        \omega(y)=\frac{1}{R(y)-\frac{1}{R(y-\epsilon/A)-\frac{1}{R(y-2\epsilon/A)-\ddots}}},
    \end{equation}
    where the precise way how to terminate the fraction is not important.
    It is useless for us, although pretty.}
    \begin{equation}
         \omega(y)=\lim_{m\to +\infty}\rho^{m}_{y-m\epsilon/A}(X_0)
    \end{equation}
    For example, we can take $X_0=\sqrt{q}$ or $X_0=0$ (since the case $q=\infty$ is definitely outside our consideration).  

    \par The asymmetry between the limits $n\to +\infty$ and $n\to -\infty$ suggests that it is better to speak of one initial rather than two boundary conditions. With this terminology, behaviour of (\ref{eqn:Rstructure-app}) reminds thermodynamical systems. Although the `microscopic' equation (\ref{eqn:Rstructure-app}) is clearly invertible, when the number of steps becomes large, a preferred orientation of the `time axis' appears. In this sense, (\ref{eqn:wrong-approx}) can be interpreted as an attempt to restore initial low entropy state from an outcome of a diffusion-like process.

    \paragraph{Error estimation}

    By now we established that at a point $y\in U^{(\infty)}$, where $R(y)$ is close to its asymptotic value $\sqrt{q}+\sqrt{q}^{-1}$, the solution $\omega(y)$ defined by (\ref{eqn:solutionomega}) is close to $\omega_0(y)$. Let us now ask what requirements should be satisfied for its continuation to the rest of the complex plane to be close to $\omega_0(y)$ too.  A physicist would immediately answer that the requirement is that on a path from $y_\infty\in U^{(\infty)}$ to $y\in \mathbb{C} \setminus U^{(\infty)}$ the recurrent scheme (\ref{eqn:rho-def-rec}) should converge to the solution before $R$ changes too much, \textit{i.e.} the steps along the path are not too big, $R$ changes not too fast and the convergence rate is not too slow. This is indeed true, but let us write this condition more rigorously. 
    
    \par Let $C_{\mathrm{sing}}\subset \mathbb{C}$ be the set containing all the singularities of $\omega_0$ and $R$, and let $U^{(0)}\supset C_{\mathrm{sing}}$. The set $C_{\mathrm{sing}}$ is a union of some curves (cuts and poles), and we can think of $U^{(0)}$ as its thin neighbourhood.
    \par Since $ D_{\omega_0(y)}$ is small only in the proximity of the branching points of $\omega_0(y)$, we can ensure that
    \begin{equation} \label{eqn:U0thickness}
        D_{\omega_0(y)}\gg \delta, \ \forall y\notin U^{(0)}
    \end{equation}
    for some small $\delta$.

 \par For any $y\notin U^{(0)} \cup U^{(\infty)}$ we can define a small neighbourhood
\begin{equation}
        U^{(y)}=\{y'\in \mathbb{C} \mid |y-y'|<d_y\},
    \end{equation}
    where $d_y$ is chosen so that
    \begin{equation}
        |R(y)-R(y')|<\delta, \ \forall y': \, |y'-y|<d_y.
    \end{equation}
It is always possible to find such $d_y$ due to the continuity of $R(y)$.
\par The size of a small neighbourhood can be estimated as
    \begin{equation}
        d_y \sim \frac{\delta}{R'(y)} .
    \end{equation}
\par These neighbourhoods cover $\mathbb{C} \setminus (U^{(0)} \cup U^{(\infty)})$, and since it is compact, we can choose finitely many points $y^{(1)}$, $\ldots$, $y^{(M)}$ so that 
    \begin{equation}
        U^{(\infty)}\cup U^{(0)}\cup \bigcup U^{(l)}=\mathbb{C},
    \end{equation}
    where $U^{(l)}=U^{(y_l)}$.

    \par Now consider any $y\in\mathbb{C}$, and let $n\in\mathbb{N}$ be so large that $y-n \epsilon/A\in U^{(\infty)}$. Let us see which neighbourhoods the line connecting these two points crosses.   
    \par We can always choose a partition 
    \begin{equation}
        n=n_1+\ldots+n_k
    \end{equation}
    so that for each $j=1,\ldots,k$ there is $l_j\in \{0,1,\ldots,M,\infty\}$ such that 
    \begin{equation}
        y-m \epsilon/A\in U^{(l_j)}\ \mathrm{if}\ \sum_{i=1}^{j-1} n_i \leq m \leq \sum_{i=1}^{j} n_i.
    \end{equation}
    In other words, if you make $\sum_{i=1}^{j-1} n_i$ steps, you will arrive at the neighbourhood $U^{(l_j)}$ and certainly will not leave it in the next $n_j$ steps. Note that if $A\to \infty$, the number of steps $n_j$ which we can compute within $U^{(l_j)}$ can be very large. Indeed,
    \begin{equation}
        n_j\sim d^{(l_j)} \frac{A}{\epsilon},
    \end{equation}
    where $d^{(l_j)} =d_{y^{(l_j)}}$.
    \par Applying (\ref{eqn:comprule}) multiple times, we can decompose the map in (\ref{eqn:solutionomega}) as
    \begin{equation}\label{eqn:rho-decompose}
        \rho_{y-n \frac{\epsilon}{A}}^{n_1+\ldots+n_k}=\rho_{y_1}^{n_1} \circ \rho_{y_2}^{n_2}  \circ  \cdots   \circ \rho_{y_k}^{n_k},
    \end{equation}
    where
    \begin{equation}
        y_m=y-\frac{\epsilon}{A}\sum_{j=1}^{m}n_j.
    \end{equation}
    Then each $\rho_{y_j}^{n_j}$ depends only on values of $R$ in $U^{(l_j)}$. For $l_j \neq 0$ variation of $R(y)$ within $U^{(l_j)}$ does not exceed $\delta$ by construction, hence the analysis developed above can be applied.  

    \par To make quantitative predictions, we have to take into account two competing requirements. On the one hand, the neighbourhoods should be big enough to contain a large enough number of steps. Namely, adding an index $j$ enumerating the neighbourhoods to $\varepsilon_{1,2}$ of the bound (\ref{eqn:eps12-est}) we have the condition
    \begin{equation} \label{eqn:cond1}
        |\mathrm{ln}(\varepsilon_1^{(j)} \varepsilon_2^{(j)})| \sim n_j |\mathrm{ln}|\omega_0(y^{(l_j)})|| \sim \frac{\delta}{R'(y^{(l_j)})} \frac{A}{\epsilon} |\mathrm{ln}|\omega_0(y^{(l_j)})||\gg 1
    \end{equation}
    ensuring that the bound (\ref{eqn:Xnest}) is close to its saturated value.
    \par On the other hand, we want the saturation radius to be relatively small too, so
    by (\ref{eqn:noise-rs-err}) and  (\ref{eqn:U0thickness}) we require
\begin{equation} \label{eqn:cond2}
    \frac{2 \delta}{D_{\omega_0}}|\omega_0| \ll |\omega_0|.
\end{equation}
Note that this is an upper bound on $\delta$, so, in contrast to (\ref{eqn:cond1}), it forces the neighbourhoods to be small.
\par To make (\ref{eqn:cond1}) and (\ref{eqn:cond2}) compatible, we have to demand that 
\begin{equation} \label{eqn:thecondition}
        \frac{\epsilon R'(y^{(l_j)})}{A  |\mathrm{ln}(|\omega_0|)|D_{\omega_0}}\ll 1.
 \end{equation}
 
    \par In other words, as expected, $\rho_{y_j}^{n_j}$ indeed maps almost all of the Riemann sphere to a small neighbourhood of $\omega_0(y^{(l_j)})$ if $A$ is large, $R$ does not change too fast, and $\omega_0$ does not approach the unit circle. It is interesting that near the stable (but not exponentially attracting) points $\omega_0=\pm 1$ the convergence is even slower than at the unstable points. This can be explained by the fact that these are the branching points of $\omega_0$, where the behaviour of the function changes drastically.

    \par Let us assume that (\ref{eqn:thecondition}) holds in every $U^{(l)}$, $l \in \{ 1, \ldots,M \}$. Can we guarantee that the solution (\ref{eqn:solutionomega}) at a generic point $y$ is close to $\omega_0(y)$ in this case?
    \par
\begin{figure}
    \centering
    \begin{tikzpicture}
        \tikzmath{\n = 3; }

        \node[draw=none,minimum size=4cm,regular polygon,regular polygon sides={\n}] (b) {};

        \foreach \x in {1,2,3}
           \coordinate[label=below left: $\mathcal{C}_{\x}$] (o) at (b.corner \x);

       \node[draw=none,minimum size=2cm,regular polygon,regular polygon sides={\n}] (c) {};

        \foreach \x in {1,3}
          \draw[dashed] (b.corner \x) -- ++(20:3cm);
    
       \foreach \x in {1,3}
          \draw[dashed] (c.corner \x) -- ++(20:3.5cm) ;
                  
        \draw[dashed] (b.corner 2) -- ++(20:6cm);

        \draw[dashed] (c.corner 2) -- ++(20:5cm);

        \coordinate (endb 1) at ($(b.corner 1) + (20:3cm)$);
        \coordinate (endb 2) at ($(b.corner 2) + (20:6cm)$);
        \coordinate (endb 3) at ($(b.corner 3) + (20:3cm)$);
        
        \coordinate (endc 1) at ($(c.corner 1) + (20:3.5cm)$);
        \coordinate (endc 2) at ($(c.corner 2) + (20:5cm)$);
        \coordinate (endc 3) at ($(c.corner 3) + (20:3.5cm)$);

     \foreach \x in {1,2,3}
        \shade[shading=axis, left color=orange, right color=white]
       (b.corner \x) -- (c.corner \x) -- (endc \x) -- (endb \x) --   (b.corner \x);

    \foreach \x in {1,2,3}
            \draw[rotate=360/4+360*(\x-1)/\n,black] (b.corner \x) arc (0:360:0.5cm and 0.03cm) ;

    \draw[->] ($(b.corner 2)+(90:0.8cm)+(200:0.5cm)$) -- ++(20:2.5cm);

    \node at  ($(b.corner 2)+(90:0.7cm)+(20:1cm)$) [above left] {$\frac{\epsilon}{A}$} ;


    
    \end{tikzpicture}
     \caption{Singular areas and their shadows}     
     \label{fig:shadows}
 \end{figure}
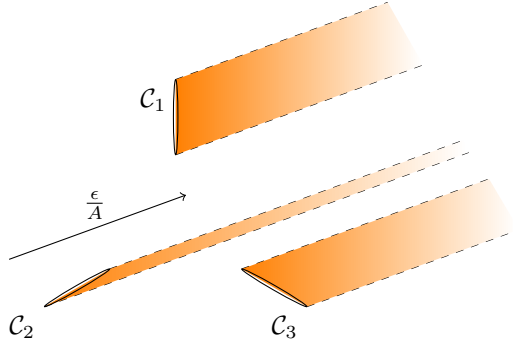

 {Let us assume that } a ray from $y$ toward the directed infinity $-\frac{\epsilon}{A}\infty$ does not cross the neighbourhood of singularities $U^{(0)}$. In other words, we assume that $y$ is outside of the shadow of the singular area that appears under imaginary parallel light rays in the $\epsilon/A$ direction (see Fig. \ref{fig:shadows}). Then the composition (\ref{eqn:rho-decompose}) can be understood as slow adjustment of stable maps, and one can inductively show that  
    \begin{equation} \label{eqn:adjustment}
        |\rho_{y_k}^{n_j}\circ \cdots \circ \rho_{y_2}^{n_2}\circ \rho_{y_1}^{n_1}(z)-\omega_0(y)|\lesssim  \frac{2 \delta  |\omega_0(y)|}{ D_{\omega_0}(y)} ,
    \end{equation}
    where
    \begin{equation}
        z= \omega(y_\infty)=\lim_{n\to +\infty}\rho^{n}_{y_\infty-n\epsilon/A}(\sqrt{q})
    \end{equation}
and $y_\infty \in U^{(\infty)}$ is a point on the ray. We already showed that $z$ is close to $\omega_0(y_\infty)$, and the key idea to show (\ref{eqn:adjustment}) is that $\omega_0(y_k)$ is always far enough from $\omega_0(y_{k+1})^{-1}$, so we never leave the stable region and can be sure that (\ref{eqn:solutionomega}) is close to $\omega_0(y)$. 

   \par  If we meet a singular neighbourhood along our way, but the point $y$ is far from the singular set $C_{\mathrm{sing}}$, we still can be \textit{almost sure} that $\omega(y)$ is close to $\omega_0(y)$. The only case when for $j$ being the smallest number such that $l_j=0$ (i.e. $y_j$ is the last singular point on the path from the infinity to $y$, see Fig. \ref{fig:shadows}) holds
    \begin{equation}\label{eqn:bad-point}
        |\rho_{y_k}^{n_j}\circ \cdots \circ \rho_{y_2}^{n_2}\circ \rho_{y_1}^{n_1}(z)-\omega_0(y_j)^{-1}|\lesssim   \frac{2 \delta}{D_{\omega_0}}|\omega_0| 
    \end{equation}
    This is a very strong condition, so, unless we choose the singularities of $R$ and the ratio $\epsilon/A$ in a very special way, it is safe to expect that 
    \begin{equation}
        |\omega(y)-\omega_0(y)|\lesssim     \frac{2 \delta}{D_{\omega_0}}|\omega_0| 
    \end{equation}
    everywhere except for the set of zero measure.

\section{Quantum Seiberg-Witten curve is perturbatively algebraic} \label{app-pertalg}
In this Appendix, we demonstrate that to all orders in perturbation theory the Quantum Seiberg-Witten equation (\ref{eqn:Rstructure}), considered with the boundary condition (\ref{eqn:omegaexp-lead}), reduces to an algebraic equation, which further reduces to a quadratic equation.
\par
We start by demonstrating that for every $l\in\mathbb{N}_0$ we have
\begin{equation}\label{eqn:structure-allgebra}
    \omega(y)+\frac{1}{\omega(y)}=R(y)+\sum_{n=1}^{l}R_n(\omega(y),y)\left(\frac{\epsilon}{A}\right)^{n}+O\left(\left(\frac{\epsilon}{A}\right)^{l+1}\right).
 \end{equation}
  where for each $n$, $R_n$ is a rational function independent of $\epsilon/A$. Moreover, we claim that for every $l$ and any $r\in\mathbb{N}$ the following holds
 \begin{equation}\label{eqn:structure-allgebra-der}
     \omega^{(r)}(y)=\sum_{n=0}^{l-1}D_n^r(\omega(y),y)\left(\frac{\epsilon}{A}\right)^{n}+O\left(\left(\frac{\epsilon}{A}\right)^{l}\right),
 \end{equation}
 where $D_n^{r}$ are again rational functions. Note that the meaning of the index $l$ in (\ref{eqn:structure-allgebra}) and (\ref{eqn:structure-allgebra-der}) is different: in the former it denotes the highest order taken into account, while in the latter it is the lowest omitted order. It is done to simplify the proof. 
 \par 
 We prove it by induction on $l$ and $r$. Clearly, (\ref{eqn:structure-allgebra}) holds for $l=0$. Moreover, we can formally say that (\ref{eqn:structure-allgebra-der}) holds for $l=0$, meaning simply that $\omega^{(r)}=\mathcal{O}(1)$, \textit{i.e.} is bounded when $\epsilon/A\to 0$. 
 \par 
 Now, we assume that both (\ref{eqn:structure-allgebra}) and (\ref{eqn:structure-allgebra-der}) hold for some $l=l_0$ and intend to prove them for $l=l_0+1$. Let us start from  (\ref{eqn:structure-allgebra-der}). Differentiating (\ref{eqn:structure-allgebra}) $r$ times with $l=l_0$  with respect to $y$ we get
\begin{equation}\label{eqn:omega-rec}
    \omega^{(r)}(y)\big(1-\frac{1}{\omega(y)^2}\big)+\big(\mathrm{lower\ derivatives}\big)=R^{(r)}(y)+\sum_{n=1}^{l_0}\frac{\dd^r}{\dd y^r}R_n(\omega(y),y)\left(\frac{\epsilon}{A}\right)^{n}+O\left(\left(\frac{\epsilon}{A}\right)^{l_0+1}\right).
\end{equation}
Here the second term in the left-hand side is a rational function of derivatives of $\omega(y)$ of lower orders. 
In the right-hand side the first term is known, while the rest are proportional to positive powers of $\epsilon/A$, so we can substitute them up to $O\left(\left(\frac{\epsilon}{A}\right)^{l_0}\right)$ without losing anything. Thus, we can use (\ref{eqn:structure-allgebra-der}) with $l=l_0$, which we have assumed to hold. Therefore, the right-hand sides can be represented as a polynomial of $\epsilon/A$ with the coefficients being the rational functions of $\omega(y)$ and $y$. Now, if $r=1$, there are no lower order derivatives in the left-hand side, so we just get desired expression for $\omega'(y)$ in the form (\ref{eqn:structure-allgebra-der}) (with $l=l_0+1$) by dividing both sides of (\ref{eqn:omega-rec}) by $1-\frac{1}{\omega(y)^2}$. It is easy to see that the case of higher $r$ follows by induction on $r$.
\par
Finally, we are ready to show that (\ref{eqn:structure-allgebra}) holds for $l=l_0+1$. For that we write
\begin{equation}\label{eqn:structure-orderl0+1}
    \omega(y)+\frac{1}{\omega(y)}+\sum_{t=1}^{l_0+1}\frac{\omega^{(t)}(y)}{t!} \left(-\frac{\epsilon}{A}\right)^{t}+\mathcal{O}\left(\left(\frac{\epsilon}{A}\right)^{l_0+2}\right)=R(y) ,
\end{equation} 
and note that it is enough to substitute the derivatives in the right-hand side up to $\mathcal{O}\left(\left(\frac{\epsilon}{A}\right)^{l_0+1}\right)$. By (\ref{eqn:structure-allgebra-der}), already proven for $l=l_0+1$, they all are rational functions of $\omega(y)$ and $y$, so the same is true for the right-hand side of (\ref{eqn:structure-orderl0+1}). This proves (\ref{eqn:structure-allgebra}) for $l=l_0+1$, and thus for any $l$ by induction. 




So, in each order of the perturbation theory we get an algebraic equation for $\omega(y)$. Moreover, it can be further reduced to a quadratic equation. To see this, note that the relation
\begin{equation}
    \omega(y)^2=R(y)\omega(y)-1+\omega(y)\sum_{n=1}^{\infty}R_n(\omega(y),y)\left(\frac{\epsilon}{A}\right)^{n},
\end{equation}
allows to reduce any polynomial in terms of $\omega(y)$ to a linear one order by order.
\paragraph{Remark.} The fact that in all orders of the perturbation theory (\ref{structure}) reduces to a quadratic equation implies that it has two solutions, corresponding to two solutions of the leading order equation (\ref{eqn:Rstructure-lead}). These solutions are selected by the boundary conditions $\lim_{y\to \infty} \omega(y)=q^{\pm \frac{1}{2}}$. Moreover, it is straightforward to see from the analysis of Appendix \ref{app-rec} that these solutions correspond to two possible ways to solve (\ref{structure}) in terms of generalised continued fraction, by expressing step by step $\omega(y)$ via $\omega(y\mp n \epsilon)$ and taking $n\to \infty$. Finally, we recall that, although only one branch of $\omega_0$ is needed for practical computations in Seiberg-Witten theory, the full Seiberg-Witten curve is a Riemann surface consisting of two sheets, corresponding to two branches of $\omega_0$. In other words, in the classical Seiberg-Witten theory the two solutions of (\ref{structure}) are equally meaningful. On the other hand, in our case only one solution has direct interpretation. It could be interesting to find an interpretation for the second solution. One of the possible hints is that formally it solves the saddle point equation with $q$ replaced by $\frac{1}{q}$. Recall that this is how the $S$-duality acts on the ultraviolet coupling $q$ (see Subsection \ref{subsubsec:modularprop}).
\paragraph{Remark.} In Subsection \ref{subsec:next} we used a modified version of the algorithm above, including a redefinition of the function $\omega$ according to (\ref{eqn:omegaredef}) in order to eliminate higher derivatives. In general, it is easy to see that the highest derivative in each order of the perturbation theory can be eliminated. Moreover, any term of the form
\begin{equation}
    \omega^{(n)}(z)h(\omega(z),\omega'(z),\ldots,\omega^{(n-1)}(z))
\end{equation}
can be replaced with an expression depending only on the derivatives of orders up to $n-1$.

    \section{Computation of integrals}\label{app-comp}
    Our goal is to compute integrals of the following types:
    \begin{equation}
       2\pi\ii \, \mathcal{I}_{\alpha}(\kappa)=\oint_{|z|=r}\left(1-\frac{1}{\kappa} \left(z+\frac{1}{z}\right)\right)^{\alpha}\frac{dz}{z},
    \end{equation}
    and 
    \begin{equation}
       2\pi\ii \, \mathcal{J}_{\alpha}(\kappa)=\oint_{|z|=r}\left(1-\frac{1}{\kappa} \left(z+\frac{1}{z}\right)\right)^{\alpha}\frac{z+\frac{1}{z}}{\left(z-\frac{1}{z}\right)^4}\frac{dz}{z}.
    \end{equation}
    Here $r\in (r_0,1)$, $r_0\in(0,1)$ is chosen so that the contour does not intersect the cut of the integrand, and $\kappa$ is a parameter which we can assume to be sufficiently large (recall that we need these integrals for $\kappa=\sqrt{q}+\frac{1}{\sqrt{q}}$, where we can assume that $q$ is sufficiently small). 
    Clearly,  the particular value of $r$ is not important.
    
    The technique we describe here can be applied to more general integrals of this type appearing in higher orders of the perturbation theory, but for the sake of simplicity we concentrate only on these two.
    \subsection{Elementary integrals}
    Our strategy will be to use Taylor expansion around $\kappa\to \infty$. We begin with the computation of elementary integrals which we will meet on this way. The integrals appearing in $\mathcal{I}_{\alpha}$ are simple:
    \begin{equation}
        \oint_{|z|=r}\left(z+\frac{1}{z}\right)^k \frac{dz}{z}=2\pi\ii \sum_{l=0}^k \frac{k!}{l!(k-l)!}\mathrm{res}_{z=0}z^{2l-k-1},
    \end{equation}
    so
    \begin{equation}
        \oint_{|z|=r}\left(z+\frac{1}{z}\right)^{2k} \frac{dz}{z}=2\pi\ii\frac{(2k)!}{(k!)^2}, \qquad   \oint_{|z|=r}\left(z+\frac{1}{z}\right)^{2k+1}=0 \ (k\in \mathbb{N}_0).
    \end{equation}

    The other one is slightly more involved. To compute
    \begin{equation}
        J_k=\oint_{|z|=r}\frac{\left(z+\frac{1}{z}\right)^k}{\left(z-\frac{1}{z}\right)^4}\frac{dz}{z}
    \end{equation}
    we introduce a parameter $t$ such that 
    \begin{equation}\label{eqn:app2-test}
        |t|<\frac{r}{2}.
    \end{equation}
    \begin{equation}
        |t|\left|z+\frac{1}{z}\right|\leq |t|\left(r+\frac{1}{r}\right)<1.
    \end{equation} 
    Here we use the fact that $r<1$. 
    Then we can write 
    \begin{equation}\label{eqn:app2-Jt}
        J(t)=\sum_{k=0}^{\infty}J_k t^k=\oint_{|z|=r}\frac{ dz}{z\left(z-\frac{1}{z}\right)^4\left(1-t\left(z+\frac{1}{z}\right)\right)}=-\oint_{|z|=r}\frac{dz}{\left(z-\frac{1}{z}\right)^4 t(z-z_-)(z-z_+)},
    \end{equation}
    where $z_{\pm}$ are the solutions of the equation
    \begin{equation}\label{eqn:app2-poles}
        z+\frac{1}{z}=\frac{1}{t}.
    \end{equation}
    Note that by (\ref{eqn:app2-test}) we have $\left|\frac{1}{t}\right|>2$, thus  (\ref{eqn:app2-poles}) has two distinct solutions, and we can always enumerate them so that $|z_{-}|<1$ (see Appendix \ref{app-rec}). Then $|z_{+}|=\frac{1}{|z_-|}>1$, hence the pole at $z_+$ definitely lies outside of the contour of integration. To see that $z_-$ lies inside the contour, observe that
    \begin{equation}
        \frac{2}{r}\leq \frac{1}{|t|}=\left|z_-+\frac{1}{z_-} \right|\leq |z_-|+\frac{1}{|z_-|}\leq 1+\frac{1}{|z_-|}\leq \frac{1}{r}+\frac{1}{|z_-|},  
    \end{equation}
    so $\frac{1}{|z_-|}\geq  \frac{1}{r}$, or $|z_-|<r$. We conclude that
    \begin{equation}
        J(t)=-2\pi \ii\, \mathrm{res}_{z=z_-}\frac{1}{\left(z-\frac{1}{z}\right)^4 t(z-z_-)(z-z_+)}=2\pi \ii \frac{1}{t(z_{-}-z_{+})^5}
    \end{equation}
    Explicitly we have
    \begin{equation}
        z_{\pm}(t)=\frac{1\pm\sqrt{1-4t^2}}{2t},
    \end{equation}
    where the square root in the right-hand sides is the principal branch, and
        \begin{equation}
        J(t)=2\pi \ii t^{4}(1-4t^2)^{-\frac{5}{2}}=
        2\pi \ii \frac{t^4}{6} \sum_{k=0}^{\infty}\frac{(2k+3)!}{k!(k+1)!}t^{2k}.
    \end{equation}
    Comparing with (\ref{eqn:app2-Jt}), we conclude that
    \begin{equation}
        J_{2k+4}=2\pi \ii \frac{1}{6}\frac{(2k+3)!}{k!(k+1)!},\ J_0=J_2=J_{2k+1}=0\ (k\in\mathbb{N}_0).
    \end{equation}
    \subsection{Computing \texorpdfstring{$\mathcal{I}_{\alpha}$ and $\mathcal{J}_{\alpha}$}{Ia and Ja}}
    Recall the Pochhammer  symbol
    \begin{equation}\label{eqn:pochh}
        (x)_n=\prod_{j=0}^{n-1} (x+j).
    \end{equation}
   Using this notation, and for sufficiently large $\kappa$ (it suffices to assume $|\kappa|>\frac{2}{r_0}$) 
    \begin{equation}
        \mathcal{I}_{\alpha}(\kappa)=\frac{1}{2\pi\ii}\sum_{k=0}^{\infty}\frac{(-\alpha)_{k}}{k!}\kappa^{-k}I_k=\frac{1}{2\pi\ii}\sum_{k=0}^{\infty}\frac{(-\alpha)_{2k}}{(2k)!}\kappa^{2k}I_{2k}=\sum_{k=0}^{\infty}\frac{(-\alpha)_{2k}}{(k!)^2}\kappa^{-2k}={}_2 F_1(-\frac{\alpha}{2},\frac{1-\alpha}{2};1;\frac{4}{\kappa^2})
    \end{equation}
    and
    \begin{eqnarray}
        \mathcal{J}_{\alpha}(\kappa)=\frac{1}{2\pi\ii}\sum_{k=0}^{\infty}\frac{(-\alpha)_{k}}{k!}\kappa^{-k}J_{k+1}=\frac{1}{2\pi\ii}\sum_{k=0}^{\infty}\frac{(-\alpha)_{2k+3}}{(2k+3)!}\kappa^{-2k-3}J_{2k+4} \nonumber \\ =
   \frac{1}{6} \sum_{k=0}^{\infty}\frac{(-\alpha)_{2k+3}}{k!(k+1)!}\kappa^{-2k-3}=   -\frac{1}{6}\frac{(\alpha-2)_3}{\kappa^3} {}_2 F_1\left(\frac{3-\alpha}{2},\frac{4-\alpha}{2},2,\frac{4}{\kappa^2} \right).
    \end{eqnarray}

\end{document}